\documentclass[iop]{emulateapj}
\usepackage{epstopdf}

\def\p{\partial}

\def\nab{\mbox{\boldmath $\nabla$}}
\def\rb{\bar{\rho}}
\def\tb{\bar{T}}
\def\sb{\bar{S}}
\def\vrr{\tilde{v}_r} % rms velocities
\def\vtr{\tilde{v}_{\theta}}
\def\vphr{\tilde{v}_{\phi}}
\def\vvr{\tilde{v}}
\newcommand{\rh}{\bar{\rho}}
\newcommand{\bdot}{\mbox{\boldmath $\cdot$}}
\newcommand{\del}{\mbox{\boldmath $\nabla$}}
\newcommand{\dv}{\mbox{\boldmath $\nabla \bdot$}}
\newcommand{\calf}{\mbox{\boldmath ${\cal F}$}}
\def\p{\partial}
\def\rbi{\frac{1}{\bar{\rho}}}
\def\Om{\mbox{\boldmath $\Omega_*$}}
\newcommand{\curl}{\mbox{\boldmath $\nabla \times$}}

\newcommand{\vort}{\mbox{\boldmath $\omega$}}
\newcommand{\uvr}{\mbox{\boldmath $\hat{\bf e}_r$}}
\newcommand{\uvt}{\mbox{\boldmath $\hat{\bf e}_\theta$}}

\def\pomega{\varpi}

\newcommand{\lrsp}{\textit{LRSP} }
\newcommand{\jfm}{\textit{JFM} }

\usepackage{color}
\definecolor{darkgreen}{RGB}{0,142,128}
\definecolor{darkblue}{RGB}{0,100,170}
\definecolor{gold}{RGB}{218,165,32}

\usepackage{hyperref}
\hypersetup{colorlinks,citecolor=blue,linkcolor=blue}

\shorttitle{On differential rotation and overshooting in solar-like stars}
\shortauthors{Brun et al.}

\begin{document}

%\title{How is the differential rotation in solar-like stars influenced by the Rossby number?}
%\title{Turbulent convection and large scale flows in solar-like stars: the role of overshooting and Rossby number}
%\title{Turbulent convection and large scale flows in solar-like stars}
\title{On differential rotation and overshooting in solar-like stars}
%\title{On the various states of rotation of solar-like stars}

%preliminary order and list of authors
\author{Allan Sacha Brun$^{1}$, Antoine Strugarek$^{3,1}$, Jacobo Varela$^{1}$, Sean P. Matt$^{2,1}$, Kyle
  C. Augustson$^{1}$, Constance Emeriau$^{1}$, Olivier Long DoCao$^{1}$, Benjamin Brown$^{4}$, Juri Toomre$^{5}$}

\affil{$^1$ AIM, CEA/CNRS/University of Paris 7, CEA-Saclay, 91191 Gif-sur-Yvette, France\\
\and $^2$ Physics and Astronomy, University of Exeter, Stocker Road, EXA4 4QL Exeter, UK  \\
\and $^3$ Astronomy Dept., University of Montreal, Montreal, Canada \\
\and $^4$  Laboratory for Atmospheric and Space Physics and Department of Astrophysical \& Planetary Sciences, University of Colorado, Boulder, Colorado 80309, USA\\
\and $^5$ JILA, University of Colorado, Boulder, CO 80309, USA
}

\begin{abstract}
We seek to characterize how the change of global rotation rate influences the overall dynamics and large scale flows arising in the 
convective envelopes of stars covering stellar spectral types from early G to late K. 
We do so through numerical simulations with the ASH code, where we consider stellar convective envelopes coupled to a radiative interior with various global properties. 
As solar-like stars spin down over the course of their main sequence evolution, such change must have a direct impact on their dynamics and rotation state. 
We indeed find that three main states of rotation may exist for a given star: anti-solar-like (fast poles, slow equator), solar-like (fast equator, slow poles), or a cylindrical rotation profile. 
Under increasingly strict rotational constraints, the latter profile can further evolve into a Jupiter-like profile, with alternating prograde and retrograde zonal jets. 
We have further assessed how far the convection and meridional flows overshoot into the radiative zone and investigated the morphology of the established tachocline. 
Using simple mixing length arguments, we are able to construct a scaling of the fluid Rossby number $R_{of} = \tilde{\omega}/2\Omega_* \sim \tilde{v}/2\Omega_* R_*$, 
which we calibrate based on our 3-D ASH simulations. We can use this scaling to map the behavior of differential rotation versus the global parameters of stellar 
mass and rotation rate. Finally, we isolate a region on this map ($R_{of} \gtrsim 1.5-2$) where we posit that stars with an anti-solar differential rotation may exist 
in order to encourage observers to hunt for such targets.
\end{abstract}


\section{Introduction} \label{sec_intro}

Most solar-like stars possess a deep turbulent convective envelope. Understanding their dynamical
and nonlinear properties is fundamental in order to characterize the transport of heat, energy and
angular momentum in these stars.  Of particular importance are the large-scale flows, like the
differential rotation and meridional circulation, that are achieved in these turbulent convective
envelopes. Indeed, such flows are considered as essential ingredients to explain stellar magnetism
\citep{Weiss:1994tz,Saar:1999dg,Jouve:2010jl}. For instance, the differential rotation can
  convert poloidal magnetic fields into toroidal fields via the so-called $\Omega$-effect
  \citep{Parker:1955km,Moffatt:1978tc}.  Such toroidal magnetic fields are possibly at the origin of
  star spots \citep{Parker:1955gc}. Furthermore, the meridional circulation can transport both the
  surface poloidal field from low latitudes to the pole as well as the magnetic field at the base of
  the convective envelope in the so-called flux transport solar models
  \citep{1989Sci...245..712W,2001ApJ...547..475S,Charbonneau:2005il,2011ApJ...733...90D,Brun:2015kca},
  some of which consider self-consistent dynamo action.

Despite substantial efforts, there are still many unknowns related to how the large-scale flows and stellar
  convection change with global stellar parameters such as rotation $\Omega_*$ and mass $M_*$. Over the years, several
observational campaigns have been pursued using either ground based or space-born instruments, to derive useful
constraints on rotation of solar like stars \citep[see for instance the recent studies by][and references
therein]{Bouvier:2013cz,Reinhold:2013eo,doNascimento:2014is,Garcia:2014ds,Reinhold:2015kx}.  What is clear from these
observational studies is that young solar-like stars are fast rotators with rotational periods on the order of days and
old stars are slow rotators, with periods closer to months. This trend of increasing rotation period, or decreasing
rotation rate $\Omega_*$, is clearly seen on the main sequence. Indeed, as was first suggested by
\citet{Skumanich:1972fq}, it appears that $\Omega_*\propto t^{-1/2}$, with $t$ the stellar age.  

We have a relatively good understanding about how the rotational braking of solar-like stars occurs. It is due to the
continuous action of a stellar wind driven by thermal pressure taking away mass and angular momentum from the aging star
\citep{Parker:1958dn,Schatzman:1962vc,Weber:1967kx,Kawaler:1988fi,Matt:2012ib}.  The relation between stellar age and
rotation has been termed {\it gyrochronology} \citep{Barnes:2003ga,barnes2010,Meibom:2015if} and is a useful marker of
stellar evolution.  Note however that in their recent asteroseismic study with data coming from the {\it Kepler}
satellite, \citet{vanSaders:2016cr} have questioned Skumanich's law and the accuracy of {\it gyrochronology} for
solar-like stars older than the Sun (e.g., $t > 4.5 Gyr$), but these techniques have their own accuracy issues
\citep{Aigrain:2015cc}.  Associated with this decreasing influence of rotation is a lower level of stellar activity
\citep{Wilson:1978is,Noyes:1984bp,Pizzolato:2003ga,Reiners:2012ug}, also called {\it magnetochronology}
\citep{Vidotto:2014ba,Folsom:2016dl}.  There is thus a positive feedback loop between rotation, magnetism and wind in
solar-like stars that is also important to characterize \citep{Brown:2014el,Meibom:2015if,Matt:2015cb,Reville:2015bb}.

It is to be expected that the global rotation rate of a star will influence its large-scale mean
flows. How those flows vary with rotation rate is still poorly known.  Observational studies of
large-scale flows in solar-like stars are rare and even less accurate than studies of the
stellar rotation rate itself. Most studies have focused on the surface differential rotation, for
the meridional circulation is too weak to be detected. However, several interesting trends for the
differential rotation in solar-like stars have been uncovered
\citep{Donahue:1996ch,Barnes:2005eb,CollierCameron:2007ce,Saar:2009tb}. First, the observed
  differential rotation $\Delta \Omega$ appears to increase with stellar mass, or more precisely
  with a higher effective temperature.  Second, some variations of the surface differential rotation
  have been found with rotation rate: $\Delta \Omega \propto \Omega_*^n$, with $n$ a positive
  exponent. However, observers currently disagree on the exact value of the exponent
$n$. \citet{Donahue:1996ch}, \citet{2003A&A...409.1017M}, \citet{Saar:2009tb} advocate for a value close to 0.6-0.7, whereas
\citet{Barnes:2005eb} and \citet{CollierCameron:2007ce} argue for a much smaller value around 0.15, implying
only a weak dependence on $\Omega_*$. Studies based on asteroseismic data have more recently started providing trends
that somewhat lie in between whose values ($n \sim 0.3$ in \citet{2015A&A...583A..65R}; see also  
\citep{Reinhold:2013eo,Garcia:2014ds,2014ExA....38..249R,2015A&A...583A..65R,2016MNRAS.461..497B}). Actually, \citet{2016MNRAS.461..497B} 
even propose that the exponent $n$ is different for each spectral types, confirming a strong 
dependency of the differential rotation amplitude with effective temperature, massive stars having larger shear. Theoretical
interpretation are thus in order to explain those seemingly different trends.

  Another physical property of key interest in solar-like stars is the extent and amount of overshooting at the
    base of their convective envelope, where surrounding layers of extra mixing are triggered by convective plumes at
    the boundaries of convection zones \citep{1978A&A....65..281R,Zahn:1991uz}.  It is quite unknown how the overshoot of convection varies
  with global stellar parameters such as mass or rotation rate, yet it has a direct impact on mixing and transport of
  chemicals and magnetic fields.  The amount and extent of overshooting is usually chosen to be proportional to the
  local pressure scale height and to be a small fraction of the convective heat flux \citep{1989A&A...210..155M,Zahn:1991uz,Browning:2004bx}. 
  Those model parameters are often calibrated using stellar isochrones to match observations of surface chemical abundances 
  (see for instance \cite{1993A&AS...98..477M}). However, it has been shown
  that the overshooting depth can be overestimated if the stellar models used to compute the isochrones are neglecting
  other physical processes such as rotational mixing \citep{Ekstrom:2012ke}. Hence, we must find complementary means to
  constrain stellar overshooting.

  It is extremely difficult to observationally infer information about the meridional circulation and overshooting
    layers in stars, and the observational data for surface differential rotation are often inadequately robust.  Hence,
    theoretical developments and numerical simulations are useful to help characterize the large-scale mean flows and to
    assess the degree of coupling between convection and radiative zones.  For instance, multi-cellular meridional
    flows, which are now confirmed to exist within the Sun using local helioseismic analysis
    \citep{Haber:2002ib,MitraKraev:2007ej,Zhao:2013kz}, have been the natural outcome of solar convection 3-D
    simulations for many years \citep{Miesch:2000gs,Brun:2002gi,Aurnou:2007cp,Kapyla:2011kr,Gastine:2013fg,Guerrero:2013hb,Featherstone:2015bv}, long before they became accepted
    as a plausible solar flow pattern by the community.

  Thus, along side observational campaigns, several groups have attempted to make progress in understanding
    rotating convection and the establishment of large-scale flows in stars through numerical simulations. Many of these
    groups have benefited from the pioneering work of Gilman \& Glatzmaier where they modeled the largest scales in the
    Sun's convection zone \citep[see the series of papers listed in][]{Glatzmaier:1982io}. For such simulations, a full
    spherical geometry is required since we wish here to characterize a star's global-scale flows. Such global-scale
    simulations of stellar convection have been performed using various numerical codes based on either finite volumes
    or on spectral methods \citep[see][and references there in for a recent review]{Brun:2015co}. These various
    anelastic codes, such as ASH \citep{Clune:1999vd}, have been benchmarked internationally \citep{Jones:2011in}. They
    generally agree on the main properties of global rotating convection when identical setups and nondimensional
    numbers are chosen. These codes have been used to simulate very different type of stars: from massive ones
    \citep{Browning:2004bx,Featherstone:2009ft,Augustson:2016wr} to dwarf stars \citep{Dobler:2006ha,Browning:2008dn}
    and of course solar-like ones
    \citep{Ballot:2007ea,Brown:2008ii,Bessolaz:2011ih,Matt:2011jl,Kapyla:2011kr,Augustson:2012bn,Guerrero:2013hb,Fan:2014ct,Karak:2015dw}. It
    is found that large-scale flows are sensitive to the intensity of the convective driving and to the rotation rate of
    the simulation. Yet almost none of these studies have systematically taken into account the coupling to a stably
    stratified interior \citep[see][for counter-examples]{Miesch:2000gs,Guerrero:2016cz} and/or looked at the influence of the aspect
    ratio of the convection zone on the resulting convection and its mean flows, as we have simultaneously done in this work.

  Another very important motivation for running global convection simulations is to understand stellar magnetism. This
  implies the need to compute non-ideal magnetohydrodynamic (MHD) simulations of magnetized convection. In particular,
  the focus of many of these simulation has been to find and understand how convective dynamo solutions can become
  cyclic and to further analyze them in terms of mean-field dynamo theory. Recently, significant progress has been made
  in that direction with many global MHD solutions now possessing a cyclic behavior. We refer to the following
    recent papers for a discussion of stellar magnetism and how it may arise \citep[see for
    instance][]{Brun:2004ji,Browning:2006ba,Ghizaru:2010im,Brown:2010cn,Brown:2011fm,Racine:2011gh,Kapyla:2012dg,Augustson:2013jj,Kapyla:2013gr,Nelson:2013fa,Fan:2014ct,Karak:2015dw,Augustson:2015er,Lawson:2015fq,Simitev:2015fc,Guerrero:2016cz}. In\citet{Varela16},
    we have started computing and studying the MHD equivalent to the 15 simulations here, which cover four mass bins and
    several rotation rates. 

  In this paper, we report on novel 3--D numerical experiments in spherical geometry designed to investigate how the
  complex, nonlinear dynamics occurring in the convective envelope of solar-like stars changes with stellar parameters
  such as mass and rotation rate. Our approach differs from other recent studies listed above by taking into account the
  dynamical influence of a deep and stable radiative interior. It is complementary to the study of
    \citet{Guerrero:2013hb}, since we further consider various aspect ratios and stellar spectral types.  We also
  propose a model that utilizes a simple mixing length theory scaling to identify the possible states of differential
  rotation for various stellar spectral types.  This model is then calibrated using the set of 15 models discussed
  below.

The paper is organized as follows. In \S2, we derive a scaling relationship for the Rossby number based on mixing length
arguments. In \S 3 we describe our equations, numerical models, and the various ingredients employed to model four
  stars of differing spectral type at a selection of rotation rates. In sections 4, 5, and 6, we discuss the properties
of convection, penetration, and the large-scale flows of our models as well as derive scaling relationships for all key
quantities. In \S 7, we perform a detailed analysis of angular, energy, and heat transport in our simulations. Finally,
we conclude in \S 8. \\

\section{Hints from mixing length theory for states of stellar differential rotation}
\label{sec:hint_dr_ML}

It is worth noting that one can already guess by using simple mixing length scaling and 1-D stellar structure models the
outcome of the simulations in terms of the overall surface differential rotation by distinguishing two states: fast vs
slow equator rotation. Our 3-D numerical simulations are key as they will help us characterizing the differential
rotation profile as a function of depth (and latitude) and how the coupling to a radiative interior may tilt the
iso-contour of omega by so-called thermal wind effect \citep{Miesch:2006iz}.

One can make an educated guess of the differential rotation state realized in a star, by evaluating the convective
velocity from mixing length arguments \citep{Kippenhahn:1994tm,Augustson:2012bn,Brun:2015kca}:

\begin{equation}\label{v_mlt}
v = c_1 \left(\frac{L_*}{\rho_{bcz}R_*^2}\right)^{1/3}
\end{equation}

\noindent with the typical values taken for the stellar luminosity, radius, and density at the base of the convection
zone listed in Table \ref{tablespectraltype} for the stellar spectral range considered in this study, and with $c_1$ a proportionality factor.
Classical stellar evolution indicates that $L_*\sim M_*^4$ and $R_* \sim M_*^{0.9}$ \citep{Kippenhahn:1994tm}. Assuming
that $\rho_{bcz} \sim M_*^n$, with $n<0$ but undetermined for now, one directly sees that $v \sim M_*^{(2.2-n)/3}$.
Regression fits to the values listed in Table 2 allow us to obtain a scaling for $\rho_{bcz}$, and so we can refine the
stellar mass dependence of $L_*$ and $R_*$. In particular, we find that $\rho_{bcz} \sim M_*^{-6.9}$,
$L_*\sim M_*^{4.6}$, and $R_* \sim M_*^{1.3}$.  Replacing these scalings in equation \ref{v_mlt}, yields $v \sim M_*^3$.
As expected, more massive stars have faster convective flows.  Knowing how $v$ is expected to scale with stellar
mass, we can now compute an approximated fluid Rossby number $R_{of} = v/2\Omega_* R_* = c_1 M_*^{1.7}/\Omega_*$ as a
function of the stellar rotation rate $\Omega_*$. We choose to use the fluid Rossby number instead of the stellar or
convective ones. We defer the reader to Appendix \,\ref{sec:rossby-numbers} for a further discussion of the various
definitions of Rossby numbers.

\begin{figure}[!htb]
\begin{center}
\includegraphics[width=0.5\textwidth]{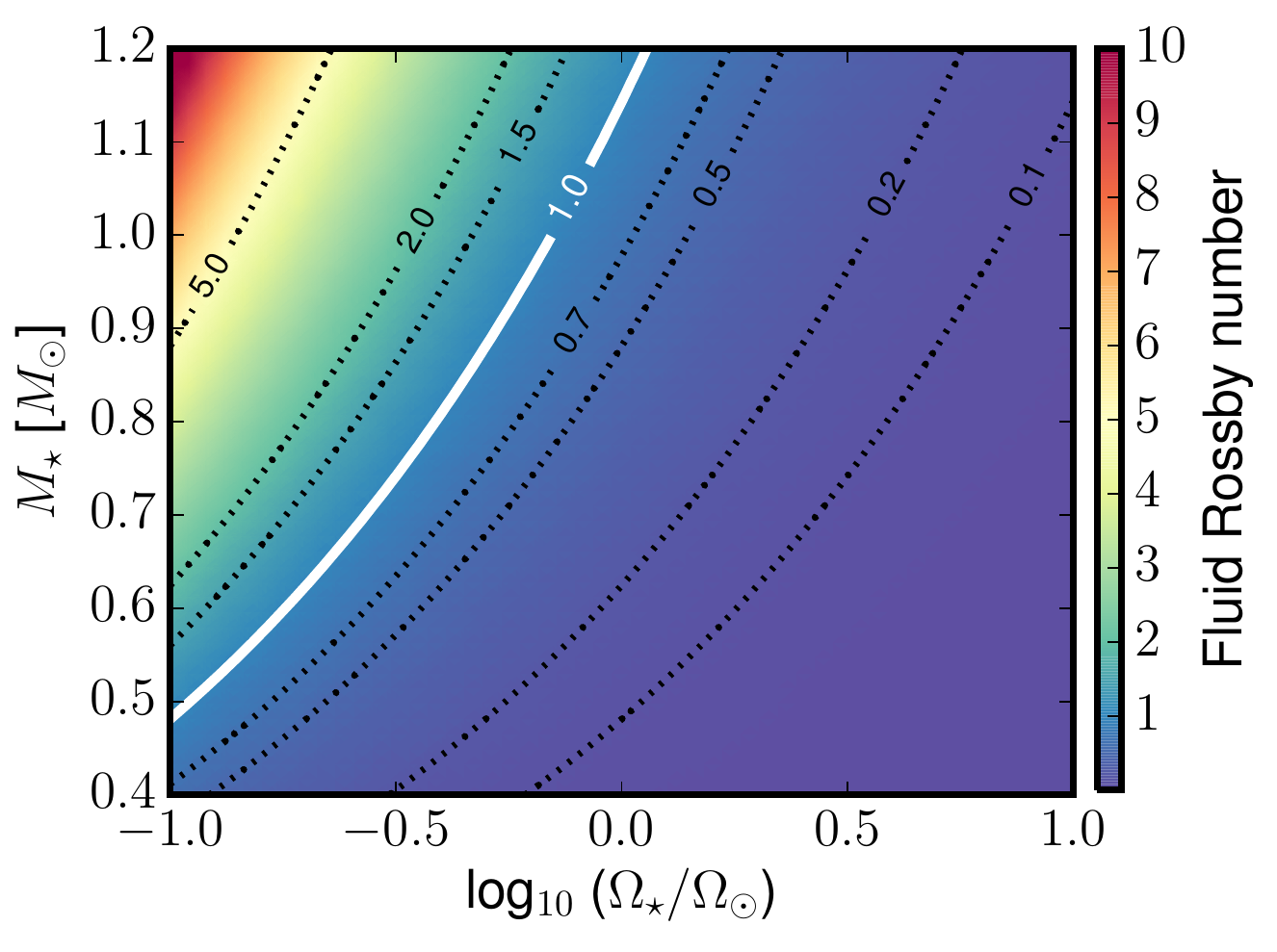}
\end{center}
\caption{Rossby number as a function of stellar rotation and mass as
  deduced from mixing length theory, assuming a constant $c_1=0.7$.}
\label{om_rossby}
\end{figure}

In Figure \ref{om_rossby}, we display the fluid Rossby number as a function of stellar rotation rate and mass. One can easily 
identify 2 regimes separated by the Rossby equals 1 white line: small Rossby number ($\lesssim 0.66$) and large Rossby number ($\gtrsim1.5$).  
  From the various studies published over the last decade
  \citep{Ballot:2007ea,Brown:2008ii,Bessolaz:2011ih,Augustson:2012bn,2013ASPC..479..285T,Gastine:2014jr,Kapyla:2014jr}, it is clear that
  simulations of rotating convective shells with small Rossby numbers often possess fast prograde flows at the equator and slowly rotating poles,  
  possibly ressembling a solar-like differential rotation.  In contrast, those simulations with a large Rossby number typically
  have retrograde flows at the equator and fast rotatin polar regions e.g. something like an anti-solar-like differential rotation. With $0.66 \lesssim R_{of} \lesssim 1.5$,
  it is more difficult to anticipate the result given the crudeness of our derivation, but one may expect that for
  $R_{of} \gtrsim 1$ the simulation will likely be slowly rotating at the equator. In the remainder of the paper we will use
  the following terminology for characterizing differential rotation profile: solar-like will mean fast equator, slow poles, anti-solar like
  will mean slow equator and fast poles, cylindrical, will mean that the iso-contours of $\Omega(r,\theta)$ are constant along cyclinders
  aligned with the rotation axis and Jupiter-like differential rotation will mean that the profile is cylindrical but non monotonic, 
  with alternance of prograde and retrogade jets.
  Analyzing this figure further, one might  expect that a 1.1 solar mass star could be anti-solar-like at a rotation rate around the solar rate.  Such a state may
  also occur for respectively 0.9, 0.7 and 0.5 solar mass stars at rotation rates around 0.6, 0.4, and 0.25 times the
  solar rate. One also sees that for a solar-mass star the solar rotation rate corresponds to a Rossby number less than
  unity and hence it likely has prograde equatorially, as observed. In the following, we will characterize further the
  profile of differential rotation and compare the outcome of nonlinear 3-D numerical simulations of rotating convection
  with this simple analysis based on the mixing length. We will in particular show that for low values of the Rossby
  number, there is a shift from a conical to cylindrical profile, where this third state is akin to Jupiter's
  alternating zonal jets.

\section{Modelling stars in 3-D with ASH}
\label{sec_3d_model_ash}

We present the simulation setup used to model the various spectral type stars considered in our study with the ASH code.

\subsection{Model Equations}
\label{sec_mod_eq}

We use the ASH code \citep[see][]{Clune:1999vd,Miesch:2000gs,Brun:2004ji} to model solar-like stars with mass ranging
from 0.5 to 1.1 $M_{\odot}$.  The domain of each of these simulations is large enough to encompass both a portion
  of the deep radiative zone and most of the overlying convective envelope that is representative of the targeted
  star. Hence, we are self-consistently capturing the nonlinear interactions that seamlessly couple those two zones. ASH
solves the full set of 3--D anelastic equations of motion in a rotating, convective and radiative spherical shell
utilizing massively-parallel computing architectures.  These equations are fully nonlinear in the velocity
variable. However, under the anelastic approximation, the thermodynamic variables are linearized with respect to a
spherically symmetric and evolving mean state having a density $\rb$, pressure $\bar{P}$, temperature $\tb$ and specific
entropy $\sb$.  Fluctuations about this reference state are denoted by $\rho$, $P$, $T$, and $S$.  The resulting
equations are \citep{Glatzmaier:1984jh,Clune:1999vd}:

\begin{eqnarray}
\nab\cdot(\rb {\bf v}) &=& 0, \\
\frac{\p {\bf v}}{\p t}+({\bf v}\cdot\nab){\bf v} &=& -\nab \pomega - \frac{S}{c_p} {\bf g} -2{\bf \Omega_*}\times{\bf v}  \label{eq:anelastic momentum}\\
- \frac{1}{\rb}\nab\cdot\mbox{\boldmath $\cal D$}&-&[\nab\bar{\pomega} + \bar{\pomega}\nab\ln\rb-{\bf g}], \nonumber \\
\rb \tb \frac{\p S}{\p t}&+&\rb \tb{\bf v}\cdot\nab (\sb+S)= \rb {\epsilon} \nonumber \\
+\nab\cdot[\kappa_r \rb c_p \nab (\tb+T)&+&\kappa \rb \tb \nab S+\kappa_0 \rb \tb \nab \sb] \\
&+&2\rb\nu\left[e_{ij}e_{ij}-1/3(\nab\cdot{\bf v})^2\right] \nonumber \, , 
\end{eqnarray}

\noindent where ${\bf v}=(v_r,v_{\theta},v_{\phi})$ is the local velocity in spherical coordinates in the frame rotating at
constant angular velocity ${\bf \Omega_*}$, ${\bf g}$ is the gravitational acceleration, $c_p$ is the specific heat per
unit mass at constant pressure, $\pomega=P/\rb$ is the reduced or kinematic pressure, $\kappa_r$ is the radiative diffusivity, 
and ${\bf \cal D}$ is the viscous stress tensor.  The components of ${\bf \cal D}$ are given by

\begin{eqnarray}
{\cal D}_{ij}=-2\rb\nu[e_{ij}-1/3(\nab\cdot{\bf v})\delta_{ij}]\, ,
\end{eqnarray}

\noindent where $e_{ij}$ is the strain rate tensor, and $\nu$, $\kappa$ and $\kappa_0$ are effective eddy diffusivities.
A volumetric heating term $\rb \epsilon$ is also taken into account to mimic generation of energy by nuclear reactions.
The nuclear reactions are modelled very simply by assuming that $\epsilon = \epsilon_0 \tb^{n_c}$.  By enforcing that
the integrated luminosity of the star match its known surface value, we can determine $\epsilon_0$ and $n_c$ as listed
in Table \ref{tablediff}. Note that only M05 and M07 series of models require that heating source term  since their
computational domain include a portion of the nuclear energy generation core.  
Here we solve the energy conserving anelastic equations, which have a momentum equation~(\ref{eq:anelastic momentum})
that takes a slightly different form than the compressible Navier-Stokes equations.  These anelastic equations 
have been shown to properly conserve energy in both convection zones and stably-stratified
regions like the tachoclines that we study here \citep{Brown:2012bd,Vasil:2013ij}.

\begin{figure}[!htb]
\begin{center}
\includegraphics[width=0.9\linewidth]{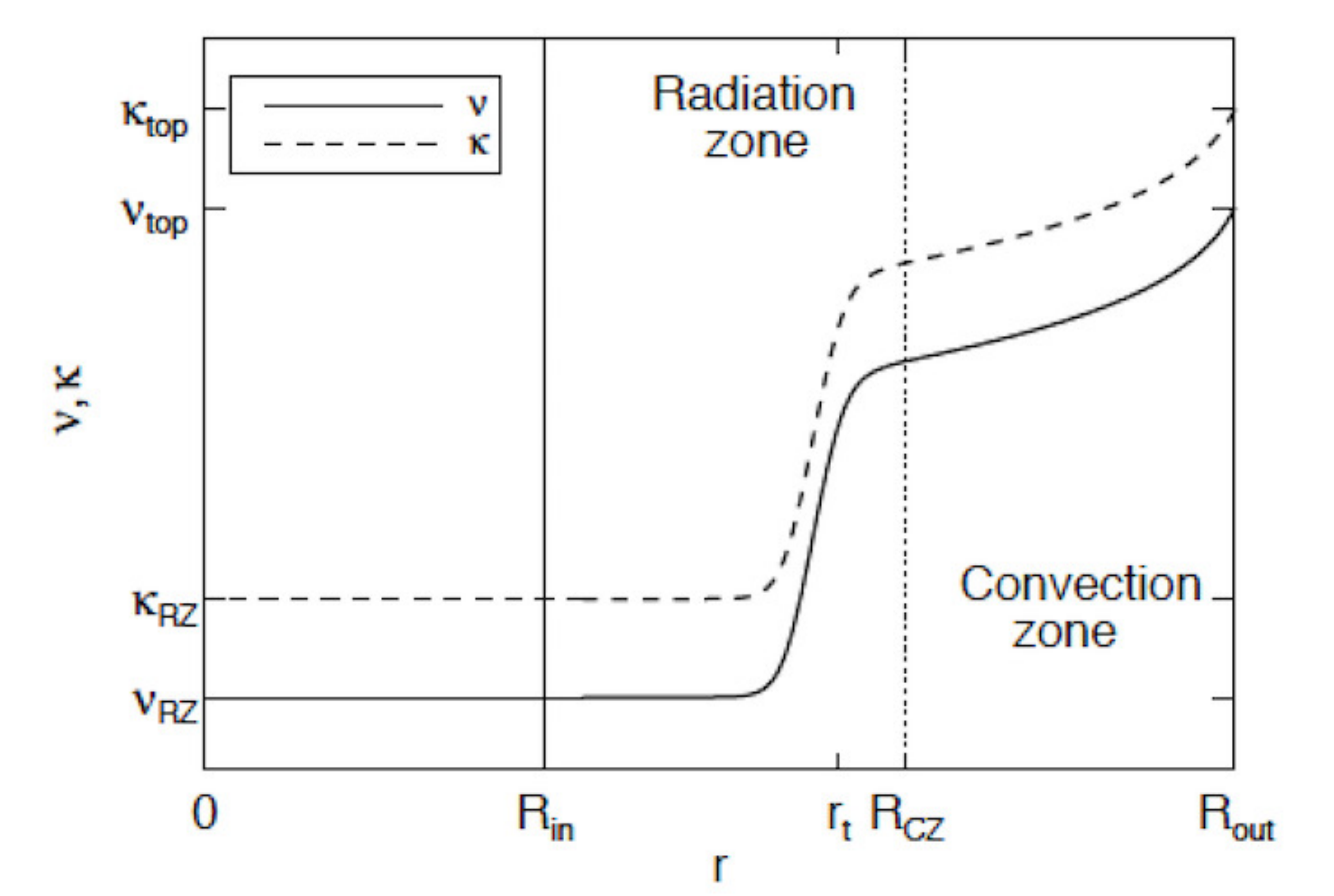}
\includegraphics[width=0.85\linewidth]{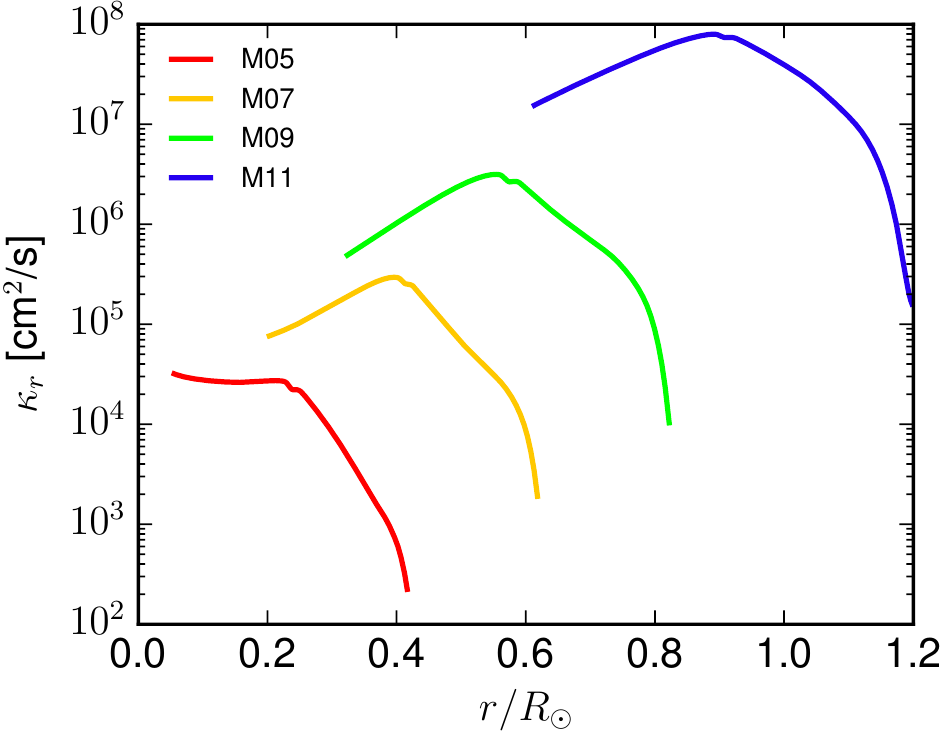}
\end{center}
\caption{\textit{Top panel}: Typical radial profile of kinematic
  viscosity and thermal diffusivity. \textit{Bottom panel}: Radiative
  diffusivity [$\kappa_r$] profiles of the four models,
  fitted from stellar models computed with the CESAM code.}
\label{diff1D}
\end{figure}

To complete the set of equations, we use the linearized equation of state
\begin{equation}
\frac{\rho}{\rb}=\frac{P}{\bar{P}}-\frac{T}{\tb}=\frac{P}{\gamma\bar{P}}-\frac{S}{c_p}\, ,
\end{equation}

\noindent where $\gamma$ is the adiabatic exponent, and assume the ideal gas law 
\begin{eqnarray}
\bar{P}={\cal R} \rb \tb\, ,
\end{eqnarray}

\noindent where ${\cal R}$ is the gas constant.  The reference state is derived from a 1--D solar structure model
\citep[][cf. \S 3.2]{Brun:2002cy} and is continuously updated with the spherically-symmetric components of the
thermodynamic fluctuations as the simulation proceeds.  It begins in hydrostatic balance so the bracketed term on the
right-hand-side of equation (3) initially vanishes.  However, as the simulation evolves, turbulent pressure drives the
reference state slightly away from hydrostatic balance.

Due to limitations in computing resources, no simulation achievable now or in the near future can hope to directly
capture all scales of stellar convection from global to molecular dissipation scales.  The simulations reported here
resolve nonlinear interactions among a large range of scales both in the convective and radiative zones.  The nonlinear
coupling of the two zones plus the use of a realistic stratification in the radiative interior is what sets these 3-D
global simulations of solar-like stars apart from previous work. Motions and waves must exist in the solar-like stars
on scales smaller than our grid resolution.  In this sense, our models should be regarded as large-eddy simulations
(LES) with parameterizations to account for subgrid-scale (SGS) motions.

\begin{figure*}[!htb]
\begin{center}
\includegraphics[width=0.9\textwidth]{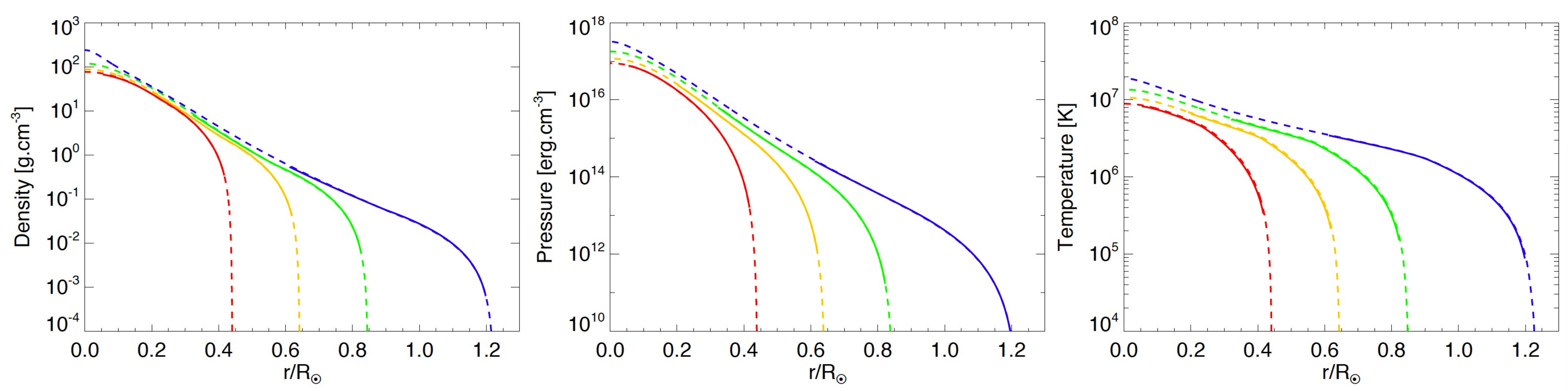}
\end{center}
\caption{Radial profile of the mean density, pressure and temperature in the four main class of models. The color
  correspond to the stellar mass (0.5 in red, 0.7 in yellow, 0.9 in green and 1.1 $M_{\odot}$ in blue).  Solid lines
  represent the 1-D structure used in the 3-D model, the dash lines the profile from the stellar structure model
  computed with the CESAM code \citep{Morel:1997gy,Brun:2002cy}.}
\label{dens1D}
\end{figure*}

\begin{figure*}[!htb]
\begin{center}
\includegraphics[width=0.9\textwidth]{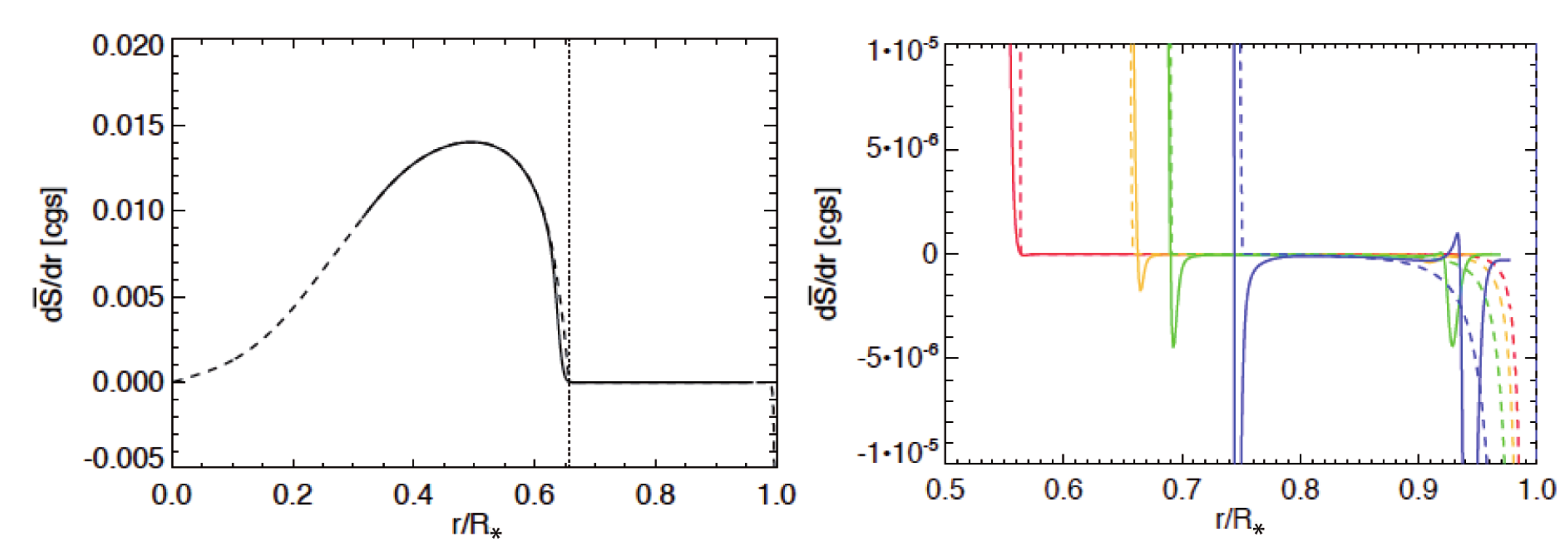}
\end{center}
\caption{Typical radial profile of the mean entropy gradient as realized in case M07R1. Solid lines represent the 1-D
  structure used in the 3-D model, the dash lines the profile from the stellar structure model computed with the CESAM
  code \citep{Morel:1997gy,Brun:2002cy}.  Right: Zoom of the mean entropy gradient at the base of the convective
  envelope for each main class of models. Color code is the same as in Figure \ref{dens1D}}
\label{dsdr1D}
\end{figure*}

Thus the effective eddy diffusivities $\nu$, $\kappa$ and $\kappa_0$ represent momentum and heat transport by motions
which are not resolved by the simulation.  They are allowed to vary with radius but are independent of latitude and
longitude, and vary only slightly with time for a given simulation as the reference density evolves.  Their amplitudes
and radial profiles are varied depending on the resolution and objectives of each simulation. In the simulations
reported on here, the radial profiles of $\nu$ and $\kappa$ are given by

\begin{equation}
  \nu(r) = \nu_{bot} + \nu_{top} f_{step}(r),\nonumber
\end{equation}

\noindent where 
\begin{eqnarray}
  f_{step}(r)&=(\rb/\rb_{top})^{\alpha}[1 - \beta]f(r), \nonumber \\
  f(r)&=0.5(\tanh((r-r_t)/\sigma_t)+1), \nonumber \\
  \beta &= \nu_{bot}/\nu_{top}=10^{-3}, \nonumber
\end{eqnarray}

\noindent and with $\nu_{top}$ in {\rm cm$^2$ s$^{-1}$} and $r_t$ and $\sigma_t$ in {\rm cm} are given in Table \ref{tablediff}
(see Appendix), $\alpha$ is -0.5 for all cases.  All models assumed a Prandtl number of 0.25, such that $\kappa$ can be
directly obtained from the amplitude and profile of $\nu$.  These tapered profiles are chosen in order to take into
account the much smaller sub-grid scale transport expected in the convectively stable radiative interior.  Their profile
is shown in Figure \ref{diff1D}.

To maintain a high degree of supercriticality of the convective instability in our simulations, we have lowered 
the diffusivities as we increase the rotation rate (see Table \ref{tablediff}). We have chosen to scale
$\nu_{top} \propto 1/\Omega_*^{0.5}$.  This is a compromise between keeping the diffusivity constant but making the
convective patterns too laminar and the exact scaling as $\propto 1/\Omega_*^{2}$ which would otherwise implies too wide
a parameter range to cover given our computer resources.

The diffusivity $\kappa_0$ is set such as to have the unresolved eddy flux carrying the stellar flux (which depends on
the spectral type considered see Table \ref{tablespectraltype}) outward at the top of the domain (see Figure
\ref{fluxbal}).  It drops off exponentially with depth in order to avoid a large inward heat flux in the stable zone
\citep[see][]{Miesch:2000gs,Brun:2011bl,Alvan:2014gx}. Of course there is some arbitrariness in choosing the exact shape
and amplitude of our diffusivity profiles and we optimize their profiles such as to limit their influence on the results
reported here.

The velocity and thermodynamic variables are expanded in spherical harmonics $Y_{\ell m}(\theta,\phi)$ for their
horizontal structure and in Chebyshev polynomials $T_n(r)$ for their radial structure \citep[see][for more details on
the numerical method and anelastic approximation]{Glatzmaier:1984jh,Clune:1999vd}.

\begin{figure*}[!htb]
\begin{center}
\includegraphics[width=0.45\textwidth]{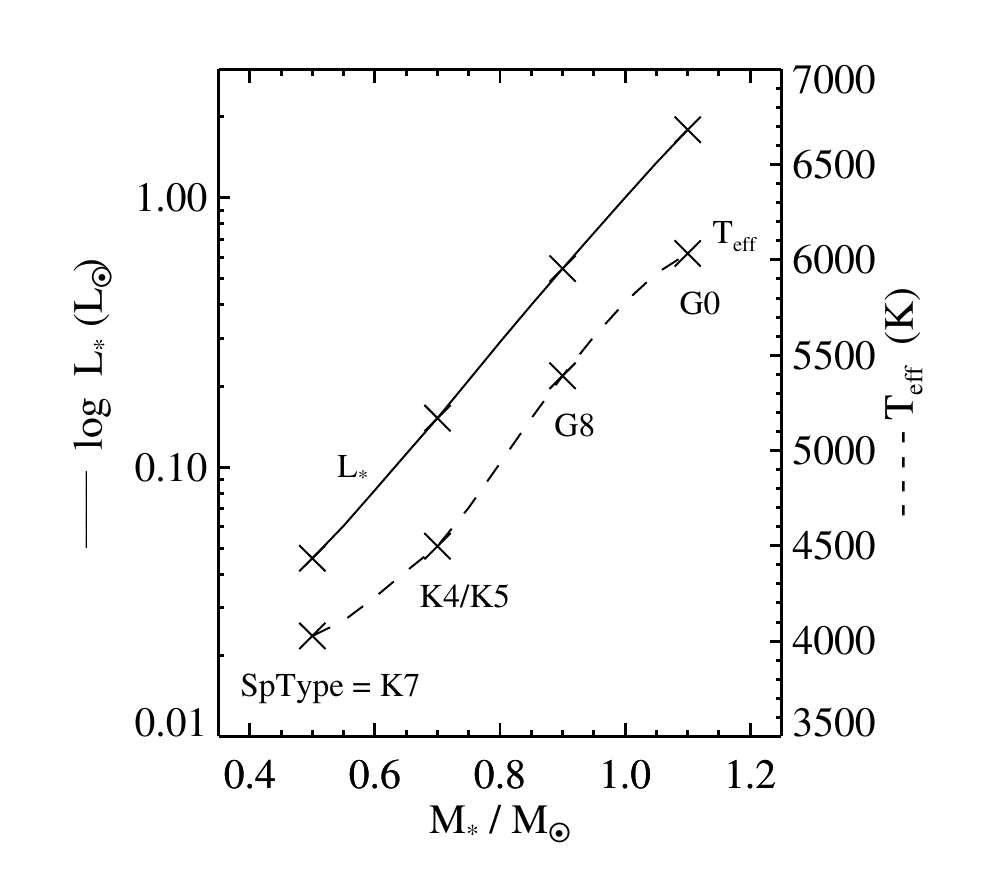}
\includegraphics[width=0.45\textwidth]{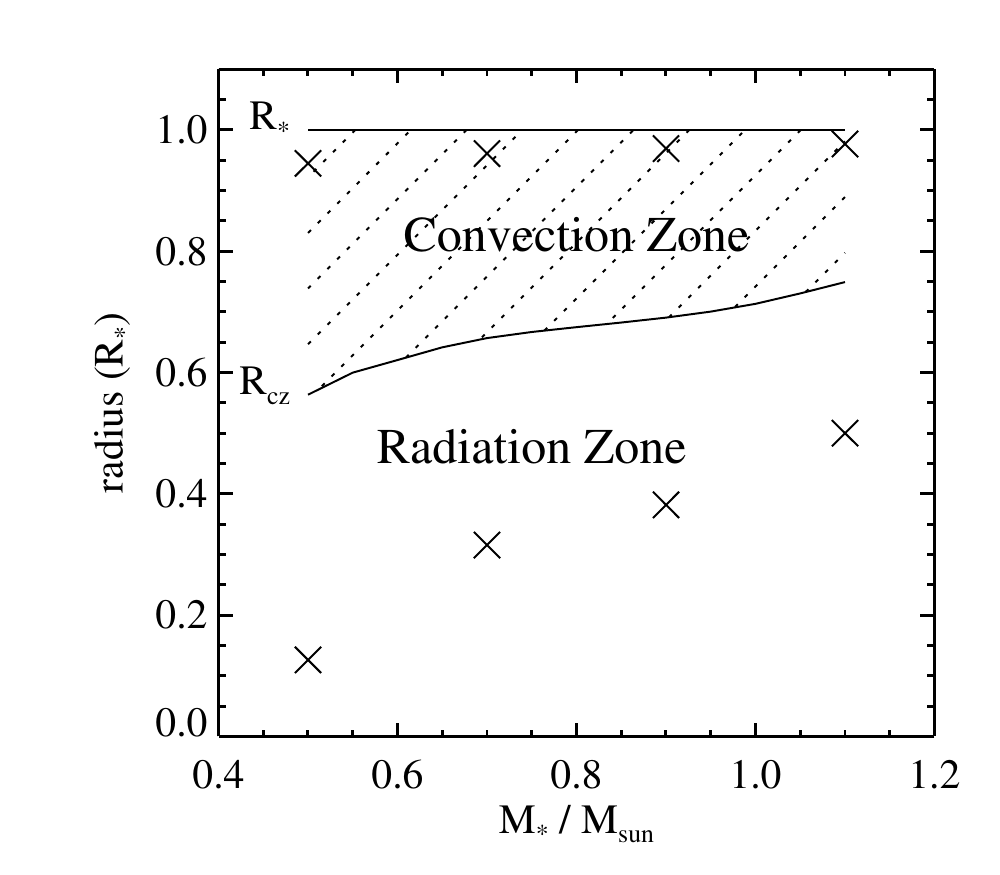}
\end{center}
\caption{Luminosity, effective temperature and depth of the convection zone in 1-D stellar model computed with the CESAM code
  \citep{Morel:1997gy,Brun:2002cy} used to set up the background state of the 3-D ASH simulations. Left: $\times$ marks 
  models chosen and lines show variation with mass. Right: $\times$ marks $R_{in}$ and $R_{out}$ for each of the 4 model masses.}
\label{LumR1D}
\end{figure*} 

\begin{table*}[!ht]
\begin{center}
\caption{Global properties on the main sequence of the 4 stars used in our ASH models}\label{tablespectraltype}
\vspace{0.2cm}
%\begin{tabular}{||p{1.8cm}*{1}{||c} cccc ||}
\begin{tabular}{ccccccccccc}
\tableline
\tableline
\\ [-1.5ex]
 Mass & Radius & $L_{*}$ & $T_{eff}$ & Sp. T. & $M_{bcz}$ & $R_{bcz}$ & $\bar{T}_{bcz}$ & $\rb_{bcz}$ & $\Delta_{cz} \rb$ & $\Delta_{f} \rb$\\ [0.5ex]
 $(M_{\odot})$ & $(R_{\odot})$ &  $(L_{\odot})$ & $(K)$ & & $(M_{\odot},M_*)$ & $(R_{\odot},R_*)$  & $(K)$ & $(g cm^{-3})$ & - & - \\ [0.8ex] 
\tableline
\tableline
\\ [-1.5ex]
 0.5 & 0.44 & 0.046 & 4030 & K7 & 0.18, 0.36 & 0.25,0.56 & $4.3\times 10^6$ & 14.0 & 42 & 193 \\
 0.7  & 0.64 & 0.15 & 4500 & K4/K5 & 0.079, 0.11 & 0.42,0.66 & $3.0\times 10^6$ & 2.1 & 50 & 605  \\
 0.9  & 0.85 & 0.55 & 5390 & G8 & 0.042, 0.046 & 0.59,0.69 & $2.6\times 10^6$ & 0.51 & 67 & 1013 \\
 1.1 & 1.23 & 1.79 & 6030 & G0 & 0.011, 0.010 & 0.92,0.75 & $1.6\times 10^6$ & 0.048 & 81 & 830 \\ [0.5ex]
\hline
 \tableline
 \tableline
\end{tabular}
\end{center}
All the listed values were computed with the CESAM stellar evolution code \citep{Morel:1997gy}.
We adopt $M_{\odot}=1.989\times 10^{33}\, g$, $R_{\odot}=6.9599\times 10^{10}\, cm$, and $L_{\odot}=3.846\times 10^{33}\, erg\cdot s^{-1}$.
The density ratios $\Delta_{cz} \rb$ and $\Delta_{f} \rb$ are evaluated by forming the ratio between the value of the 
density respectively at the base of the convection and the top of the domain and at the bottom and the top of the domain.
\end{table*}

Given that the convective and radiative zones are nonlinearly and dynamically coupled, internal waves can easily be
excited by the pummeling of convective plumes on top of the radiative interior \citep{Brun:2011bl,Alvan:2014gx}.  The
Brunt-V\"{a}is\"{a}l\"{a} frequency $N$ of the models are very close to that deduced from 1--D stellar models computed
with the CESAM code \citep[][see Figure \ref{dsdr1D}]{Morel:1997gy,Brun:2002cy}. The transition at the base of the convective envelope 
has just been slightly soften, as can be seen in the Figure \ref{dsdr1D} when comparing solid and dash lines.  We are thus expecting the
propagation of the internal waves to be realistic, aside from the shallower cavity due to our choice of $r_{bot}\neq 0$
and the enhanced thermal and viscous diffusion present in the model that translates into an enhanced damping
\citep{Zahn:1997ui}. While present in the simulations, we will not study in details the internal waves and their spectra
in this paper. We choose instead to focus our study on convection and the generation and maintenance of the large-scale
mean flows and how they vary as a function of spectral type. As shown in \citet{Alvan:2014gx}, fully spherical models
are more adequate to realistically model internal waves and gravity modes, but they are much more expensive to run,
which makes a comprehensive parameter study impractical.
  
In order to ensure that the mass flux remains divergenceless, we use a toroidal--poloidal decomposition as:

\begin{eqnarray}
{\rb\bf v}=\nab\times\nab\times (W \hat{\bf e}_r) +  \nab\times (Z \hat{\bf e}_r)\, .
\end{eqnarray}

This system of hydrodynamic equations requires 8 boundary conditions in order to be well-posed. Since assessing the
angular momentum redistribution in our simulations is one of the main goals of this work, we have opted for torque-free
velocity conditions:

\begin{enumerate}
\item impenetrable top and bottom: $v_r=0|_{r=r_{bot},r_{top}}$
\item stress free top and bottom: \\
$\frac{\p}{\p r}\left(\frac{v_{\theta}}{r}\right)=\frac{\p}{\p r}\left(\frac{v_{\phi}}{r}\right)=0|_{r=r_{bot},r_{top}}$
\item constant entropy gradient at top and bottom: $\frac{\p \sb}{\p r}=a|_{r=r_{bot}} \mbox{ and }b|_{r=r_{top}}$ 
\end{enumerate}

the values of $a$ and $b$ depend on the modelled star (see Table \ref{tablediff}).

\begin{table}[!ht]
\begin{center}
\caption{Parameters for the Stellar ASH Models}\label{tableash}
\vspace{0.2cm}
%\begin{tabular}{||p{1.8cm}*{1}{||c} cccc ||}
\begin{tabular}{cccccc}
\tableline
\tableline
\\ [-1.5ex]
 Mass & $R_{bot}$ & $R_{top}$ & $R_{cz}$ & Name & Rotation $\Omega_*$ \\ [0.5ex]
 $(M_{\odot})$ & $(R_*)$ & $(R_*)$ & $(R_*)$ & & $(\Omega_{\odot})$ \\   [0.8ex]
\tableline
\tableline
\\ [-1.5ex]
  0.5 & 0.13 & 0.95 & 0.56 & M05 S & 0.125 \\
  & & & & M05 R1 & 1 \\
  & & & & M05 R3 & 3 \\
  & & & & M05 R5 & 5 \\ [0.5ex]
\hline
\\ [-2ex]
  0.7 & 0.32 & 0.97 & 0.66 & M07 S & 0.3 \\
  & & & & M07 R1 & 1 \\
  & & & & M07 R3 & 3 \\
  & & & & M07 R5 & 5 \\ [0.5ex]
\hline
\\ [-2ex]
  0.9 & 0.38 & 0.97 & 0.69 & M09 S & 0.5 \\
  & & & & M09 R1 & 1 \\
  & & & & M09 R3 & 3 \\
  & & & & M09 R5 & 5 \\ [0.5ex]
 \hline
 \\ [-2ex]
  1.1 & 0.50 & 0.97 & 0.75 & M11 R1 & 1 \\ 
  & & & & M11 R3 & 3 \\
  & & & & M11 R5 & 5 \\ [0.5ex]
\hline
 \tableline
 \tableline
\end{tabular}
\end{center}
$R_{bot}$, $R_{top}$ and $R_{cz}$ correspond respectively to the bottom, top and base of the convection envelope radii.
\end{table}

\subsection{Numerical Experiments}
\label{sec_num_exp}

Our numerical model is a simplified portrayal of convection and radiative zones in solar like-stars: typical values
deduced from 1-D stellar evolution models are taken for the heat flux, rotation rate, mass and radius, and a perfect gas
is assumed \citep{Morel:1997gy}.  The anelastic reference state is based on a 1--D standard stellar structure model
discussed in \citet{Brun:2002cy}. We list the main stellar parameters in Table \ref{tablespectraltype}.  We also show in
Figures \ref{dens1D} and \ref{dsdr1D} the density, pressure, temperature and entropy gradient radial profiles used 
for the four stellar spectral types studied.  We initialize the
reference state of the 3--D model by specifying the entropy gradient $d\sb/dr$ and gravitational acceleration $g$ based
on the 1--D model.  The steep negative entropy gradient near the stellar surface is artificially suppressed to avoid the
driving of small-scale convective motions that cannot be resolved in this model.  We then solve the equation of
hydrostatic balance for the reference density $\rb$ with a Newton-Raphson method, assuming an ideal gas equation of
state and using the density profile of the 1--D structure model as an initial guess to initiate the iterative solve.
The resulting reference state is close to the stellar structure model, with slight departures due to the modified
entropy profile in the convection zone and the ideal gas equation of state.  Similarly, the radial profile of the
radiative diffusivity $\kappa_r$ is based on the 1--D stellar structure model, slightly adjusted to accommodate the
small changes between the reference state and the structure model and later on to compensate for the inward enthalpy
flux at the base of the convection zone. The radial profiles of $\kappa_r$
are shown in Figure \ref{diff1D} for all four stellar models.

\begin{figure*}[!ht]
\begin{center}
\includegraphics[width=0.9\textwidth]{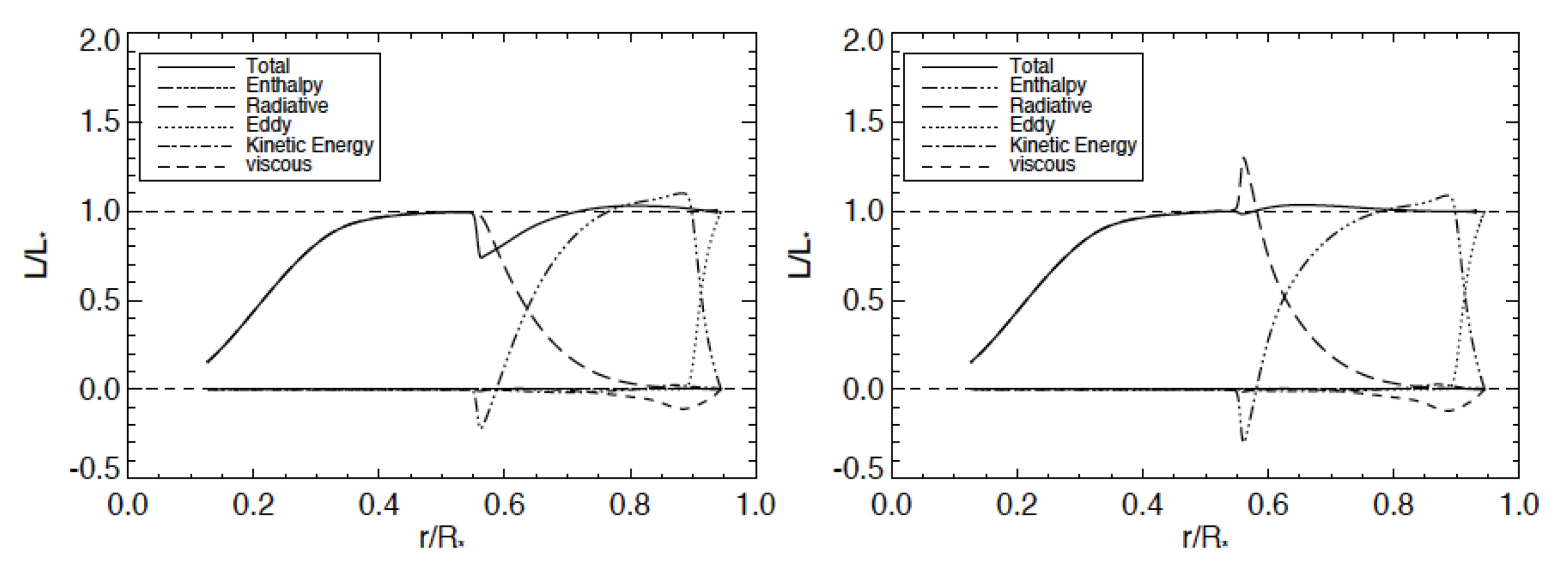}
\end{center}
\caption{Radial flux balance for model $M05R1$ before and after the adjustment of the radiative diffusivity
  $\kappa_{rad}$ in order to ease the thermal relaxation of the simulations and compensate for convective overshooting
  at the base of the convective envelope.}
\label{fluxbal}
\end{figure*}

Thus the departure of the reference entropy gradient from the stellar structure model near the top of the convection
zone evident in Figure \ref{dsdr1D}b is largely imposed.  However, the departure near the base of the convection zone is
established by the convection itself, as downflow plumes deposit low entropy material just before entering the stable
radiative zone.  The initial reference state in this region follows more closely the 1--D structure model.

The computational domain for each of the 4 stellar masses considered is listed in Table \ref{tablediff} and shown in
Figure \ref{LumR1D}.  All models use a numerical resolution of
$N_r \times N_{\theta} \times N_{\phi}= 770 \times 256 \times 512$. The depth of the convection zone $r_{bcz}$ is
defined by the change of sign of the initial mean entropy gradient $d\sb/dr$, and it is listed in Table \ref{tableash}.
This depth is slightly modified by the convective motions as the simulation evolves and matures. We also list the
density contrasts both in the convective envelope and over the whole radial extent of the domain. We note that several 
scale heights are present in the convective envelope and that overall the models have large density contrasts.  
The resolution at the base of the convection zone is $\delta r = 1.5\times 10^{-3}\, R_*$.

\section{Convection in spherical shells of various aspect ratios and rotation rates}
\label{sec_conv_state}

 Once the 1-D structure of the model is established, a 3-D random perturbation of the entropy
  field is introduced such as to trigger the convective instability. Our simulations begin with a
  Rayleigh number of the order of $\sim 10^6$ to $10^7$, which is substantially larger than the
  critical Rayleigh number that is typically about $10^4$ for the value of the Taylor numbers used
  here \citep[see][]{Jones:2009ko}. Following a linear phase of exponential growth, the convective
instability non linearly saturates by reducing the entropy gradient in the bulk of the domain except
for intense thermal boundary layers created near the surface and at the base of the convective
envelope. After several convective overturning times the radial transport of energy reaches an
equilibrium as shown in Figure \ref{fluxbal}. As can be seen, the enthalpy flux, due to the
correlation between radial velocity and temperature fluctuations, is dominant throughout the
convective zone. It peaks near the surface where the negative (radially-inward) kinetic energy
  flux forces it to become locally greater than the stellar luminosity. At the bottom of the
convective envelope, we note the presence of a negative enthalpy flux associated with overshooting.
This requires some adjustment of the radiative diffusivity near the base of the convective envelope
is necessary to ensure the full transport of the stellar luminosity, as is shown in the right panel
of Figure \ref{fluxbal}. Such adjustment is done for all models \citep[as in][]{Brun:2011bl}.  We will discuss in more details
the properties of the overshooting layer in section \S \,\ref{sec:oversh-solar-like}.

In the model shown in Figure \ref{fluxbal}, the luminosity $L(r)$ increases from the lower boundary until it reaches
$L_*$. This is due to the presence of a volumetric heating source that mimics the production of internal energy by
nuclear reactions.  Only the stellar models with masses of 0.5 and 0.7 $M_{\odot}$ have such a heating source term as
the models are deep enough that they include part of the nuclear core in their radiative zone.

\subsection{Convective patterns}
\label{sec_conv_patterns}

\begin{figure*}[!htb]
\begin{center}
\includegraphics[width=0.9\textwidth]{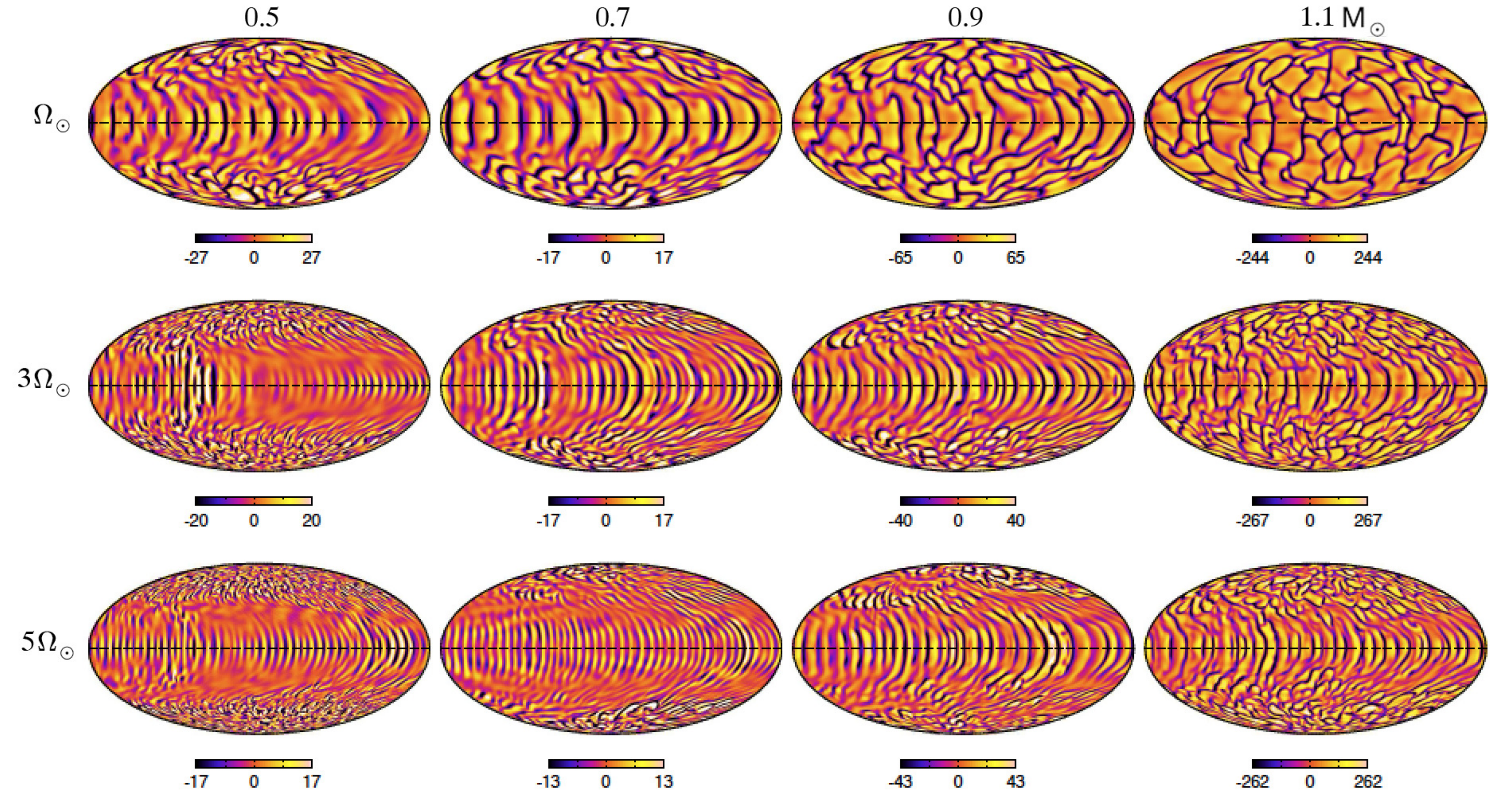}
\end{center}
\caption{Convection patterns near the top of the simulation domain are shown as the radial velocity for four masses
    in the range 0.5, 0.7, 0.9 and 1.1 (from left to right) and for three rotation rates: one,
    three, and five times the solar rate (from top to bottom).}
\label{conv}
\end{figure*}

The development of the convective instability in the spherical shells leads to the classical network
of downflow lanes surrounding broad upflows as can be seen in Figure \ref{conv}. As the rotation
rate increases (from top to bottom) we note that a higher azimutal wavenumber $m$ characterizes the
convection. We clearly see that there are more downflow lanes near the equator, which are narrower,
and smaller convective patches in the polar caps. At the fastest rates convection can be
longitudinally modulated at low latitude, even exhibiting the so-called \textit{active nests} of
convection \citep{Brown:2008ii}. Such active nests dominate the local transport processes and imprint their motions onto the top of
the radiative zone. Except for the localized convective patterns, their overall behavior
is consistent with models having a $R_{of} < 1$, e.g. Reynolds stresses accelerate
the equator (see Sections \ref{sec_dr_prof} and \ref{sec_amom}).
% Such active nests dominate the
%   local transport processes. They imprint their motion into the top
% of the radiative zone. Overall we do not find such convective states to
% depart from the scaling laws identified in our parameter study and described
% in subsequent sections.

% To our knowledge, this is the first time such active nests of 
% convection have been obtained with an underlying stably-stratified region, rather than fixed boundaries. 
Notice that the low-latitude downflow lanes are bent into a
  banana-like shape, with its retrograde tails occurring at mid-latitudes. This is consistent with
  the statistical correlation of the radial and longitudinal motions that produce an efficient
  angular momentum transport through Reynolds stresses, as will be discussed in \S
  \,\ref{various_states_dr} and \S \,\ref{sec_amom}. The slowly rotating cases show a modest
  deflection in the opposite direction due to the prograde, high-latitude shear of the differential
  rotation that modifies the structure of the convective motions. At the low rotation rates of
cases M05 S, M07 S, M09 S (not shown), and M11 R1, the convective patterns look more alike at all latitudes,
much like a \textit{soccer-ball}, and the mean flows are more randomly oriented. As demonstrated in
\citet{Chandrasekhar:1961gu,2009ApJ...702.1078B}, at low rotation rate a dominant $\ell=1$ mode
develops, as is clearly seen in the fluctuating temperature field. It wobbles around, making
longitudinal averages hardly meaningful.

As anticipated in our introduction of a Rossby number scaling based on mixing length theory
arguments, our numerical simulations show that for higher mass the convective \textit{rms} velocity
increases. For instance, going from about 10 $\mathrm{m s^{-1}}$ in a $0.5 \,M_\odot$ star to
hundreds of meters per second in a $1.1\, M_\odot$ star (see Table \ref{Table 3b}). 
The convective velocity amplitude realized in the 3-D models are higher than those evaluated
using mixing length theory. This difference (of order 10-20\%) in the model presented in this study is due to the 
inward kinetic energy flux found in the models that results in a slightly overluminous convective enthalpy flux and
hence larger convective velocity (which we recall scales like $\sim \sqrt[3]{L_*}$, see section 2). 
This is a well documented physical mechanism in 3-D compressible turbulent convection whose 
amplitude depends on both turbulence degree and stratification, see for instance discussion in 
\citep{1991ApJ...370..282C,2009ApJ...702.1078B}.
We also note that as the rotation rate is increased the fluctuating \textit{rms} velocity decreases due to
  a slightly decreased super-criticality of the system, whereas the full \textit{rms} velocity increases due
to the presence of a differential rotation. When considering only the fluctuating \textit{rms}
velocity, the flow field is found to be close to isotropy ($\tilde{v}'_r,\, \tilde{v}'_\theta$ and
$\tilde{v}'_\phi$ of the same order). Note that for $\tilde{v}_r$ and $\tilde{v}_\theta$, the weak
meridional circulation does not make much difference between the full and fluctuating components of
the \textit{rms} velocity. This is of course not true for $\tilde{v}_{\phi}$, which increases
  due to the differential rotation.

We also see that as the mass increases (from left to right in Figure \ref{conv}), the convective
flows are less constrained by rotation.  This is clearly seen by comparing their respective Rossby
number.  In Table \ref{tablenondimnb}, we list three different flavors of this important
  number (see Appendix), and they each show that, for an identical rotation rate, the more massive
  stars possess the largest Rossby number, and thus the lowest level of rotational influence. Still,
  for a fixed Rossby number, we found that models of the more massive stars exhibit smaller
  convective patterns, which is likely due to their shallower convective zones.

The time dependence of the convective flow is very rich with continuous emergence, merging, cleaving
of the convective cells and strong vortical downflows at the interstices of the downflow lanes for
the low Rossby number cases. A clear advection to the right at low latitudes and to the left at high
latitudes (e.g., respectively faster and slower than the global rotation rate) is evident for the cases rotating 
at the intermediate regime that possess a monotonic
differential rotation. For the slowly rotating ones, convection is more isotropic and less
influenced by mean flows. In the case with the highest rotation rate (Rossby number $\le$ 0.1), the
advection by the mean flow is less systematic due to the alternating prograde and retrograde jets
(see \S \,\ref{various_states_dr}).

\begin{table}[!ht]
{\footnotesize
\begin{center}
\caption{Representative Velocities}\label{Table 3b}
%\begin{tabular}{1.1\linewidth}{@{\extracolsep\fill}||p{0.8cm}||p{0.4cm}p{0.4cm}p{0.4cm}p{0.4cm}p{0.4cm}p{0.4cm}||}
\begin{tabular}{ccccccccc}
\tableline
\tableline
\multicolumn{1}{c}{}&\multicolumn{8}{c}{Mid Convective Zone}  \\
\multicolumn{1}{c}{Case}& $\vrr$ & $\vtr$ & $\tilde{v}_{\rm pol}$ & $\tilde{v}_{\rm mc}$ & $\vphr$ & $\vphr'$ & $\vvr$ & $\vvr'$ \\
\tableline
\tableline
\multicolumn{1}{c}{M05 S} & 15 & 14 & 20 & 3.3 & 25 & 18 & 32 & 27 \\
\multicolumn{1}{c}{M05 R1} & 8 & 8 & 11 & 1.1 & 40 & 8 & 41 & 14 \\
\multicolumn{1}{c}{M05 R3} & 6 & 7 & 9 & 0.5 & 53 & 7 & 54 & 11 \\
\multicolumn{1}{c}{M05R5} & 5 & 5 & 7 & 0.3 & 46 & 5 & 46 & 9 \\
\multicolumn{1}{c}{M07 S} & 27 & 26 & 37 & 3.2 & 43 & 36 & 57 & 50 \\
\multicolumn{1}{c}{M07 R1} & 22 & 21 & 30 & 1.3 & 77 & 24 & 83 & 39 \\
\multicolumn{1}{c}{M07 R3} & 17 & 18 & 25 & 0.9 & 116 & 23 & 119 & 34 \\
\multicolumn{1}{c}{M07 R5} & 15 & 16 & 22 & 0.3 & 131 & 20 & 133 & 30 \\
\multicolumn{1}{c}{M09 S} & 61 & 47 & 77 & 4.7 & 70 & 63 & 105 & 99 \\
\multicolumn{1}{c}{M09 R1} & 50 & 46 & 68 & 2.5 & 100 & 52 & 121 & 85 \\
\multicolumn{1}{c}{M09 R3} & 44 & 44 & 62 & 1.4 & 240 & 53 & 248 & 82 \\
\multicolumn{1}{c}{M09 R5} & 27 & 38 & 47 & 1.9 & 259 & 47 & 263 & 66 \\
\multicolumn{1}{c}{M11 R1} & 149 & 133 & 200 & 22.2 & 225 & 133 & 300 & 238 \\
\multicolumn{1}{c}{M11 R3} & 103 & 96 & 141 & 3.6 & 404 & 107 & 428 & 176 \\
\multicolumn{1}{c}{M11 R5} & 83 & 82 & 117 & 3.0 & 605 & 93 & 616 & 149 \\
 \tableline
 \tableline
\end{tabular}
\end{center}}

In all cases, temporal averages at mid-layer depth in convection zone of rms components of velocity
$\vrr$, $\vtr$, $\tilde{v}_{\rm pol}=\sqrt{\vrr^2+\vtr^2}$,
$\tilde{v}_{\rm mc} =\sqrt{\left\langle v_r \right\rangle_\phi^2 + \left\langle v_\theta \right\rangle_\phi^2 }$,
$\vphr$ and of speed $\vvr$, and of fluctuating velocities $\vphr'$ and $\vvr'$ (with temporal and azimuthal mean
subtracted), all expressed in m$\,$s$^{-1}$.
\end{table}

\begin{table}[!ht]
\begin{center}
\caption{Non-dimensional numbers and Stellar, Convective and Fluid Rossby numbers \label{tablenondimnb}
}
\vspace{0.2cm}
%\begin{tabular}{||p{1.8cm}*{1}{||c} cccc ||}
\begin{tabular}{cccccccc}
\tableline
\tableline
\\ [-1.5ex]
 Name & $R_{e}$ & $R_{a}$ & $T_{a}$ & $P_e$  & $R_{of}$ & $R_{oc}$ & $R_{os}$ \\ [0.5ex]
 & & $(10^6)$ & $(10^6)$ &  & & & \\ [0.8ex] 
\tableline
\tableline
\\ [-1.5ex]
 M05 S & 107 & 0.80 & 0.09 & 14.8 & 1.77 & 1.32 & 2.44 \\
 M05 R1 & 131 & 9.98 & 33.5 & 18.7 & 0.35 & 0.82 & 0.16 \\
 M05 R3 & 179 & 54.1 & 907.7& 24.4 & 0.16 & 0.64 & 0.04 \\
 M05 R5 & 189 & 97.7 & 4202.5 & 26.2 & 0.09 & 0.51 & 0.02 \\ [0.5ex]
\hline
\\ [-2ex]
  M07 S & 58 & 0.12 & 0.041 & 7.9 & 1.23 & 0.73 & 1.94 \\
  M07 R1 & 72 & 0.96 & 1.66 & 10.2 & 0.42 & 0.62 & 0.39 \\
  M07 R3 & 109 & 5.81 & 44.9 & 13.6 & 0.17 & 0.51 & 0.10 \\
  M07 R5 & 124 & 27.0 & 208.1 & 15.5 & 0.11 & 0.66 & 0.05\\ [0.5ex]
\hline
\\ [-2ex]
  M09 S & 60 & 0.13 & 0.07 & 9.3 & 1.29 & 0.82 & 1.79 \\
  M09 R1 & 64 & 0.43 & 0.41 & 9.4 & 0.67 & 0.74 & 0.73 \\
  M09 R3 & 106 & 2.54 & 10.9 & 14.2 & 0.28 & 0.60 & 0.21 \\
  M09 R5 & 110 & 5.48 & 50.9 & 11.2 & 0.14 & 0.53 & 0.08 \\ [0.5ex]
 \hline
 \\ [-2ex]
  M11 R1 & 63 & 0.18 & 0.08 & 9.9 & 1.40 & 2.45 & 1.80 \\ 
  M11 R3 & 81 & 1.13 & 2.1 & 11.9 & 0.54 & 2.05 & 0.41 \\
  M11 R5 & 89 & 2.86 & 9.6 & 12.4 & 0.34 & 1.96 & 0.20 \\ [0.5ex]
\hline
 \tableline
 \tableline
\end{tabular}
\end{center}
In all cases the Prandtl number $P_r = \nu/\kappa$ = 0.25.  The Taylor number is defined as
$T_a = 4 \Omega_*^2 L^4/\nu^2$, where $L = r_{top}-r_{bcz}$ for each case.  Also listed are the {\it rms} Reynolds number
$R_e=\vvr' L/\nu$, the Rayleigh number $R_a=(-\p\rho/\p S)\Delta S g L^3/\rho \nu \kappa$, the P\'eclet number $P_e=\vrr L/\kappa$, 
the fluid Rossby number $R_{of}=\tilde{\omega}/2\Omega_*$, the convective Rossby number 
$R_{oc}=\sqrt{R_a/T_a P_r}$ and the stellar Rossby number $R_{os} = P_{rot}/\tau_{conv}$ (see Appendix). 
All numbers have been evaluated at mid depth in the convection zone. A Reynolds number evaluated with the maximum 
speed achieved in the domain will be at least 4 times larger.
\end{table}

\subsection{Overshooting in solar-like stars}
\label{sec:oversh-solar-like}

At the base of the convective envelope of all the models presented in this work, a shallow region of
mixing develops. These overshooting regions exist because the downward plumes do not immediatdiately
halt their descending motions as they go through the swift transition in stratification, which is
from unstable to convection (negative entropy gradient) to the stable radiative interior (positive
entropy gradient). Indeed, these vortical downward-flowing structures require a finite
  distance before they overturn, since they have some incoming inertia as they cross the transition
  point. This distance depends upon the stratification, degree of turbulence, the rotation rate, and
  the thermal diffusivity
  \citep{Zahn:1991uz,Brummell:2002dj,Browning:2004bx,Rempel:2004ek,Rogers:2006ks,2011IAUS..271..317C,Brun:2011bl,Masada:2013fc}.
As they decelerate due to the action of buoyancy breaking, they induce extra mixing and turbulent
flows across this finite layer. The superadiabatic to subadiabatic change in the stratification
implies that the correlation between convective velocity and temperature fluctuations should
reverse. This is illustrated in Figure \ref{fluxbal}, where the enthalpy flux is clearly negative at
the base of the convective envelope.  Since the amount of overshooting depends on the
  stiffness of the stratification of the radiative interior, this has motivated our choice of using
  a realistic stratification that is directly deduced from 1-D stellar evolution models. Theoretical
  studies have revealed that the P\'{e}clet number ($Pe=v l/\kappa$, with $l$ here a typical size of the plumes) 
  of the individual plumes is the key quantity to assess the properties of the overshooting \citep[see for
    instance][]{Zahn:1991uz,Brummell:2002dj}. Flows with a small $Pe$ number tend to overshoot,
  leading to a stratification that remains convectively stable. In contrast, flows with a high
  P\'{e}clet number require that the stratification become nearly adiabatic, being locally modified
  by efficient turbulent mixing, and the convection extends deeper into the radiative zone. Hence
  the use of convective penetration to describe this regime.  It is clear that the degree of
turbulence of our convective zone is mild, and hence our P\'{e}clet number small. In real solar-like
stars, the P\'{e}clet number of turbulent plumes is much higher. We can thus expect that our
simulations give an upper limit to the amount of penetration occurring in solar-like stars.

%Still we
%believe that our simulations are sophisticated enough (by including the key physical mechanisms) to
%be a fair description of what is occurring in these stars.

\begin{figure}[!htb]
\begin{center}
\includegraphics[width=0.5\textwidth]{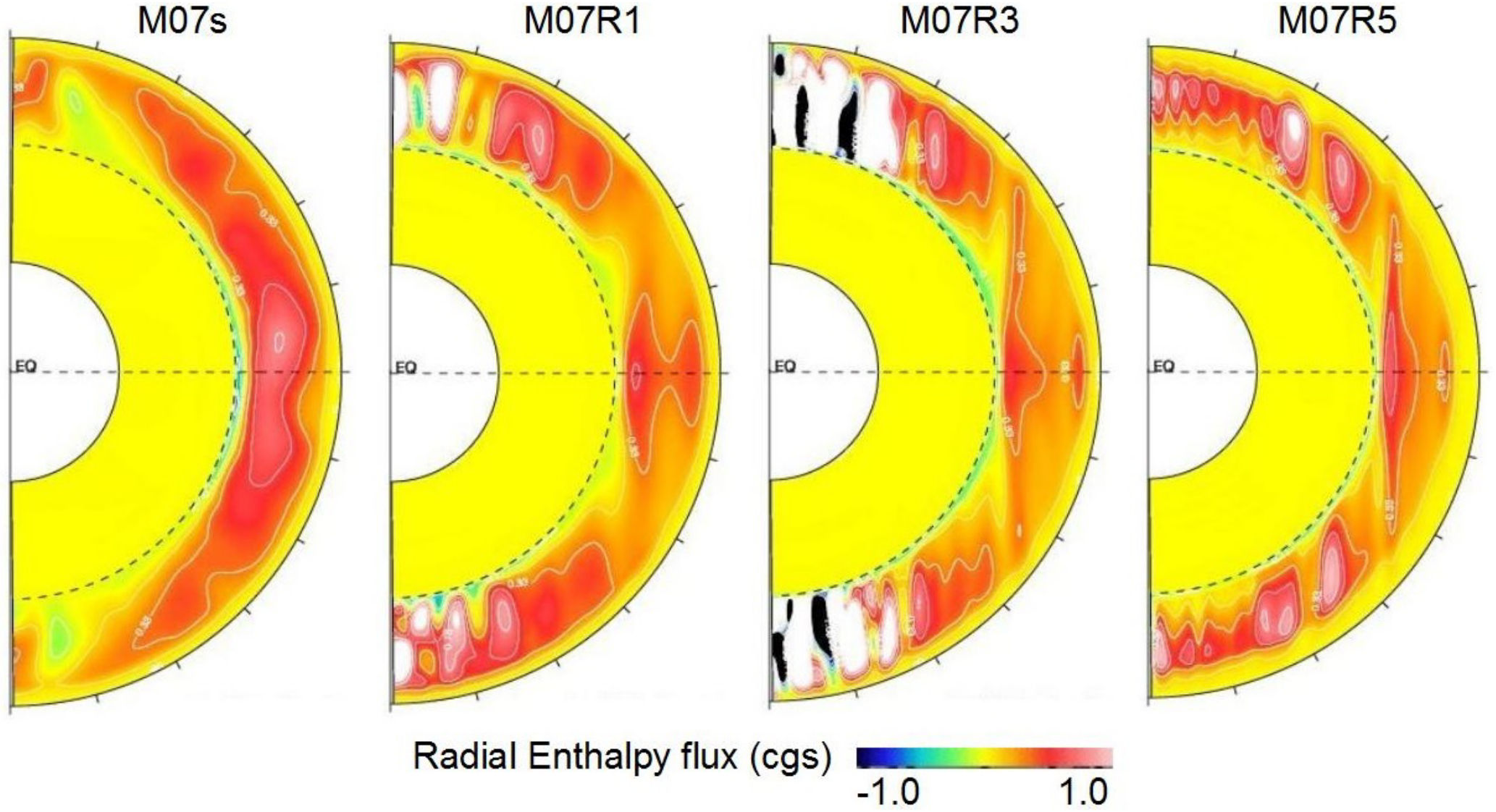}
\includegraphics[width=0.5\textwidth]{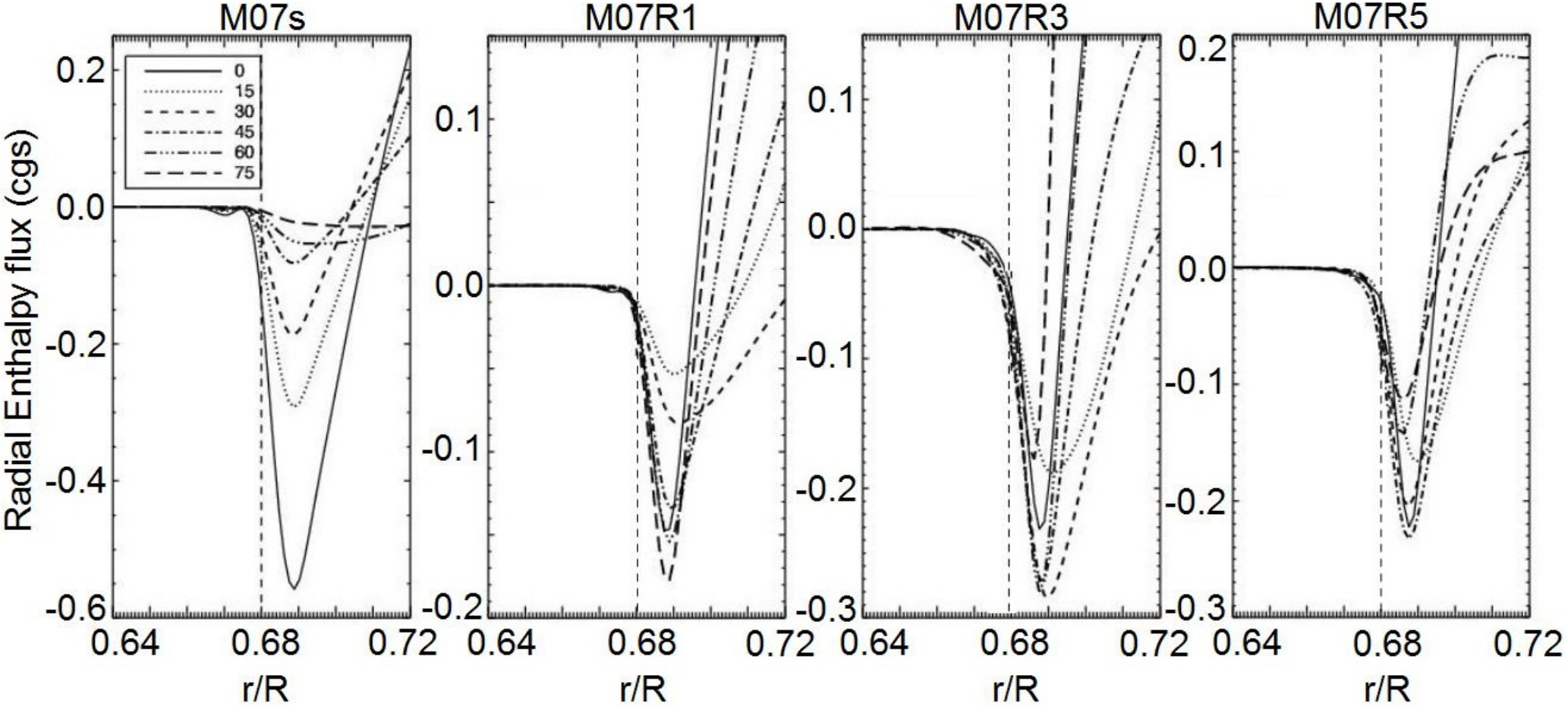}
\end{center}
\caption{\textit{Upper panels:} Temporal and longitudinal average of the radial enthalpy flux
  profiles for the model M07 with different rotation rates. \textit{Lower panels:} Radial cuts from
  the equator to the latitude 75$^{\circ}$ in 15$^{\circ}$ intervals and between the radii $0.64$
  and $0.72$ $r/r_{*}$ of the temporally and longitudinally averaged radial enthalpy flux for the
  same models.}
\label{Radial_E_flux}
\end{figure}

\begin{figure}[!htb]
\begin{center}
\includegraphics[width=0.5\textwidth]{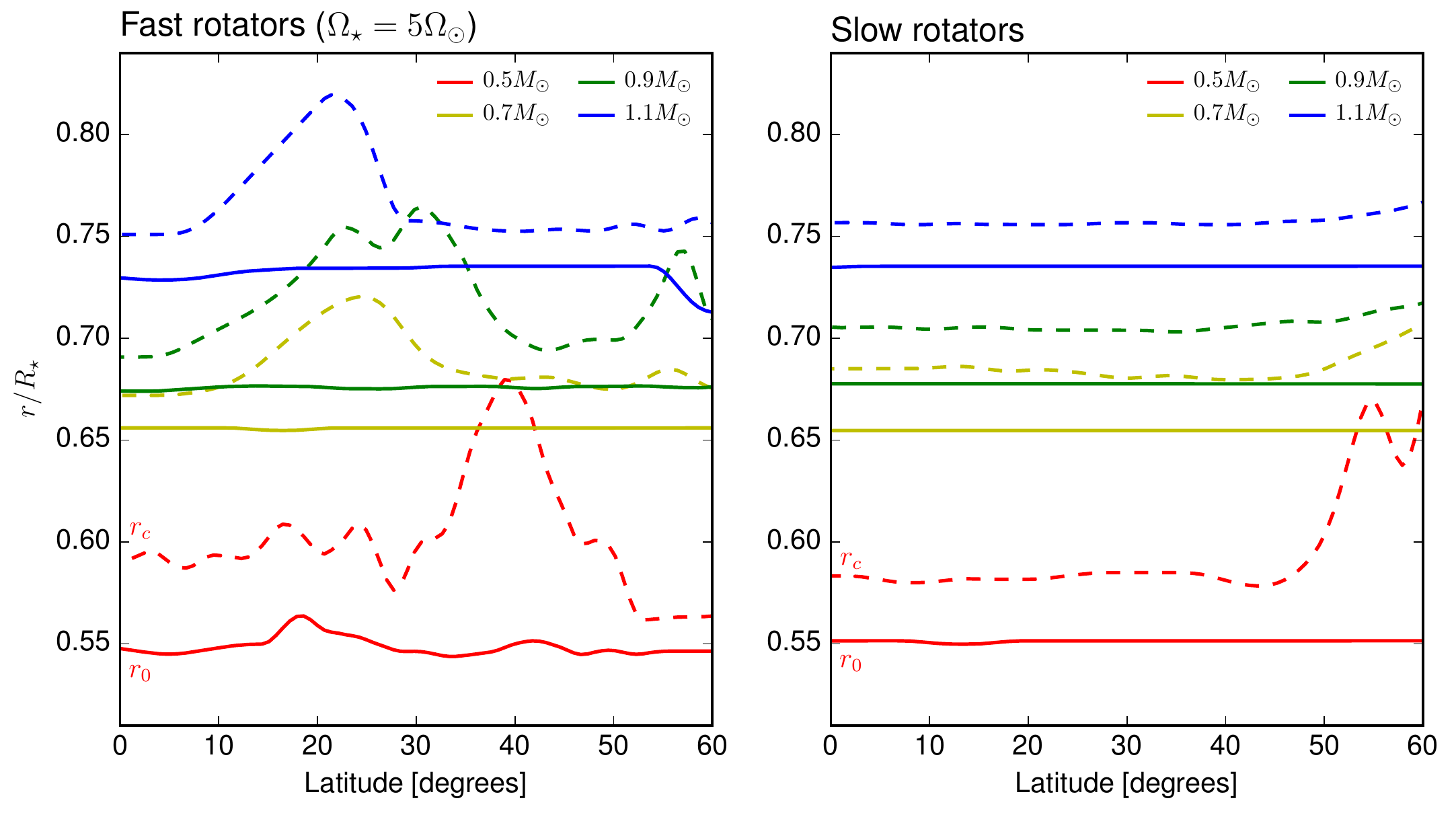}
\end{center}
\caption{Evolution of $r_{0}$ (solid lines) and $r_{c}$ (dashed lines) between the equator to
    the latitude 60$^{o}$ for fast rotators
    ($\Omega_\star=5\Omega_\odot$ left panel), and slowly rotating
    models (right panel). The masses of the models are color-coded as
    in Figure \ref{LumR1D}.
    }
\label{dov_eq_lat}
\end{figure}

In the upper panels of Figure \ref{Radial_E_flux}, we display the temporally and longitudinally
averaged radial enthalpy flux profiles for models M07 with different rotation rates. In these
meridional cuts we can see that the enthalpy flux is mostly concentrated in the convective envelope,
and it is predominantly positive at all latitudes. Some inhomogeneities are apparent. They are
likely due to the moderate degree of turbulence of the simulations, as discussed in
\citet{Miesch:2000gs}. A negative enthalpy flux is observed at the base of the convection zone
(delineated by the dashed black line). As the rotation rate increases (left to right), the radial
enthalpy fluxes maxima drifts from the equator to the poles. In the lower panels, we show radial
cuts of the radial enthalpy flux in the northern hemisphere. On these panels, we identify the
  location where the overshoot begins between the convective and radiative regions as the radial
  position ($r_c$), which is where the radial enthalpy flux crosses zero. The radial location where
  the overshooting ends is the radial position ($r_0$), which is where the radial enthalpy flux is
  only a $10 \%$ of its local minima \citep{Brun:2011bl}. The overshooting motions show a clear
  dependence upon latitude. This is quantified in Figure \ref{dov_eq_lat}, where $r_{0}$ (solid
  lines) and $r_{c}$ (dashed line) are displayed between the equator to the latitude 60$^{\circ}$
  for the fast rotators in our sample (left panel) and the slow
  rotators (right panel). The colors label the masses of the models,
  as in previous figures. The overshoot region is wider near the poles in the
  slowly rotating models than it is in the solar-like cases. The fast
  rotating cases exhibit an interesting localized
  increase of the overshooting depth at smaller and smaller latitude
  when the mass of the star increases. % For models with the solar rotation rate and
  % low masses (M05 and M07 cases), it is displaced toward the stellar surface at middle and low
  % latitudes. This is also true for models with three times the solar rotation rate and larger masses
  % (M07, M09 and M11 cases).
  The shape of the overshooting region is, as a result, sensitive to the 
  rotation rate of the star, with slow rotators favoring a wider
  overshooting region near the poles and fast rotators at mid-to-low latitude.

\begin{figure}[!htb]
\begin{center}
\includegraphics[width=0.5\textwidth]{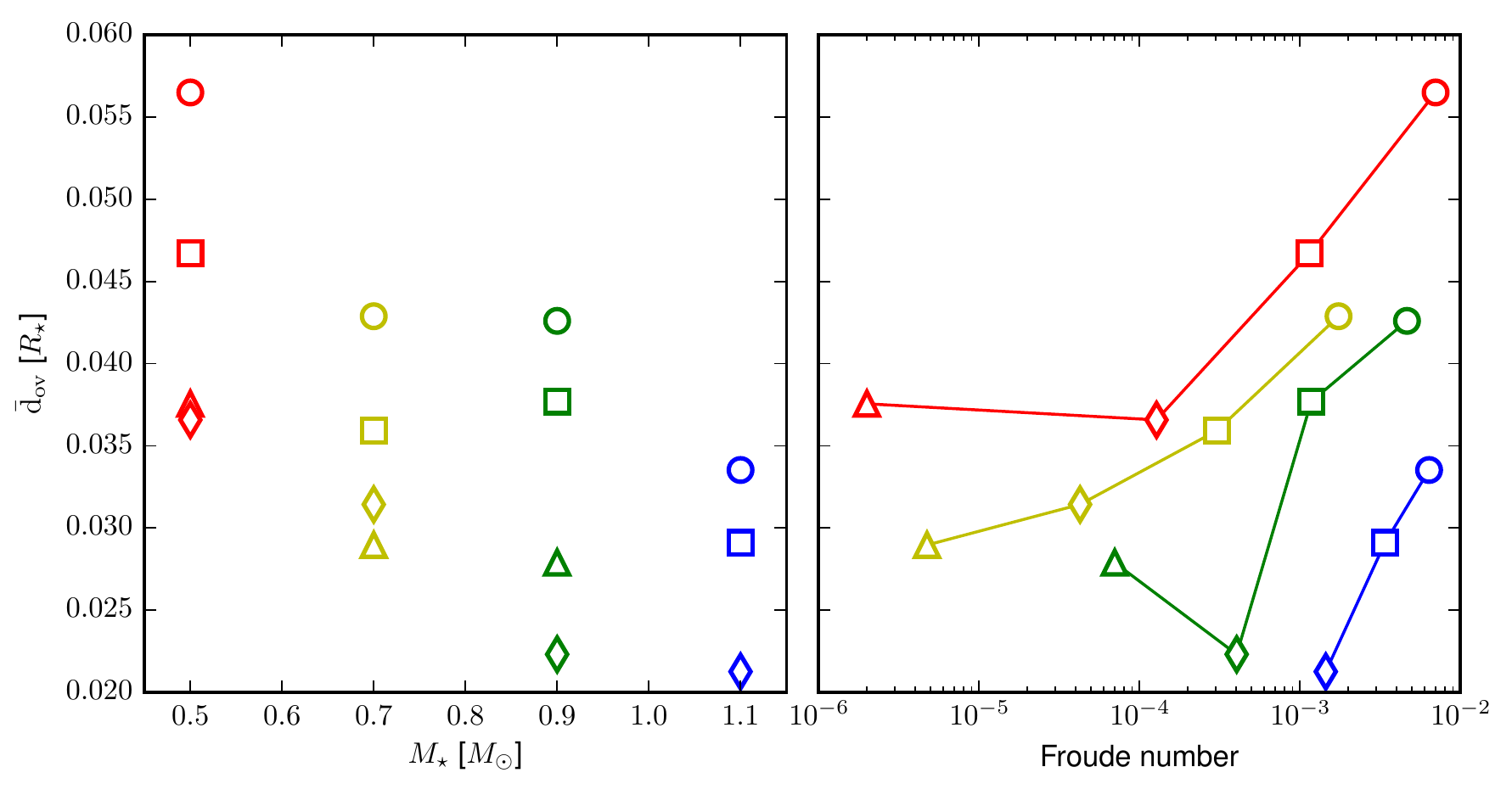}
\end{center}
\caption{Overshooting region width $\bar{\rm d}_{\rm ov}$ as a function
  of mass (left) and Froude number (right), normalized to the stellar radius. The rotation rates of the models are
  labeled by the symbols (S models are triangles, $\Omega_\odot$ models are diamonds, $3\,\Omega_\odot$ models are
  squares and $5\, \Omega_\odot$ models are circles), and the masses are color-coded. % On the right panel the star with
  % the same rotation rate (or same rotation state for the anti-solar-like models) are linked by a dashed black line.
}
\label{Trends_dov_full}
\end{figure}

We define the average overshooting depth $\bar{{\rm d}}_{\rm ov}$ by averaging the difference
  between $r_{c}$ and $r_{0}$ between the equator and latitude 55$^\circ$. We limit ourselves to
  these relatively modest latitudes to avoid any spurious averaging effects associated with the
  large temporal variations of the enthalpy flux at high latitudes. We display $\bar{{\rm d}}_{\rm
  ov}$ as a function of stellar mass and Froude number in Figure \ref{Trends_dov_full}, where
the rotation of the modelled star is indicated by the different symbols, and its mass by their
color. The normalized overshooting depth generally decreases with mass for a given rotation rate, as
is clearly seen on the left panel. The Froude number is defined as
$F_r=\left(2\Omega_\star/N\right)^2$, where $N$ is the
Brunt-V\"{a}is\"{a}l\"{a} frequency.
When the rotation rate is held constant, increasing the Froude number
corresponds to a decrease of the Br\"unt-Vaisala frequency and thus
eases the overshooting of convective plumes impacting the stable
region. We indeed observe a clear trend (see left panel in Figure
\ref{Trends_dov_full}) with an overshooting depth
increasing when the Froude number increases in all of our models. We
finally note that some modulation of the overshooting depth may occur
when magnetic fields are taken into account and will be fully
characterized in a subsequent study \citep[see][]{Varela16}.
% All models show an increased normalized overshoot depth with the
% Froude number. The models with a solar-like differential rotation exhibit a
% decrease of the normalized overshoot depth with the fluid Rossby number, and this trend is reversed
% for anti-solar models at high Rossby numbers. Finally, we note
%   that the estimate of the overshoot depth is robust near the equator,
% but becomes quickly time-variable as we go to higher
% latitudes. As a result, the trends identified in the right panel of
% Figure \ref{Trends_dov_full} need to be considered with caution, and
% we defer the detailed study of the latitudinal and temporal variation
% of the enthalpy flux to a subsequent publication.

\section{Various states of Differential Rotation}
\label{various_states_dr}

We now turn to a discussion regarding how differential rotation in rotating convection zone is
established in various solar-like stars in order to interpret recent observational surveys
\citep{Saar:2009tb,Reinhold:2013eo,doNascimento:2014is,Garcia:2014ds,Reinhold:2015kx}.

\begin{figure*}[!htb]
\begin{center}
\includegraphics[width=0.8\textwidth]{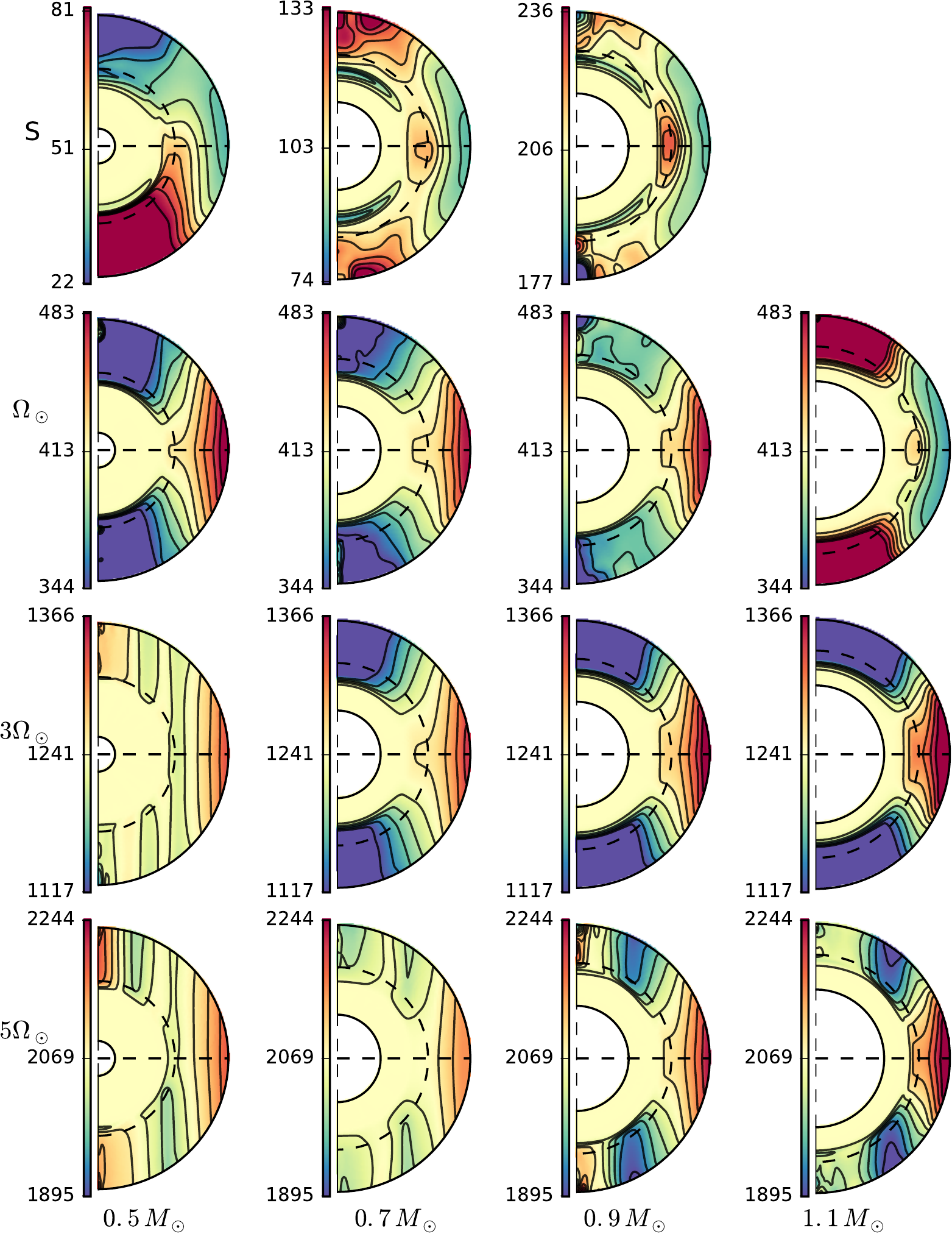}
\end{center}
\caption{The differential rotation realized in our simulations for four masses 0.5, 0.7, 0.9 and
  1.1$M_{\odot}$, and for four rotation rates: 'S' models and one, three, and five times the solar
  rate.}
\label{omega}
\end{figure*}

\subsection{Differential rotation profiles and amplitudes} 
\label{sec_dr_prof}

As is evident in Figure \ref{omega}, various states of differential rotation have been achieved in
our parametric study.  Differential rotation profiles and amplitudes are found to be influenced by
both rotation rate and spectral type. First we see that for each given row (for respectively 'S'
models, one, three, and five times the solar rotation rate going from the top to the bottom of the
figure), the angular velocity patterns change. Moreover, we observe the existence of three main
states of differential rotation for which the location of prograde and retrograde longitudinal flows
differ from one another. For instance, these three states are found near the solar rotation rate
(second row), where there is an anti-solar-like profile (slow equator/fast poles) for the 1.1
$M_{\odot}$ model, a solar-like profile (fast equator/slow poles) for the 0.9 and 0.7 $M_{\odot}$
models, and a Jupiter-like profile (alternating zonal jets and a cylindrical angular velocity
profile) for the 0.5 $M_{\odot}$ model at 3 times solar (third row). 
For the 0.5, 0.7 and 0.9 $M_{\odot}$ models, we also find anti-solar-like profiles when further reducing 
the rotation rate below the solar rate (models 'S' in the upper row). In particular, the 0.9 $M_{\odot}$ 
model shows an anti-solar rotation profile if the rotation rate is half the solar rotation rate, 0.7 $M_{\odot}$ 
for $1/4$ solar rotation rate, and 0.5 $M_{\odot}$ for $1/8$ solar rotation rate, hence for rotation rates 
smaller than those deduced in \S \,\ref{sec:hint_dr_ML} using mixing length. Looking more closely 
at these simulations we notice the retrograde flow near the surface extends up to the tangent cylinder 
in models M07S and M09S. At the equator a zone of rapid rotation at the base of the convection zone 
is present in these two models. At high latitudes, the differential rotation exhibits fast flows akin to polar vortices
already described in the literature \citep{Brown:2008ii,Featherstone:2015bv}. Case M05S rotates so
slowly that the longitudinal average is not well defined, hence the asymmetric profile observed in
the upper left panel of Figure \ref{omega}. This is due to a global dipolar mode of the convective
flows which imprint itself on the overall dynamics.

As discussed in the introduction, the behavior of the differential rotation can be understood to be
a result of the change in the amplitude of the Rossby number of the models. These three main states:
anti-solar $(R_{of} >1)$, solar-like $(0.3 < R_{of} < 0.9)$, jupiter-like or cylindrical-banded
$(R_{of} \lesssim 0.3)$ are encountered in our series of models. For those fastest rotating cases,
the cylindrical differential rotation expected from Taylor-Proudman contraints transits from a
monotonic behaviour from equator to pole into a banded structure of alternating jets. Those jets are
commonly seen in planets like Jupiter and Saturn. Their spacing can be related to the compressible
Rhines scale $\lambda_g$ as discussed in details in \citet{Gastine:2014da}. In models M05R3 and
M05R5, $\lambda_g/2$ is found to be of the order of $15\%$ to $20\%$ of the stellar radius which is
in qualitative agreement with the banded structure seen in Figure \ref{omega} for these two cases.

These results are also compatible with global 3D MHD simulations performed by other authors to model
differential rotation and stellar magnetism in the convection zone
\citep{Miesch:2006iz,Ghizaru:2010im,Racine:2011gh,Kapyla:2011kr,Augustson:2015er,Karak:2015dw},
particularly for solar like stars
\citep{Brun:2004ji,Brown:2010cn,Brown:2011fm,Brun:2011bl,Varela16}. These studies pointed out the
large magnetic temporal variability and the critical effect of stellar rotation and mass on magnetic
field generation through dynamo mechanism, which for some parameter regimes leads to cyclic activity
\citep{Gilman:1981cf,Gilman:1983dx,Nelson:2013fa,Kapyla:2013gr,Augustson:2013jj,Guerrero:2016cz,Augustson:2015er}.
The definition of these three main states allows a fast and straightforward identification of the
expected magnetic temporal variability of the solar like stars, for instance the stars with
anti-solar-like rotation profiles should exhibit smaller magnetic temporal variability than stars
with solar like rotation profile, because the magnetic field regeneration via convective motions
dominates the regeneration via differential rotation, leading to non oscillatory $\alpha^{2}$
dynamos instead of oscillatory $\alpha-\Omega$ dynamos \citep[see, e.g.,][]{Varela16}.

\subsection{Tachoclines}
\label{sec_tacho}

By considering the coupling between the convective envelope and the stable radiative interior for
each modelled stars, we observe the natural development of a transition layer between the two zones for various physical quantities
such as velocity, entropy and temperature fluctuations, and the overall dynamics. Among these
variations, that of the rotation profile is crucial. In the Sun, this transition has been named
\textit{tachocline} \citep{Spiegel:1992tr}, and it is thought to play an important role in the
organization of the eleven years cycle \citep{Charbonneau:2005il,Brun:2015kca}. How such
  tachoclines evolve in other solar-like stars with different global parameters is largely unknown,
  so we will assess that dependence here.

\begin{figure}[!htb]
\begin{center}
\includegraphics[width=0.95\linewidth]{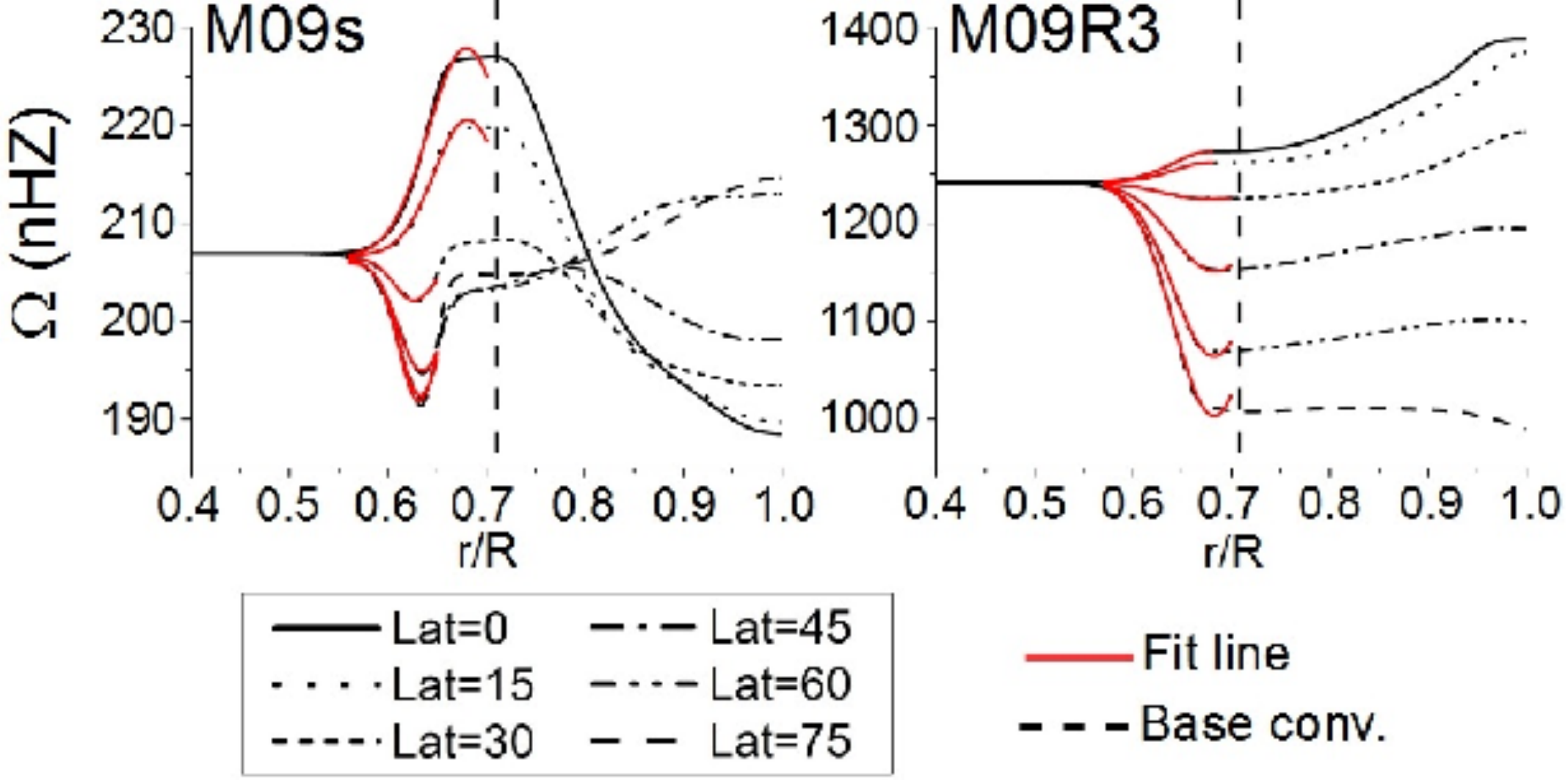}
\end{center}
\caption{Radial cuts of the temporal and longitudinal average of the angular velocity from the
  equator to the latitude 75$^{\circ}$ each 15$^{\circ}$ and between $0.4$ and $1.0$ $r/r_{*}$. Note
  that $r/r_{*} = 0.4$ is not necessarily $r = r_{bcr}$. The dashed vertical line on each panel shows 
  the base of the convective layer.}
\label{Radial_cut}
\end{figure}

The overall shape of the tachoclines achieved in our simulations is shown in Figure
\ref{Radial_cut}, with radial cuts at several latitudes of the temporal and longitudinal average of
the angular velocity for two representative cases. To help quantify these shapes, we fit the
rotation profile in the tachocline with

\begin{equation}
\Omega = \Omega_* + Ae^{-\frac{1}{2}\left[\frac{(x - x_{c})}{w}\right]^2}\, ,
\end{equation}

\noindent where parameter $A$ represents the amplitude of the differential rotation in the
tachocline, $x_{c}$ its inward/outward drift, and $w$ its thickness. The fitted profile for models
M09s and M09R3 are indicated in red in Figure \ref{Radial_cut}.

\begin{figure}[!htb]
\begin{center}
\includegraphics[width=0.8\linewidth]{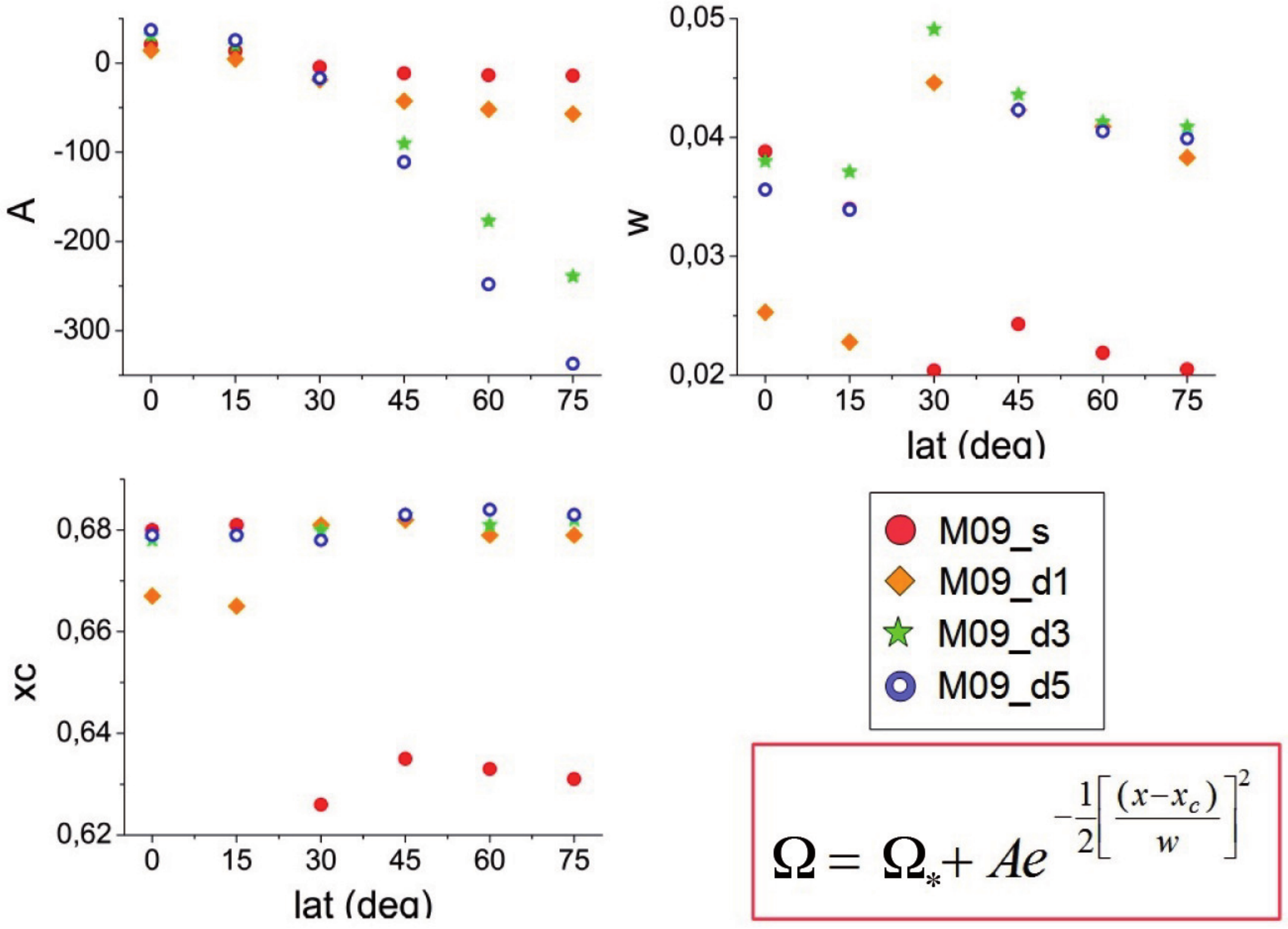}
\includegraphics[width=0.8\linewidth]{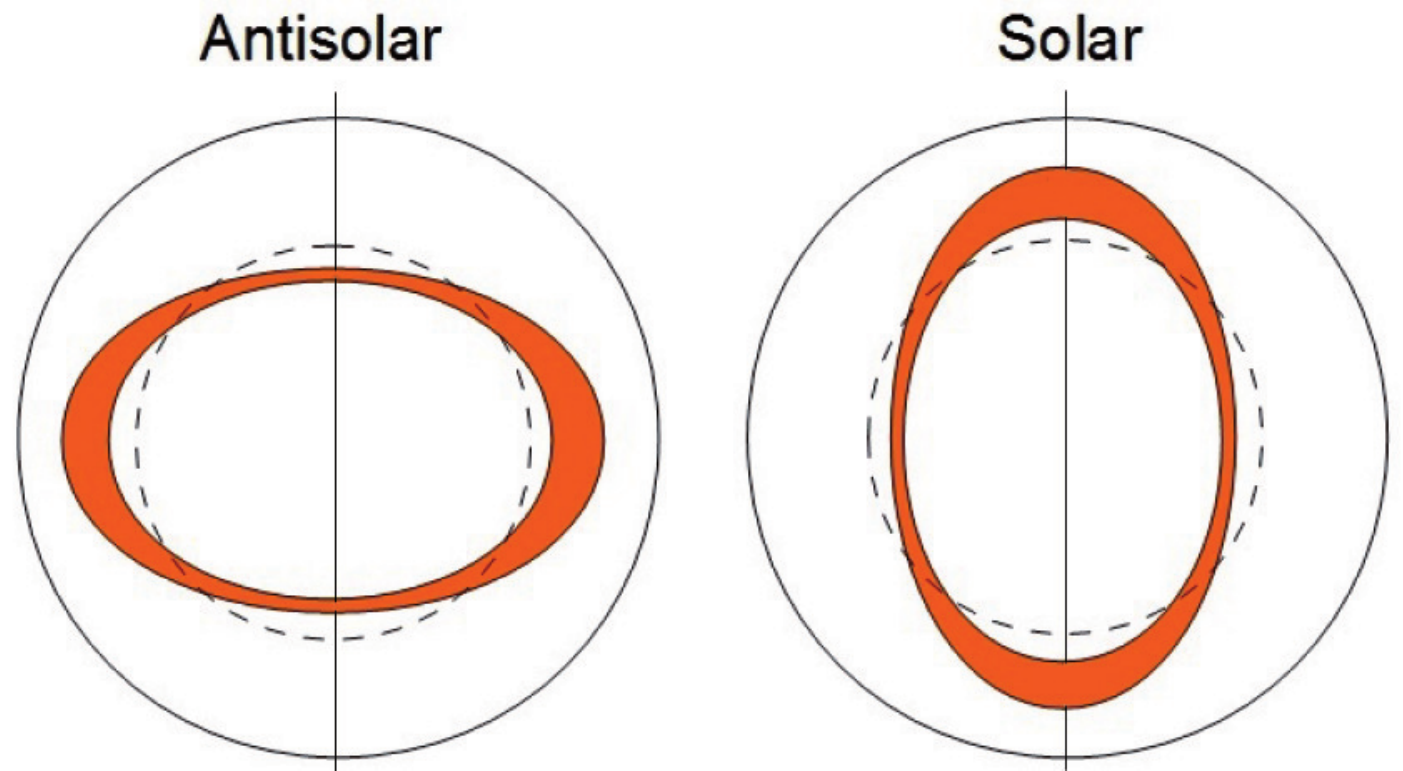}
\end{center}
\caption{\textit{Top panels:} Fit of the rotation in the region of the regime change for models
  M09. \textit{Bottom:} Schematic of the tachocline shape in anti-solar-like (oblate shape) and
  solar-like (prolate shape) cases.}
\label{Rot_M09}
\end{figure}

Applying this methodology to all models, we obtain the following trends, summarized in Figure
\ref{Rot_M09} for models M09. The amplitude $A$ of the differential rotation is found to increase
with the rotation rate. This effect is predominantly seen at high latitude beyond the tangent
cylinder. It may be due to the smaller lever arm in this region with respect to the equatorial
one. The location $x_c$ of the tachocline shows a drastic change of behaviour between anti-solar (red
circles) and solar-like cases. Indeed one notices it is closer to the surface at low latitude and
deeper at mid-to-high latitudes whereas it is the opposite in the solar-like cases. The width $w$ of
the tachocline, in part controlled by our choice of diffusivities, still exhibit a similar clear
trend. The anti-solar cases possess a thicker tachocline at low latitudes compared to higher
latitudes. Again, this trend reverses for the solar-like cases. The overall shape of the tachocline
is summarized in the lower panels of Figure \ref{Rot_M09}.

In order to disentangle the effect of viscous stresses from dynamical effects linked to rotation, we have further analyzed (not
shown) a series of models with approximately the same rms Reynolds number ($\sim$ 100) spanning a fluid
Rossby number from 0.28 to 0.84. By considering a constant Reynolds number for all cases we assume that 
the same degree of turbulence and convective advection with respect to viscouss effect are realized in the simulations and that 
we can even better isolate the role of rotation. The observed trends in shape, location and amplitude are confirmed
with those models, confirming they originate from dynamical effects rather than being viscously
controlled. 

\subsection{Scaling laws}
\label{sec_scaling_laws}

\begin{figure}[!htb]
\begin{center}
\includegraphics[width=0.65\linewidth]{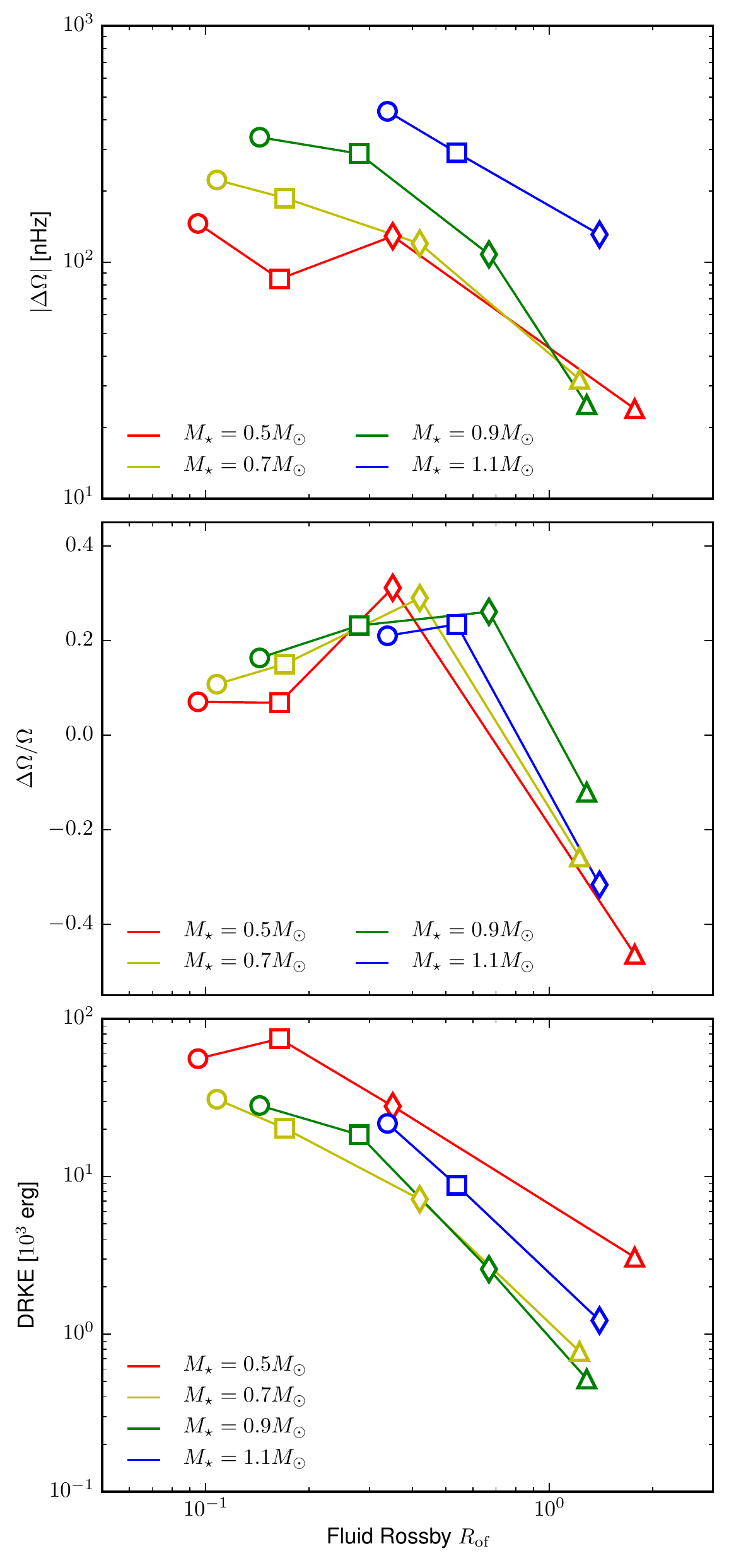}
\end{center}
\caption{Variation of the differential rotation amplitude as a function of mass and fluid Rossby
  number. \texttt{Top:} Differential rotation between the equator and 60$^{\circ}$ latitude. The
  masses are color-coded (see legend), and the different rotation rate ares respectively labeled by
  circles (5 $\Omega_\odot$), squares (3 $\Omega_\odot$), diamonds ($\Omega_\odot$), and triangles
  (models 'S'). \textit{Middle:} Normalized differential rotation (same layout as the upper
  panel). \textit{Bottom:} Kinetic energy of the differential rotation (same layout as the upper
  panel)}
\label{DeltaOmega_trends}
\end{figure}

We define the mean latitudinal contrast of differential rotation $\Delta \Omega$ as the difference
taken at the top of the domain of the azimuthally and temporally averaged profile of $\Omega$
between the equator and latitude $60^\circ$. As a result, a positive $\Delta \Omega$ denotes a
solar-like differential rotation with an equator rotating faster than the higher latitudes. The
values of $\Delta \Omega$ are reported in Table \ref{tablenondimnb2} for all our models. We show
them in the upper panel of Figure \ref{DeltaOmega_trends} as function of the fluid Rossby number
$R_{of}$, coloured by mass and labeled by rotation rate. The latitudinal differential rotation
  generally drops with the fluid Rossby number and increases with the mass of star, and it possibly
  undergoes a transition in the anti-solar cases when $R_{of} > 1$. The
normalized differential rotation (middle panel) $\Delta\Omega/\Omega_*$ tends to show saturation at
low Rossby number, and the anti-solar cases exhibit a transition similar to
\citet{Featherstone:2015bv}. Note that due to a slightly different definition of the Rossby number,
in \citet{Featherstone:2015bv} the transition occurs around $R_{of}\sim 0.1$ while in our case it
occurs around $R_{of}\sim 1$. Finally, this transition is less clear in the differential rotation
kinetic energy (lower panel), which is defined by

\begin{equation}
  \label{eq:DRKE_def}
        {\rm DRKE} = \frac{1}{2}\int \bar{\rho} \left\langle v_\phi\right\rangle^2 r^2\sin\theta{\rm d}r
        {\rm d}\theta {\rm d}\phi\, .
\end{equation}
  
Indeed, the kinetic energy of the differential rotation decreases smoothly with the fluid Rossby
number, more simulations with a larger Rossby number would be needed to confirm this lack of
transition in the kinetic energy.

We use our set of models to fit its dependency upon mass and fluid Rossy number (as well as mass and
rotation rate) and obtain
  \begin{eqnarray}
  \label{eq:dOm_Mass_Rossby}
  |\Delta \Omega| &\propto&
                            \left(\frac{M_\star}{M_\odot}\right)^{1.93 \pm 0.42}
                            \left(R_{of}\right)^{-0.76 \pm 0.13}\, ,\\
  \label{eq:dOm_Mass_Omega}
                  &\propto& \left(\frac{M_\star}{M_\odot}\right)^{0.73\pm 0.39}
                            \left(\frac{\Omega_\star}{\Omega_\odot}\right)^{0.66 \pm 0.10}\, .
\end{eqnarray}

If we do not retain the anti-solar cases in our regression fit, we have:
\begin{eqnarray}
  \label{eq:dOm_Mass_Rossby2}
  |\Delta \Omega| &\propto&
                            \left(\frac{M_\star}{M_\odot}\right)^{1.78 \pm 0.34}
                            \left(R_{of}\right)^{-0.44 \pm 0.16}\, .
\end{eqnarray}

The observational trends for the differential rotation rate exhibit either a similar dependency with rotation rate 
\citep[$\Delta\Omega \propto \Omega_\star^{0.7}$, see][]{Donahue:1996ch,Saar:2009tb} or 
a significantly lower one \citep[$\Delta\Omega \propto \Omega_\star^{0.15}$,
  see][]{Barnes:2005eb,Reinhold:2013iz}, making a direct comparison difficult. 
One explaination could be the lack of dynamo generated magnetic fields in this series of hydrodynamic models. 
Such magnetic field should feedback on the global balance establishing the large scale differential rotation pattern (see \S
\,\ref{sec_amom}). Indeed, a preliminary study of the cases presented in this work that also include
a dynamo field finds a lower exponent for the dependency upon $\Omega$
\citep[see][]{Varela16}, improving the comparison of our theoretical models with observational
trends.   
  
The trend with mass while retaining the same positive variation with $M_*$ also differs somewhat from the observational
trends ($\Delta\Omega \propto M_\star^{5.6}$ in \citealt{Barnes:2005eb,Reinhold:2013iz},
$\Delta\Omega \propto M_\star^{5.4}$ in \citealt{CollierCameron:2007ce}) .  
We attribute these differences in part to the non perfect relationship between mass and effective temperature in the observational data 
and also to a lower upper bound in stellar mass in our study compared to the observational data. 
Indeed in \citet{Augustson:2012bn} we have simulated more massive F stars 
up to 1.4 $M_{\odot}$ and found a larger dependency of the differential rotation with stellar mass. 
Hence it is likely that the differential rotation contrast increases more steadily with mass between 1.1 and 1.4 $M_{\odot}$
given the shallow convective envelop in these stars than it does for lower masses 
\citep[see also the mean field computations of][that confirms the trends found in 3-D simulations]{2007AN....328.1050K,2011A&A...530A..48K}. More recent observations tends to
also show a change of slope around the stellar spectral type F \citep[see for instance][]{Reinhold:2013eo}.

%\kyle{What about the huge mass discrepancy? Can this be reconciled? Perhaps by appealing to
%  the degeneracy between $\Delta \Omega$ and $\Delta T$ in some observing techniques.}

\begin{figure*}[!htb]
\begin{center}
\includegraphics[width=0.6\textwidth]{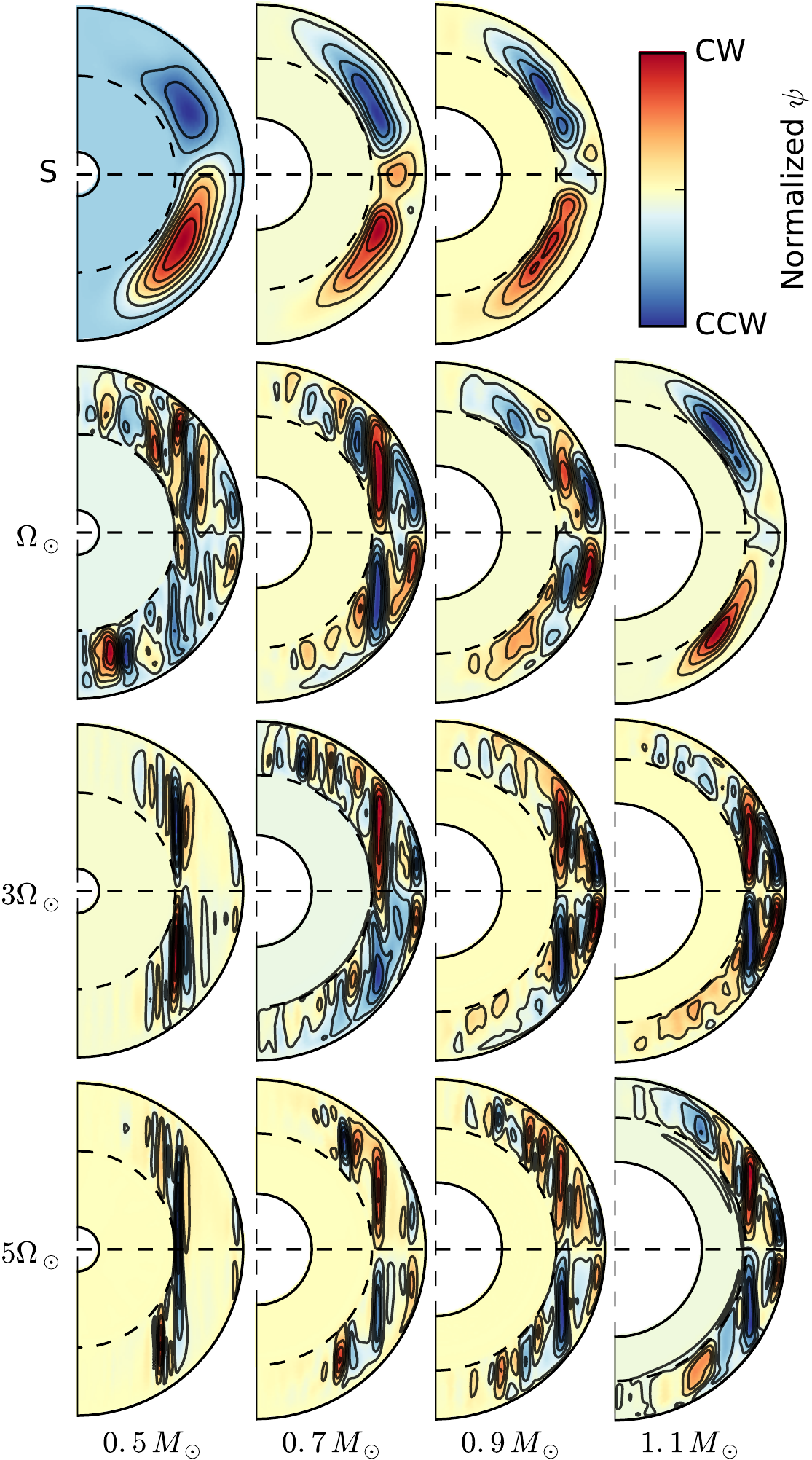}
\end{center}
\caption{Meridional circulation realized in our 3-D simulations. The layout is the same as in Figure
  \ref{omega}, and CW stands for clockwise meridional circulation while CCW stands for
  counter-clockwise meridional circulation.}
\label{mc}
\end{figure*}

\section{Meridional Circulation}
\label{sec_mc}

Meridional circulation results from the slight imbalance between the terms acting
  predominantly in geostrophic balance. These are the horizontal pressure gradients, Reynolds
  stresses, curvature terms and Coriolis force, in a purely hydrodynamical setup. As we
change the aspect ratio and the rotation rates of our convective shells, we expect the relative
amplitude of these terms to change as well, resulting in different meridional circulation profiles.

\subsection{Profiles and Amplitudes}
\label{sec_prof_and_amp}

In solar-like star simulations, we expect the meridional circulation in convective envelope to be weak because
geostrophy is mostly satisfied \citep{Pedlosky:1987vt,Ballot:2007ea,Brown:2008ii,Augustson:2012bn}. In Figure \ref{mc}
we display the meridional circulations realized in the fifteen models as contours of the meridional streamfunction
$\Psi$, defined as in \citet{Miesch:2000gs}:

\begin{equation}
  \label{eq:psi_def}
r \sin\theta\langle \rb v_r \rangle = -\frac{1}{r}\frac{\p \Psi}{\p \theta}\, \mbox{ and } r \sin\theta\langle \rb v_\theta \rangle = \frac{\p \Psi}{\p r}.  
\end{equation}

We note two main trends: solar-like models have multi-cellular flow structures, and the anti-solar
ones possess mostly unicellular meridional circulations per meridional quadrant. As the Rossby number is decreased (from
right to left), the number of cells increases, in particular near the polar cap. The amplitude of
the meridional circulation in the convective envelope is of order of meters per second (see Table
\ref{Table 3b}), being weaker for the low mass stars compared to the massive ones (as already
discussed in \S \,\ref{sec_conv_state}). In the radiative interior, this flow is extremely weak, the
radial velocity dropping by several orders of magnitude. This results in a penetration of the
meridional circulation of less than $3$ to $5\%$ of the stellar radius. The flow in the anti-solar
cases is directed poleward in both hemispheres at the surface, with a return flow at the base of the
convection zone. This is also true for the upper meridional cells in the faster rotating cases.

A more direct way to understand the maintenance of the meridional circulation is to consider the angular momentum
balance. By splitting the right hand-side term between a global net torque coming from the difference between Reynolds
and viscous stresses and from the advection of angular momentum by the meridional circulation (and by assuming
stationarity), we get:

\begin{equation}\label{eq:gp}
\left<\rh {\bf v}_m \right> \bdot \del {\cal L} = {\cal T} ~~~, 
\end{equation}

\noindent where

\begin{equation}\label{eq:amom}
{\cal L} = \varpi^2 \Omega = 
\varpi \left(\left<v_\phi\right> + \varpi \Omega_*\right) ~~~,
\end{equation}

\noindent with ${\cal T}$ the global net torque (whose expression will be made more explicit in
section \ref{sec_amom}), and $\varpi = r \sin\theta$ the moment arm.

 If the global net torque ${\cal T}$ is zero, then there is no meridional circulation. In
  stars, we do not expect the viscous stresses to play a major role and hence the meridional
  circulation arises in order to compensate for the angular momentum transport due to convection
  (Reynolds stresses), provided that the magnetic effects are negligible. Since we are considering
  purely hydrodynamical cases and the viscous stresses necessarily contribute to the angular
  momentum transport in our simulations, the meridional circulation develops as a response to the
  net torque exerted by both the Reynolds stresses and the viscous diffusion of the differential
  rotation. The responding circulation to an applied torque is due to a physical mechanisms called
gyroscopic pumping that generalizes Ekman pumping
\citep{McIntyre:2007ww,Garaud:2010kd,Brun:2011bl,Miesch:2011cg,Featherstone:2015bv}.

\subsection{Scaling laws}
\label{sec_scal_laws_2}

\begin{figure}[!htb]
\begin{center}
\includegraphics[width=0.9\linewidth]{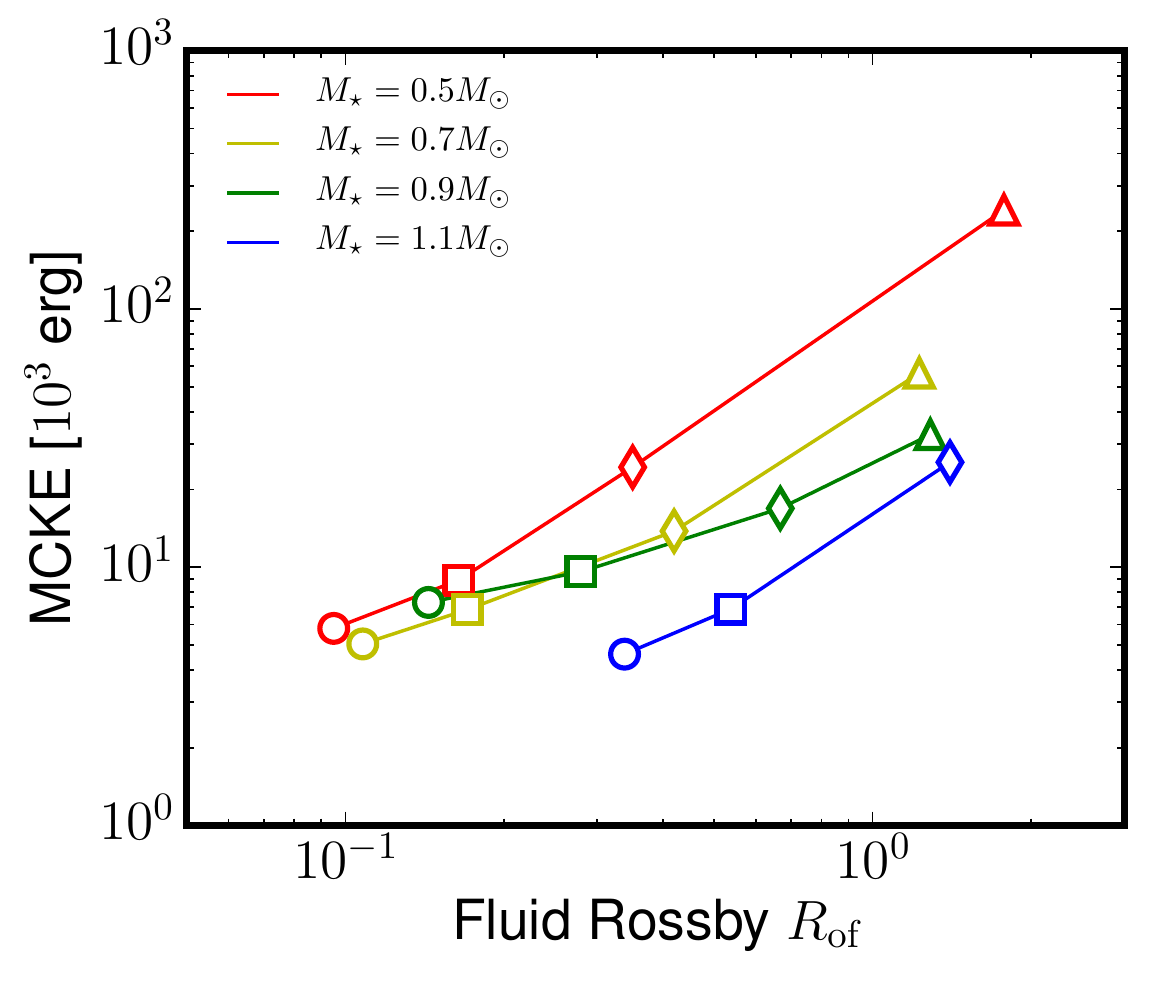}
\end{center}
\caption{Meridional circulation kinetic energy as a function of the fluid Rossby number $R_{of}$. The layout is the same
  as in Figure \ref{DeltaOmega_trends}.  }
\label{MCKE_trends}
\end{figure}

We quantify the strength of the meridional circulation by calculating its associated kinetic energy
defined as

\begin{equation}
    \label{eq:MCKE_def}
    {\rm MCKE} = \frac{1}{2}\int \bar{\rho}\left(\left\langle v_r\right\rangle^2 +  \left\langle v_\theta\right\rangle^2\right)  r^2\sin\theta{\rm d}r {\rm d}\theta {\rm d}\phi\, ,
\end{equation}

\noindent where $\left\langle\right\rangle$ stands for the azimuthal average and MCKE has been further averaged in time
over 100 days towards the end of each simulations. We display in Figure \ref{MCKE_trends} the trends of MCKE as a function of fluid Rossby number. We notice that the energy
contained in the meridional circulation increases with Rossby number, and decreases with mass. The
meridional circulation energy also increases with the Reynolds number (not shown here), which is
naturally expected as the meridional circulation results from the imbalance between the turbulent
Reynolds stresses and the way it advects angular momentum (see \citealt{Featherstone:2015bv} and \S
\ref{sec_prof_and_amp}). Finally, we use our set of models to fit the dependency of MCKE upon mass
and fluid Rossby number and obtain
  
\begin{eqnarray}
  \label{eq:MCKE_Mass_Rossby}
  {\rm MCKE} &\propto&
                 \left(\frac{M_\star}{M_\odot}\right)^{-1.90 \pm 0.30}
                 \left(R_{of}\right)^{1.05 \pm 0.09}\, .
\end{eqnarray}

\section{Analyzing the Dynamics}
\label{sec_dynamics}

In order to better understand the various dynamical states analyzed in the previous sections, we now
turn to a more quantitative analysis of the dynamics. We will first look at the angular momentum
redistribution in convective shells achieved in our simulations (\S \,\ref{sec_amom}), then on the
thermal wind balance (\S \,\ref{sec_TWB}), and finally of the energy exchange and maintenance of the
differential rotation (\S \,\ref{sec_energy_echange}).

\begin{figure*}[!htb]
\begin{center}
\includegraphics[width=0.49\textwidth]{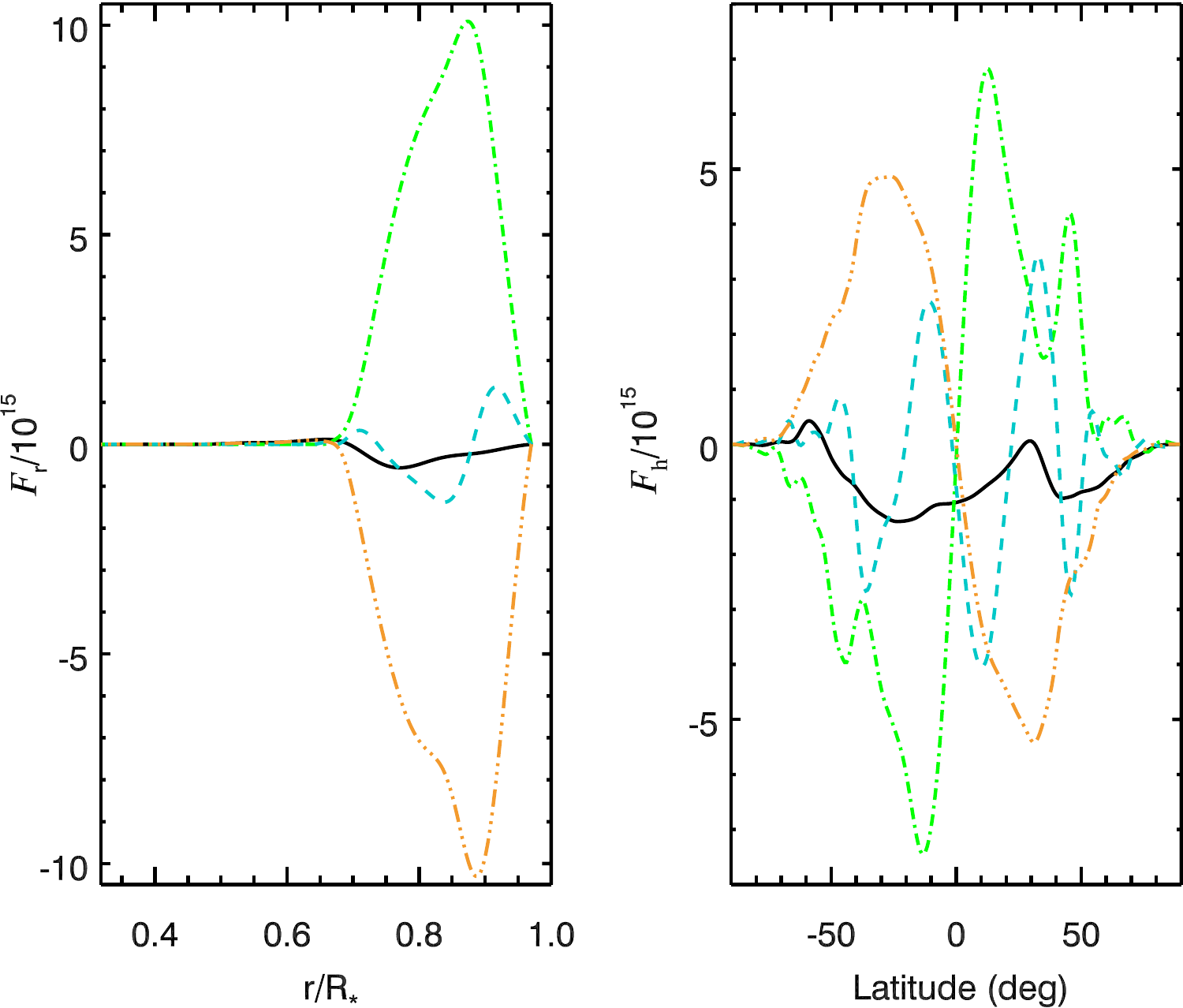}\hfill
\includegraphics[width=0.49\textwidth]{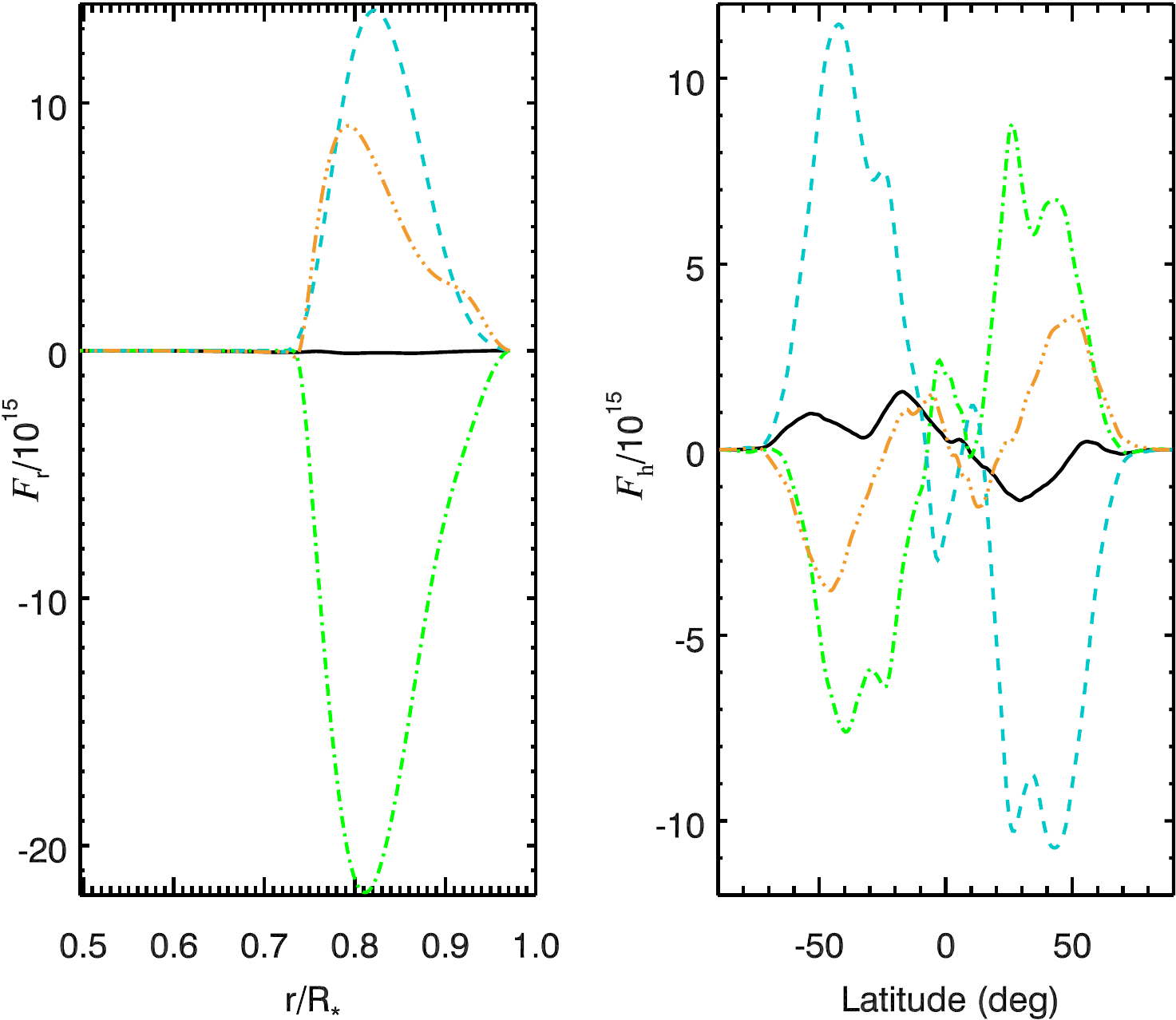}
\end{center}
\caption{Radial and latitudinal balances of the angular momentum fluxes in models M07R1 and M11R1. Reynolds stresses
are shown in green/dahs dot lines, meridional circulation in cyan/dash lines and the viscous stresses in orange/dash-triple dot lines.
The total balance is show as the solid black line.}
\label{Amombal}
\end{figure*}

As a preliminary analysis, we turn to Table \ref{tablenondimnb2} where various global quantities
characterizing the dynamics achieved in our simulations are listed. We note that the kinetic energy
contained in the convective shell increases with the rotation rate due to a global increase in the
energy contained in the differential rotation (DRKE). By contrast, the energy contained in the
convective motion (CKE=KE-DRKE-MCKE) is found to play a lesser role and decreases in overall
amplitude and significantly in proportion. This is likely due to the propensity of rapidly rotating
convective sphere to predominantly inject energy into the longitudinal flows. The kinetic energy
contained in the meridional circulation is always small, and it globally decreases in amplitude to
become almost negligible at the fastest rotation rates. In agreement with the DRKE, and as discussed
in \S \,\ref{various_states_dr}, the absolute angular velocity contrast in latitude from 0 to
60$^\circ$ ($|\Delta \Omega|$) is found to increase with rotation rate, whereas the relative
contrast decreases (see Figure \ref{DeltaOmega_trends}). We will come back to the temperature and
entropy contrasts in \S \,\ref{sec_TWB}.

\begin{table*}[!ht]
\begin{center}
\caption{Kinetic energies normalized by the shell volume, Latitudinal angular velocity, Temperature and Entropy latitudinal contrasts \label{tablenondimnb2}
}
\vspace{0.2cm}
%\begin{tabular}{||p{1.8cm}*{1}{||c} cccc ||}
\begin{tabular}{cccccccc}
\tableline
\tableline
\\ [-1.5ex]
 Name & KE & DRKE & MCKE & CKE  & $\Delta\Omega$ & $\Delta T$ &  $\Delta  S$ \\ [0.5ex]
 & $10^6 erg$ & $10^6 erg$ & $10^3 erg$ & $10^6 erg$ & $nHz$ & $K$ & $10^3 erg/g/K$  \\ [0.8ex] 
\tableline
\tableline
\\ [-1.5ex]
 M05 S & 16.0  & 3.07  (19.2\%)   &   241.4  (1.5\%)  & 12.7  (79.3\%) & -24 & -1.9 & -0.005 \\
 M05 R1 & 31.2 & 27.9 (89.4\%)  &   24.4  (0.1\%) & 3.28 (10.5\%) & 129 & 2.9 & 0.2 \\
 M05 R3 & 78.8  & 74.8 (94.9\%)  &   8.92  (0.01\%) & 4.01 (5.09\%) &  85 & 2.8 & 0.2 \\
 M05 R5 & 58.4  & 55.9 (95.7\%)  &   5.80  (0.01\%)  & 2.48 (4.29\%) & 146 & 12.9 & 0.9 \\ [0.5ex]
\hline
\\ [-2ex]
  M07 S &  3.59   &   0.775  (21.5\%)  &    56.4  (1.6\%)  & 2.76 (76.9\%) & -32 & -2.8 & -0.01 \\
  M07 R1 & 8.89  & 7.17   (80.6\%)   &   13.8   (0.2\%)  & 1.71 (19.2\%) & 120 & 2.5 & 0.2 \\
  M07 R3 & 21.9  & 20.2   (91.7\%)  &    6.85  (0.1\%) & 1.81  (8.2\%) & 187 & 13.8 & 1.1 \\
  M07 R5 & 34.9  & 30.9   (88.7\%)   &   5.05  (0.01\%) & 3.94 (11.29\%) & 223  & 27.8 & 2.3 \\ [0.5ex]
\hline
\\ [-2ex]
  M09 S & 3.25 &  0.519  (16.0\%)  &   32.6  (1.0\%)  & 2.69  (83.0\%) & -25 & -2.5 & -0.1 \\
  M09 R1 & 4.59  & 2.58  (56.2\%)   &   16.9   (0.4\%) & 1.99 (43.4\%) &108 & 8.4 & 0.9 \\
  M09 R3 & 19.9  & 18.4   (92.1\%)    &  9.61  (0.05\%) & 1.57 (7.85\%) & 288 & 57.5 & 5.9 \\
  M09 R5 & 5.49  & 28.2  (51.5\%)   &   7.31  (0.01\%)  & 26.6 (48.49\%) & 338 & 94.9 & 9.75 \\ [0.5ex]
 \hline
 \\ [-2ex]
  M11 R1 & 3.08 & 1.22  (39.6\%)   &   25.5   (0.8\%) & 1.84 (59.6\%) & -131 & -10.2 & -1.52 \\ 
  M11 R3 & 9.96 &  8.79  (88.3\%)   &   6.89  (0.07\%)  & 1.16 (11.63\%) & 291 & 133.5 & 20.7 \\
  M11 R5 & 22.8  & 21.7  (94.9\%)   &   4.61  (0.02\%)  & 1.16 (5.08\%) &  435 & 328.9 & 52.1 \\ [0.5ex]
\hline
 \tableline
 \tableline
\end{tabular}
\end{center}
\end{table*}

\subsection{Angular momentum balance}
\label{sec_amom}

The differential rotation and meridional circulation profiles in our simulations are established and maintained through
the transport of momentum and energy by convective motions that are influenced by the rotation, stratification, and
spherical shell geometry.

Following \citet{Elliott:2000kp} and \citet[][hereafter BT02]{Brun:2002gi}, an equation for the angular
momentum transport can be deduced from the $\phi$-component of the momentum equation:

\begin{equation}
\rb\frac{\p L}{\p t} = \mbox{\bf{$\tau$}}
\end{equation}

\noindent with $L=r \sin\theta v_{\phi}$ the specific angular momentum and $\mbox{\bf{$\tau$}}$ the
net torque applied to the convective envelope.  Assuming a statistically stationary state, applying
longitudinal and temporal averages and writing the net torque as a divergence of a flux one gets
\citep[see also][]{Pedlosky:1987vt}:

\begin{equation}\label{eq:gp2} 
\dv \calf = \dv \left( {\cal F}_r \uvr + {\cal F}_\theta \uvt \right) = 0 
\end{equation}  

\noindent where ${\cal F}_r(r,\theta)$ and ${\cal F}_\theta(r,\theta)$ represent the mean radial and
latitudinal angular momentum fluxes whose expressions are given by:

\begin{eqnarray}
{\cal F}_r&=&\hat{\rho}r\sin\theta \left[ {\cal F}_{r,V}+{\cal
              F}_{r,R}+{\cal F}_{r,MC} \right] \, ,\\
{\cal F}_{\theta}&=&\hat{\rho}r\sin\theta \left[{\cal
                     F}_{\theta,V}+{\cal F}_{\theta,R}+{\cal
                     F}_{\theta,MC} \right]\, .
\end{eqnarray}

${\cal F}_{r,V}$ (resp. ${\cal F}_{\theta,V}$) is the flux associated to viscous transport, ${\cal F}_{r,R}$
(resp. ${\cal F}_{\theta,R}$) that related to Reynolds stresses and ${\cal F}_{r,MC}$ (resp.  ${\cal F}_{\theta,MC}$)
represents the angular momentum flux due to meridional circulation. Their detailed expression can be found in BT02 and
in subsequent publications. As was done in BT02 we then integrate respectively each flux over colatitude and radius to
assess the net flux through a sphere of varying radius and through cones of varying inclination:

\begin{eqnarray}\label{eq:integf_r}
I_r(r)&=&\int_0^{\pi} {\cal F}_r(r,\theta) \, r^2 \sin\theta
\, d\theta \; \mbox{ , } \\ \label{eq:integf_theta}
I_{\theta}(\theta)&=&\int_{r_{bot}}^{r_{top}} {\cal
F}_{\theta}(r,\theta) \, r \sin\theta \, dr \, ,
\end{eqnarray}

These integrated fluxes are presented in Figure \ref{Amombal} for our cases M07R1 and M11R1, and have been averaged over
5 rotation periods. For cylindrical cases such as M05R3,R5 the balance is very similar to that
shown for M07R1. For simplicity we drop the letter $I$ when discussing the individual contribution of the flux. Since
a statistically stationary state is realized in our simulations, the sum of all fluxes must be close to zero as there
are no net torques left.

We start by discussing the angular momentum balance realized in case M07R1. We note that in the
radial direction the prograde Reynolds stresses act to accelerate the equator opposed mainly by the
viscous stresses and to a lesser extent by the meridional circulation.  As discussed in
\citet{Brun:2011bl} and \S \,\ref{sec_mc}, the meridional circulation is a response to the net
torque applied by the sum of the Reynolds and viscous stresses, it helps reaching a stationary
state.  We see that it is both positive and negative, reflecting the presence of multiple
cells. Convection is thus carrying angular momentum such as to accelerate the equator and slow down
the deep layers. Turning to the latitudinal flux balance, here too the Reynolds stresses are found
to carry angular momentum towards the equator (positive/negative in northern/southern hemisphere
respectively), with a peak value near $15^{\circ}$ and a secondary peak near the latitude of the
tangent cylinder. The viscous stresses are poleward as they tend to erase the differential rotation
in the convective envelope. The role of the meridional circulation is mostly to transport angular
momentum poleward at low latitude, with a small counter cell higher up. Overall the balance is well
realized as shown by the sum being nearly equal to zero in both radial and latitudinal balances.

Turning now to case M11R1, we see that in the radial direction the Reynolds stresses are now
transporting angular momentum inward, hence slowing down the surface and speeding up the deep
layers. The viscous stresses and meridional circulation have very similar amplitudes and profiles
and transport angular momentum outward. In this case, the flux associated with meridional
circulation is positive at all depths and relatively large, reflecting the mostly unicellular
profile of the meridional flow (illustrated in Figure \ref{mc}). In the radial direction, the
  angular momentum balances in cases M07R1 and M11R1 are thus of an opposite sign, in agreement with
  their differential rotation being respectively solar-like and anti-solar-like. In the latitudinal
  direction, the situation is not as dissimilar as one could have anticipated. Indeed, the Reynolds
  stresses have not changed sign with respect to case M07R1. What has changed is the amplitude and
  profile of the meridional circulation, with the viscous stresses now aiding the Reynolds stresses
  instead of opposing them. This is again due to the sign of the differential rotation. With a fast
  pole and a slow equator, the viscous stresses tend to slow down the pole and speed up the
  equator. Through gyroscopic pumping, this results from a meridional circulation profile that
  transports angular momentum poleward such that a stationary state is established. These subtle
  differences between solar-like and anti-solar-like cases are in agreement with the results from
  \citet{Featherstone:2015bv}, which were obtained without an underlying stable zone. As can be seen
  by the solid curve, the latitudinal angular momentum fluxes are not strictly balancing one another for case
  M11R1. This is in contrast to the three other panels, and it is due to the slow equilibration of
  this case.

\subsection{Thermal wind balance}
\label{sec_TWB}

\begin{figure*}[!htb]
\begin{center}
\includegraphics[width=0.9\textwidth]{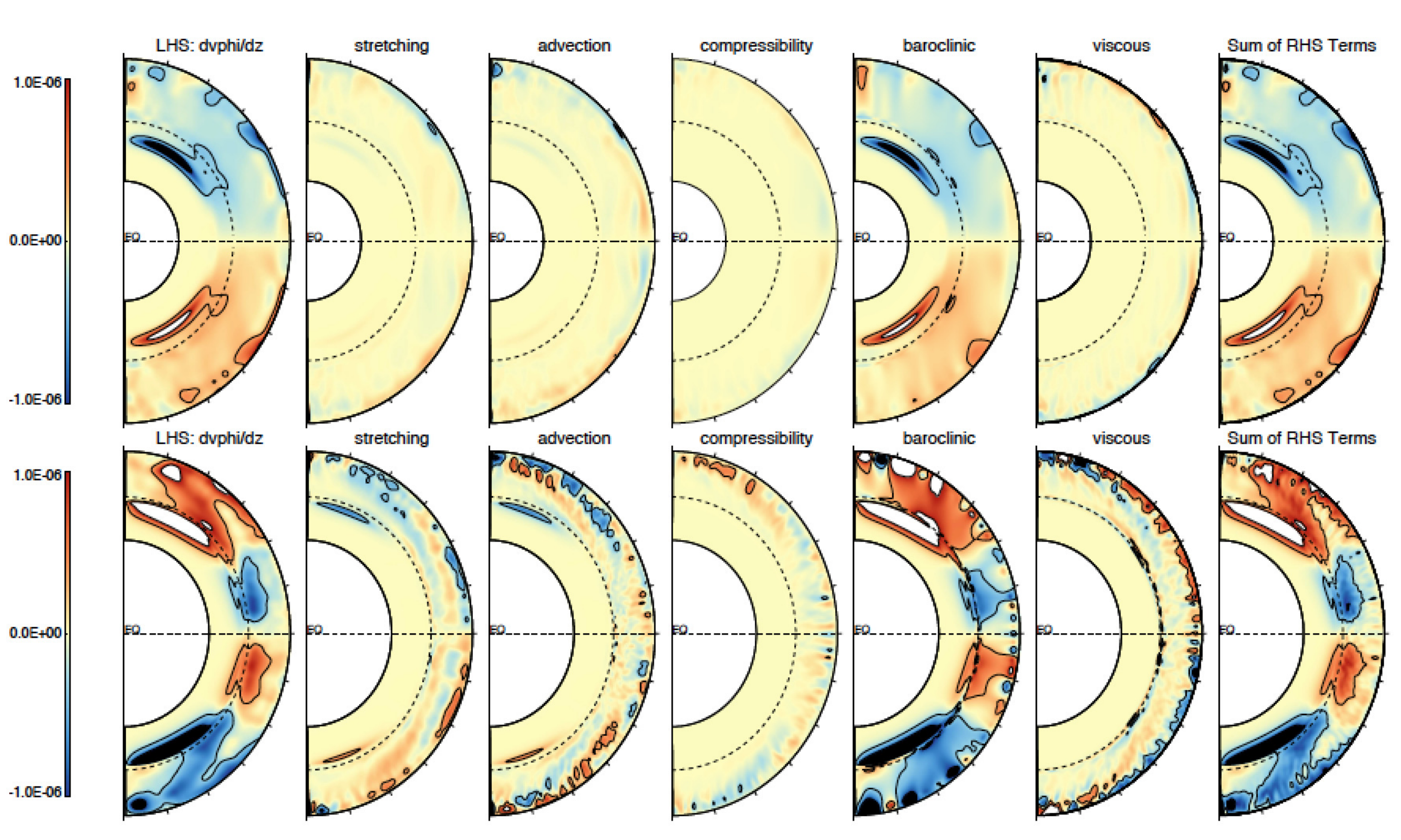}
\end{center}
\caption{Meridional cuts of the temporal and azimuthal average of the full thermal wind balance
  equation in models M07R1 and M11R1 shown as contour plots.  The mathematical expressions of the
  various terms can be found in the Appendix.}
\label{TWbal}
\end{figure*}

\begin{figure*}[!htb]
\begin{center}
\includegraphics[width=\textwidth]{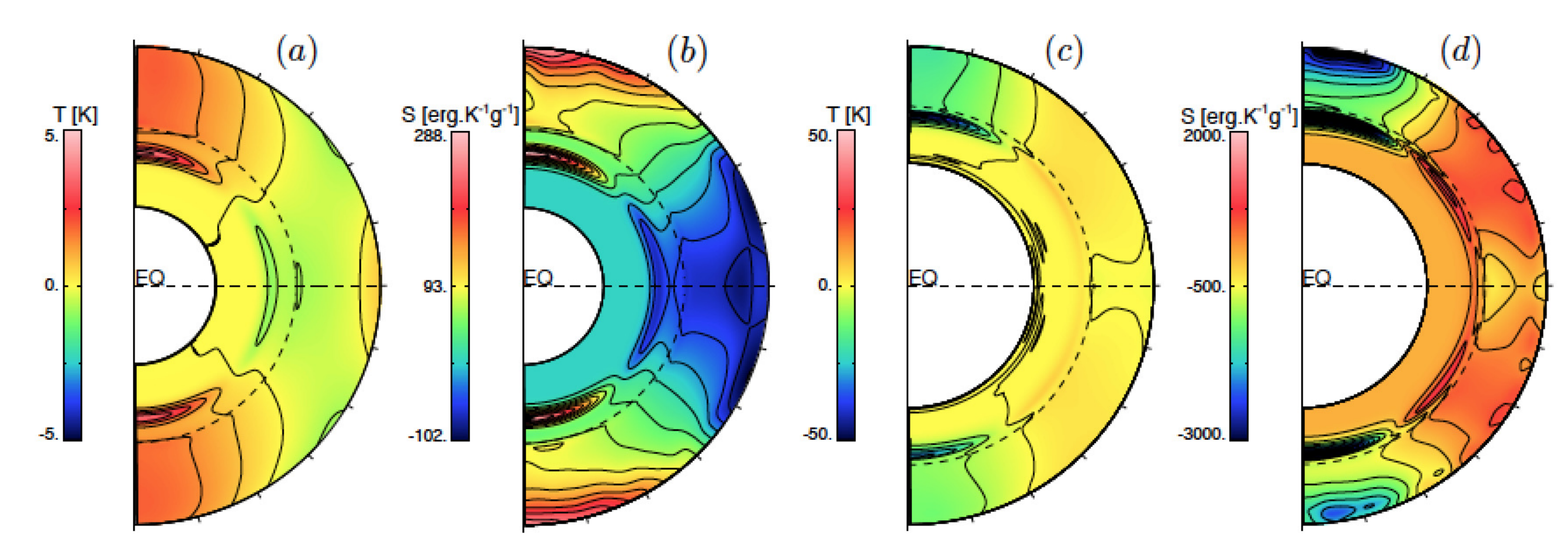}
\includegraphics[width=0.49\textwidth]{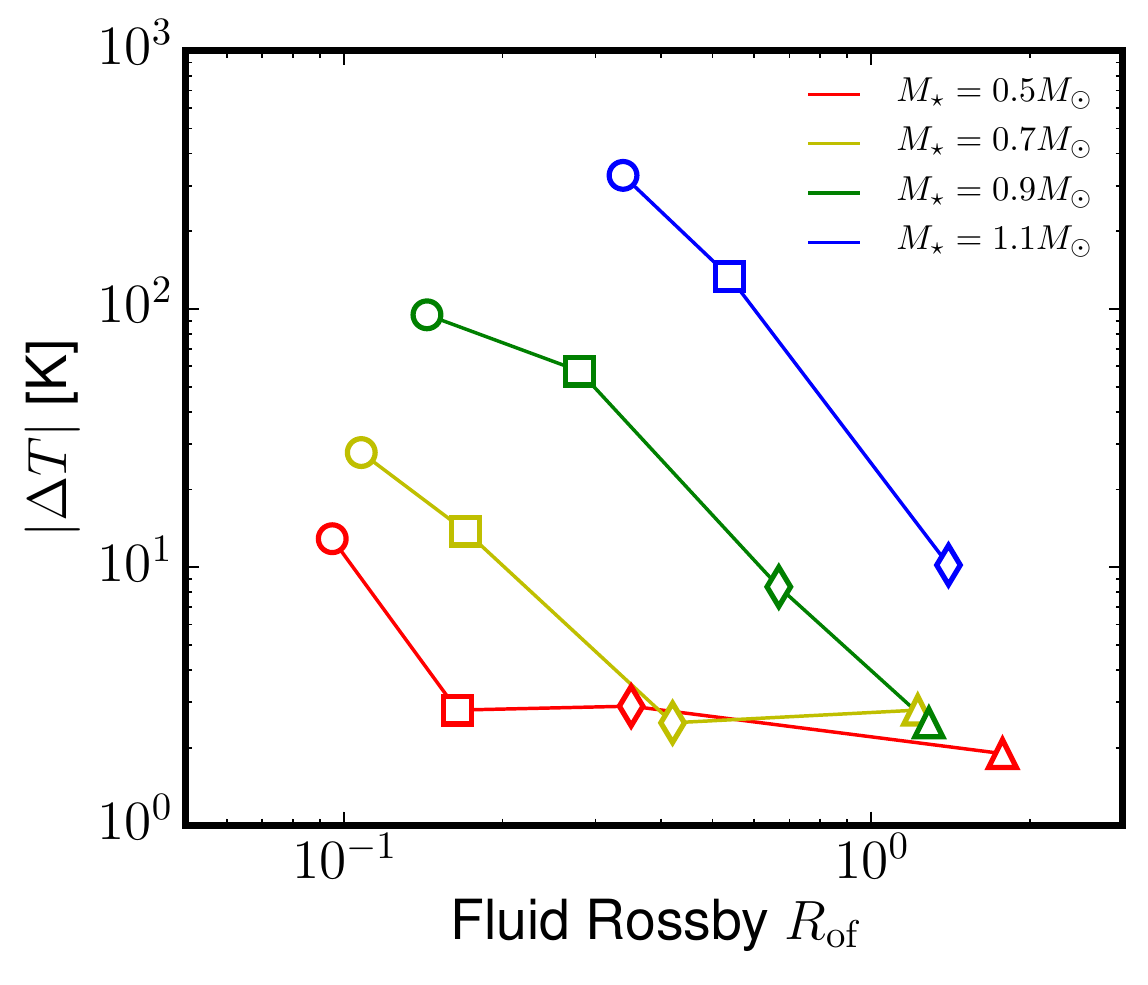}
\includegraphics[width=0.49\textwidth]{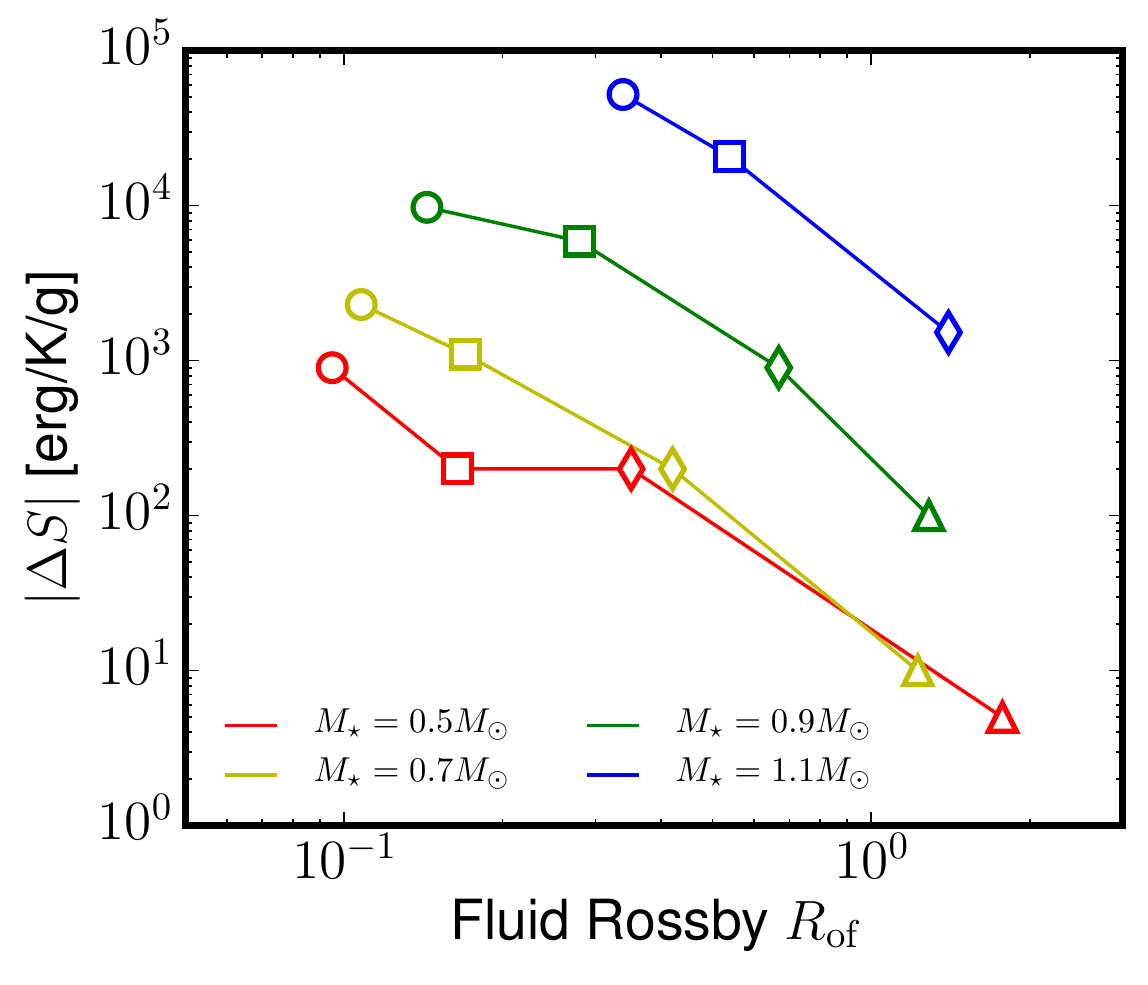}
\end{center}
\caption{Top row: Meridional cuts of the temporal and azimuthal average of the temperature and
  entropy fluctuations shown as contour plots in models M07R1 (a,b) and M11R1 (c,d).  Bottom row:
  Trends of the absolute temperature and entropy contrasts from equator to 60$^\circ$ degrees
  latitude with respect to fluid Rossby number for the four stellar masses considered. Symbols and
  colors are the same as in Figure \ref{DeltaOmega_trends}.}
\label{TSfluct}
\end{figure*}

We now turn to analyze the role played by thermal effects and heat redistribution in the dynamical
balance realized in our 3-D stellar models.  As published in \citet{Brun:2010ia}, a general
meridional force balance equation can be derived that reveals the subtle role of all processes in
maintaining a non-cylindrical rotation profile that differs from the ``classical'' thermal wind (TW)
balance \citep{Durney:1999du,Brun:2002gi,Balbus:2009id}.  It is straightforward to use our numerical
simulation to evaluate what are the dominant terms and how this meridional force balance comes
about. Its derivation is summarized in Appendix \,\ref{sec:thermal-wind-balance} and can be
compactly written as

\begin{eqnarray}
  \label{eq:TW_summary}
  2\Omega_*\frac{\p \langle v_{\phi}\rangle}{\p z} &=& \mathcal{S} + \mathcal{A} + \mathcal{C} +
  \mathcal{B} + \mathcal{V}\, ,
\end{eqnarray}

\noindent where the various right-hand side terms are:

\begin{itemize}
\renewcommand{\labelitemi}{$\bullet$}
\item $\mathcal{S}$ describes the stretching and tilting of the vorticity due to velocity gradients;
\item $\mathcal{A}$ describes the advection of vorticity by the flow;
\item $\mathcal{C}$ describes the change of vorticity due to compression;
\item $\mathcal{B}$ is the so-called baroclinic contribution from the latitudinal entropy gradient
  and the product of the radial background entropy gradient with the latitudinal pressure
  gradient. The former dominates when the stratification is nearly adiabatic;
\item $\mathcal{V}$ accounts for the viscous diffusion of vorticity.
\end{itemize}

Under the assumption that the convection zone is nearly adiabatic and hydrostatic, that the fluid
Rossby number $R_{of}$ is small, and that viscous stresses can be neglected, Equation
(\ref{eq:TW_summary}) simplifies to:

\begin{equation}\label{eq:TW}
\centering
\frac{\p \langle v_{\phi}\rangle}{\p z}=\frac{g}{2 \Omega_* r c_p}\frac{\p \langle S\rangle}{\p \theta}\, .
\end{equation}

This is the ``classical'' thermal wind equation. It states that baroclinicity can break the
Taylor-Proudman constraint of $\p \langle v_{\phi}\rangle/{\p z}=0$, implying a cylindrical rotation
profile \citep{Zahn:1992vi}. This is due to the fact that baroclinic torques suppress the
Coriolis-induced meridional circulation that would otherwise tend to establish a cylindrical state
of rotation.  It is instructive to use our numerical simulations to evaluate the role played by all
the terms of the zonal vorticity equation identified above and to discuss the nature of the
meridional force balance.

In Figure \ref{TWbal}, we display the full thermal wind balance achieved in two models having
respectively a prograde and retrograde differential rotation profile, e.g. M07R1 and M11R1. For each
case the {\it lhs} and the various terms composing the {\it rhs} are shown, using the same color
table and minimum and maximum bounds.  The first point to notice is that for both models {\it rhs
  $=$ lhs} to a high degree of fidelity, meaning that they have achieved an equilibrium and a
well-relaxed state.  Turning to M07R1 (top row of Figure \ref{TWbal}), we see that the dominant term
is the baroclinic one. It is mostly negative (positive) in the northern (southern) hemisphere as is
the {\it lhs} and possesses elongated island features in the radiative zone. In that later zone, the
``classical'' thermal wind balance is realized. In the convective envelope, in particular near
  the surface, this is less the case. Stretching ($\mathcal{S}$) and advection ($\mathcal{A}$)
  terms, and to a lesser extent the viscous term ($\mathcal{V}$), contribute to the overall
  balance. The first two of those terms have the same sign in each hemisphere as the baroclinicity
  ($\mathcal{B}$), whereas $\mathcal{V}$ has the opposite sign. This confirms the role played by
  convective motions in the meridional balance, and its departure from a strict classical thermal
  wind balance. This can be easily understood by the fact that $R_{of}$ is not very small in that
  model (see Table \ref{tablenondimnb}).

This departure from a strict TW balance is even more apparent in model M11R1 shown in the bottom row
of Figure \ref{TWbal}. With $R_{of} > 1$, the full TW balance must be considered as most terms are
expected to contribute to the {\it rhs}. The baroclinic term shows the largest and most systematic
contribution to the overall balance. In that case too, it dominates the balance in the radiative
interior and possesses a predominantly positive contribution from the low to mid-latitudes up to the
poles. Case M07R1 has the opposite response, showing one important difference between prograde and
retrograde cases. This is directly linked to the different entropy fluctuation profiles realized in
the models, as we will discuss just after in commenting Figure \ref{TSfluct}.  Nevertheless,
$\mathcal{S}$, $\mathcal{A}$, $\mathcal{C}$ and $\mathcal{V}$ terms now contribute everywhere in the
convective envelope and not only near the surface of the simulation domain. Note that in that slowly
rotating case, $\mathcal{S}$ and $\mathcal{A}$ have an opposite sign with respect to $\mathcal{B}$
and contribute a little in the radiative zone due to the development of turbulence in the
overshooting region by the more vigorous convective downdrafts. In both models (this is also true for the other cases), 
$\mathcal{S}$  and $\mathcal{A}$ have in the large the same sign. This is consistent with the analysis of the latitudinal
  angular momentum transport performed in the previous section, where their sign was found to remain
  the same when transiting from $R_{of} < 1$ to $R_{of}> 1$. This is an important property of
  prograde and retrograde models, and it holds for all the cases considered here. Such an invariance
  indicates that the turbulent latitudinal Reynolds stresses do not change sign as the rotation rate
  is varied, whereas the entropy and temperature fluctuations do.

We conclude from this TW balance analysis that for low $R_{of}$ the baroclinic term remains the key
player in tilting the iso-contours of $\Omega$ with some contribution
from turbulence. However this
becomes inefficient if $R_{of}$ becomes too small $< 0.1$ as is the case for models M05R3 and M05R5
that are mostly cylindrical. This is in part due to the increased influence of rotation that cannot
be fully compensated by thermal effects as we will see when commenting upon the trends shown in
Figure \ref{TSfluct}.  For high $R_{of}$ cases the contribution of all terms composing the {\it rhs}
is more balanced, but their cumulative action also tends to slightly bend the iso-contours of
$\Omega$.  However, in such cases, the Taylor-Proudman constraint does not play as important a role
as in the very low Rossby number cases.

\begin{figure*}[!htb]
\begin{center}
\includegraphics[width=0.9\textwidth]{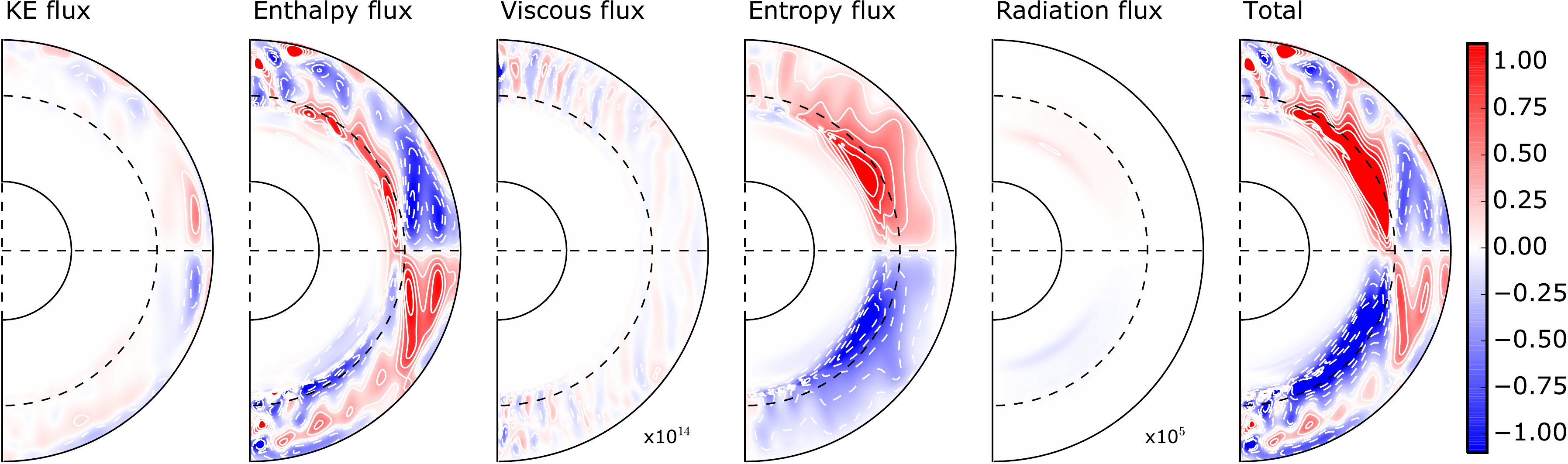}
\includegraphics[width=0.9\textwidth]{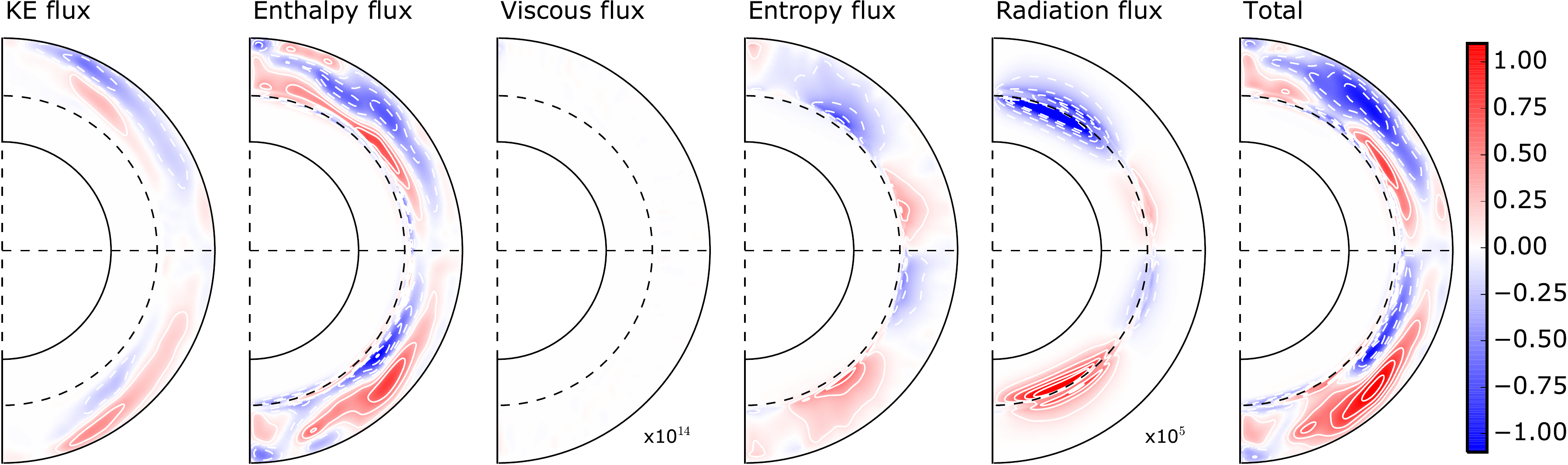}
\end{center}
\caption{Meridional cuts of the temporal and azimuthal average of the latitudinal heat transport in
  M07R1 (top row) and M11R1 (bottom row) shown as contour plots. The viscous and radiation fluxes have further been multiplied by
  respectively $10^{14}$ and $10^{5}$ in order to be visible.}
\label{HeatLat}
\end{figure*}

In order to understand further how the baroclinic term arises, it is useful to look at meridional
cuts of the entropy and temperature fluctuations averaged over time and longitudes as shown in
Figure \ref{TSfluct} (top row). We note first that the fluctuations of $S$ and $T$ are symmetric
with respect to the equator for both cases (this is also true for all models except M05s) and that
their latitudinal variations are of opposite sense. In M07R1, the poles are warm and the equatorial
regions cool, whereas it is the reverse for M11R1.  Such a latitudinal gradient of entropy (and
temperature) is consistent with the baroclinic term $\mathcal{B}$ shown in Figure \ref{TWbal} and
discussed earlier. The amplitude of the variations is also quite different, with case M11R1 having
fluctuations about a factor of 10 larger. In M07R1, $\Delta T \sim 10 K$ whereas it is around $100
K$ in M11R1.  Given the value of the background temperature at the base of the convective envelope
of each cases ($\sim 3-6 MK$), these variations can be seen as tiny but there are large enough to
drive large baroclinic torques due to the large heat capacity of stellar plasmas. The larger
variations seen in the M11R1 case is in agreement with the more vigorous convective flows realized
in the simulation due to the significantly larger stellar luminosity that needs to be carried out in
this star with respect to a 0.7 $M_{\odot}$ star (see MLT discussion in \S 2).  In the radiative
interior we further see that the variations are the largest where the rotation profile transits from
differential to uniform, e.g. in the tachoclines that possess the largest radial shear as discussed
in \citet{Brun:2011bl}. We can understand how such states are established in our models by looking
at the latitudinal heat balance as we will discuss soon, after commenting on Figure \ref{HeatLat}.

However, we first discuss how the latitudinal entropy and temperature fluctuations contrast vary
with mass and rotation rate by considering the 15 models of this study. In the bottom row of Figure
\ref{TSfluct}, we show how $\Delta T$ and $\Delta S$ vary with $R_{of}$ for the four masses
considered here. We chose to compute $\Delta T$ and $\Delta S$ at the surface and between 0 and
60$^\circ$ of latitude and to plot their absolute value to avoid artificial behavior due to their
change of sign as $R_{of}$ is varied.  We note that for faster rotation rates both entropy and
temperature contrasts increase, reaching amplitudes of several hundred $K$ for the temperature.
Likewise, we find that the fluctuations grow with increasing stellar mass, which is in
  concordance with the more increasingly intense convection achieved in those more massive stars.
  As seen in \citep{Brown:2008ii,Augustson:2012bn}, the latitudinal entropy and temperature
  contrasts have a strong mass dependence and a weaker one on rotation (or Rossby number). These
  trends can be summarized by deriving scaling relationships for $\Delta T$ and $\Delta S$, which
  are obtained through multi-parameter regression fits to the data shown in Figure \ref{TSfluct}.
  We show the resulting scalings and their uncertainties for two cases: one including the anti-solar
  cases and the other not. We see that those cases influence the exponents of the scalings, but not
  the overall trends.

\noindent With the anti-solar cases:

\begin{eqnarray}
  \label{eq:dS_Mass_Rossby}
  \Delta S &\propto&
                 \left(\frac{M_\star}{M_\odot}\right)^{7.17 \pm 0.91}
                 \left(R_{of}\right)^{-1.88 \pm 0.28}\, , \\  
  \label{eq:dT_Mass_Rossby}
  \Delta T &\propto&
                 \left(\frac{M_\star}{M_\odot}\right)^{4.43 \pm 0.84}
                 \left(R_{of}\right)^{-1.03 \pm 0.26}\, .
\end{eqnarray}

\noindent Without the anti-solar models:

\begin{eqnarray}
  \label{eq:dS_Mass_Rossby_2}
  \Delta S &\propto&
                 \left(\frac{M_\star}{M_\odot}\right)^{6.81 \pm 1.02}
                 \left(R_{of}\right)^{-1.29 \pm 0.47}\, , \\
  \label{eq:dT_Mass_Rossby_2}
  \Delta T &\propto&
                 \left(\frac{M_\star}{M_\odot}\right)^{5.92 \pm 0.90}
                 \left(R_{of}\right)^{-1.32 \pm 0.41}\, .
\end{eqnarray}

The latitudinal variation of the thermodynamics variables $S$ and $T$ is established by anisotropic
heat transport in the convective envelope.  This is an expected behavior in a rotating convective
envelope, as the Coriolis force varies with respect to latitude and so does its influence on the
convective motions \citep[see for instance
][]{Durney:1989kq,Durney:1999du,Elliott:2000kp,Miesch:2005wz,Brun:2009by,Kapyla:2011kr}.  Hence,
along with the radial heat transport realized in the convective shell of our models, as discussed in
\S 4, there is a latitudinal heat transport that is worth studying in greater detail.  As done in
\citet{Elliott:2000kp} and \citet{2009ApJ...702.1078B}, we can decompose the latitudinal heat
transport into various terms involving diffusion and advection processes.  In Figure \ref{HeatLat},
we display, using the same color table and minimum and maximum bounds, the five terms contributing
to the latitudinal transport of heat after having converted and normalized them to the adequate
stellar luminosity. We note that the main terms are the latitudinal enthalpy and entropy fluxes,
with a minor contribution from the kinetic energy flux. In this context, the viscous and radiative
fluxes are negligible. To see their structure, they have been multiplied by very large factors (e.g., $10^{14}$ and $10^{5}$ respectively) 
in Figure \ref{HeatLat}. Hence, the balance is mainly between the entropy flux and the enthalpy flux, which
are mostly of opposite sign. The leading contribution comes from the correlation of the fluctuating
latitudinal velocity and the temperature fluctuations to yield an enthalpy flux that efficiently
transports the heat, hence establishing the fluctuations seen in Figure \ref{TSfluct}. Once these
profiles have been established, the latitudinal entropy flux develops to establish a balance.
Turning to case M07R1, we see that the entropy flux tends to cool down the polar regions and heat
the equatorial region, and it is maximum at the base of the convection zone. For case M11R1 (lower
panels), the balance is such that the entropy flux mostly warms up the poles. In a small equatorial
region, the entropy flux changes sign and tends to warm up this zone in agreement with Figure
\ref{TSfluct}. In that case, the role of the enthalpy flux is less clear, but it tends to also
balance the entropy flux even though the system has not yet reached a complete equilibrium. As a
summary, we see that the heat transfer and the associated thermodynamic perturbations in the solar
and anti-solar cases differ, yielding hot poles and cool equator for the solar-like case, and the
reverse for the anti-solar ones.

\subsection{Energy exchange and maintenance of differential rotation}
\label{sec_energy_echange}

It is also useful to understand the energy exchanges maintaining the differential
rotation. Following \citet{Rempel:2005fk}, the kinetic energy balance for the differential rotation
can be written

\begin{equation}
  \label{eq:ke_balance}
  \partial_t\left({\rm DRKE}\right)= Q_C^r + Q_C^l + Q_{R}^r + Q_{R}^l + Q_A
  + Q_V + Q_{\rm curv}\, ,
\end{equation}
where the various terms represent the work of the Coriolis force, Reynolds stresses, ${\rm DRKE}$ advection and viscous
forces. Their expression is given by \citep{Rempel:2005fk} and defined in Appendix \,\ref{sec:kinet-energy-balance}.

The Coriolis force relates to energy transfers between the differential rotation and meridional
circulation. The non-linear advection is decomposed into Reynolds stresses associated with the
radial ($Q_r^r$) and latitudinal ($Q_r^l$) profiles of the differential rotation, and the
differential rotation advection $Q_A$. Finally, we regroup all the contributions from the viscous
stress tensor into $Q_V$ and the geometric terms into $Q_{\rm curv}$.

\begin{figure}[!htbp]
\begin{center}
\includegraphics[width=0.9\linewidth]{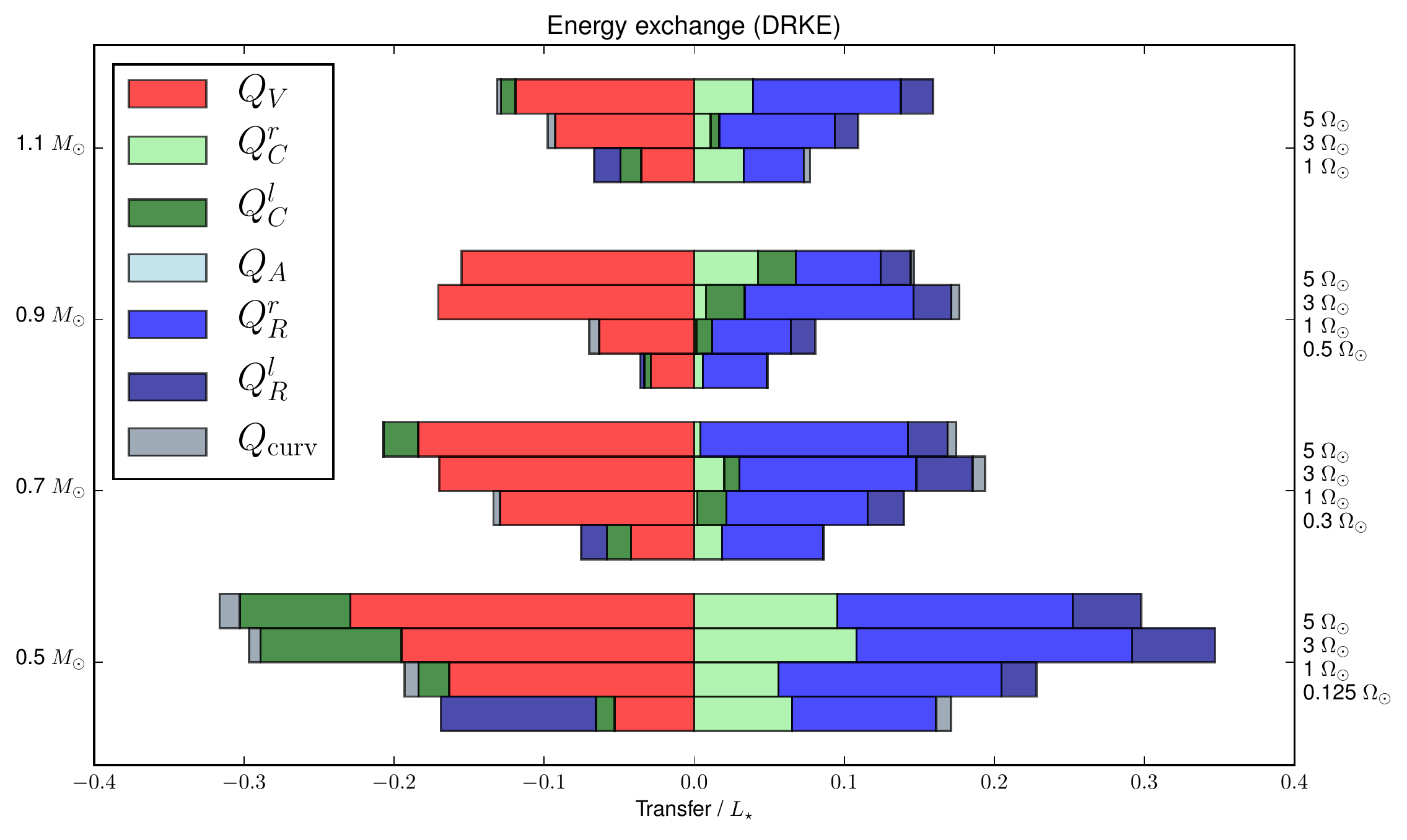}
\end{center}
\caption{Differential rotation kinetic energy balance (Equation \ref{eq:ke_balance}) as a function of stellar mass and rotation, normalized to the stellar luminosity.}
\label{fig:energy_exchange}
\end{figure}

We display in Figure \ref{fig:energy_exchange} the various contributions to the kinetic energy
balance of the differential rotation for our 15 models (the numerical values can be found in Table
\ref{tab:tab_DRKE_exchange}). In all models, the differential rotation is mainly maintained by a
balance between the non-linear Reynolds stresses and the viscous stress. The Reynolds stress
contribution itself is strongly dominated by the radial component. It only moderately increases with
rotation except for models with $M_\star=0.9\,M_\odot$. In less massive stars, the convective flows
maintaining the differential rotation correspond to a higher fraction of the stellar luminosity than
in more massive stars. This behaviour can be naively expected from the simple scaling laws derived
in Section \ref{sec:hint_dr_ML}, namely $\rho_{\rm bcz}v^2\sim M_\star^{-0.9}$. Finally, the
advection of the differential rotation $Q_A$ is completely negligible in all cases.

It is also instructive to note that for the 4 models possessing an anti-solar differential
rotation (M05s, M07s, M09s and M11R1), the latitudinal transport of energy by Reynolds stresses is negative and the
differential rotation is mainly sustained by the radial transport of angular momentum by Reynolds
stresses. In these models the Rossby number is higher than 1
(see Table \ref{tablenondimnb}), the convective motion are much less vortical than in the other
cases and this directly impacts how the Reynolds stresses transfer kinetic energy to the
differential rotation.

Following \citet{Rempel:2005fk,Rempel:2006eh}, we can finally identify the amount of
  energy transfered to the large-scale differential rotation with
  $|Q_V|$. We note in Table \ref{tab:tab_DRKE_exchange} that $|Q_V|$
  increases with rotation rate and decreases with mass. For instance
  the model with $M_\star=0.5\,R_\star$ and
  $\Omega_\star=5\Omega_\odot$ converts about $23\%$ of the stellar
  energy flux into the large-scale differential
  rotation. Note that this model is twenty times less luminous than
  the Sun, which means that only $1\%$ of a solar luminosity is converted
  into differential rotation (second column in Table
  \ref{tab:tab_DRKE_exchange}). Model M11 R5 converts the most energy
  (in absolute value) into differential rotation with $|Q_V| \sim
  21\% L_\odot$. We recall that $|Q_V|$ is only a proxy of
  the converted energy, as the differential rotation is maintained due to the
  angular momentum transfer by the turbulent Reynolds stress (see
  \S \,\ref{sec_amom}).

\begin{table}[!ht]
\begin{center}
  \caption{Kinetic energy exchanges\label{tab:tab_DRKE_exchange}}
\vspace{0.2cm}
\begin{tabular}{lcccccc}
\tableline
\tableline
\\[-1.5ex] & \multicolumn{2}{c}{$Q_V$} & $Q_C^r$ & $Q_C^l$ & $Q_R^r$ & $Q_R^l$ %\\ [0.1ex]
\\[-0ex] & $[L_\star]$ & $[L_\odot]$ & $[L_\star]$ & $[L_\star]$ & $[L_\star]$  & $[L_\star]$  \\ [0.8ex]
\tableline
\tableline
\\[-1.5ex]
M05 S  & -0.053 &       -0.002 &   0.065 &  -0.012 &   0.096 &  -0.104 \\
M05 R1 & -0.163 &       -0.007 &   0.056 &  -0.021 &   0.149 &   0.023 \\
M05 R3 & -0.195 &       -0.009 &   0.108 &  -0.094 &   0.184 &   0.055 \\
M05 R5 & -0.229 &       -0.011 &   0.095 &  -0.074 &   0.157 &   0.046 \\
% M05 S  & -0.053 &   0.065 &  -0.012 &   0.096 &  -0.104 \\
% M05 R1 & -0.163 &   0.056 &  -0.021 &   0.149 &   0.023 \\
% M05 R3 & -0.195 &   0.108 &  -0.094 &   0.184 &   0.055 \\
% M05 R5 & -0.229 &   0.095 &  -0.074 &   0.157 &   0.046 \\
[0.5ex]
\hline
\\ [-2ex] 
% M07 S  & -0.042 &   0.019 &  -0.016 &   0.067 &  -0.017 \\
% M07 R1 & -0.129 &   0.002 &   0.019 &   0.094 &   0.024 \\
% M07 R3 & -0.170 &   0.020 &   0.010 &   0.118 &   0.037 \\
% M07 R5 & -0.184 &   0.004 &  -0.023 &   0.138 &   0.026 \\
M07 S  & -0.042 &       -0.006 &   0.019 &  -0.016 &   0.067 &  -0.017 \\
M07 R1 & -0.129 &       -0.020 &   0.002 &   0.019 &   0.094 &   0.024 \\
M07 R3 & -0.170 &       -0.026 &   0.020 &   0.010 &   0.118 &   0.037 \\
M07 R5 & -0.184 &       -0.028 &   0.004 &  -0.023 &   0.138 &   0.026 \\
[0.5ex]
\hline
\\ [-2ex] 
% M09 S  & -0.029 &   0.006 &  -0.004 &   0.043 &  -0.003 \\
% M09 R1 & -0.063 &   0.001 &   0.011 &   0.052 &   0.016 \\
% M09 R3 & -0.170 &   0.008 &   0.026 &   0.113 &   0.025 \\
% M09 R5 & -0.155 &   0.042 &   0.025 &   0.056 &   0.020 \\
M09 S  & -0.029 &       -0.016 &   0.006 &  -0.004 &   0.043 &  -0.003 \\
M09 R1 & -0.063 &       -0.035 &   0.001 &   0.011 &   0.052 &   0.016 \\
M09 R3 & -0.170 &       -0.093 &   0.008 &   0.026 &   0.113 &   0.025 \\
M09 R5 & -0.155 &       -0.085 &   0.042 &   0.025 &   0.056 &   0.020 \\
[0.5ex]
\hline
\\ [-2ex] 
% M11 R1 & -0.035 &   0.033 &  -0.014 &   0.040 &  -0.018 \\
% M11 R3 & -0.093 &   0.011 &   0.006 &   0.077 &   0.015 \\
% M11 R5 & -0.119 &   0.039 &  -0.010 &   0.098 &   0.022 \\
M11 R1 & -0.035 &       -0.063 &   0.033 &  -0.014 &   0.040 &  -0.018 \\
M11 R3 & -0.093 &       -0.166 &   0.011 &   0.006 &   0.077 &   0.015 \\
M11 R5 & -0.119 &       -0.213 &   0.039 &  -0.010 &   0.098 &   0.022 \\
\hline
\tableline
\tableline
\end{tabular}
\end{center}
\end{table}

\section{Discussion and Conclusions}
\label{sec_conclusion}

In this study we have focused our analysis on the characterization of the angular velocity profiles
in the convective envelope of solar-like stars of various masses and rotation rates.  Starting from
a mixing length argument, we have shown that a Rossby number can be derived based on fundamental
stellar parameters such as mass and rotation rate that gives an important insight into the expected
differential rotation profile. Three main categories have been identified: anti-solar-like,
solar-like and cyclindrical/Jupiter-like. These categories depend strongly on the Rossby number of the
simulations, being large for anti-solar cases and smaller than $R_{of} < 0.1$ for Jupiter-like
profiles.  

Solar-like rotation profiles are found for intermediate $R_{of}$ values that are less
than unity. Such differential rotation states with conically-tilted isocontours of $\Omega$ are
achieved only if the baroclinic term plays a dominant role in the thermal wind balance, as discussed
in \S 7. As the rotation rate is increased and the Rossby number decreases toward 0.1, the angular
velocity profile tends to become more cylindrical, retaining its monotonic behavior with fast
equator and slow poles. Then for even lower Rossby numbers, the alternating prograde and retrograde
jets become increasingly apparent. Eventually, they tend toward Jupiter and Saturn's surface angular
velocity profiles. A good guess of the number of jets in such cases can be obtained by computing the
compressible Rhines scale \citep[see e.g.][]{Gastine:2014da}.  

Of course the values quoted for the
Rossby number are somewhat dependent on the definition used. Observers tend to favor the stellar
Rossby number, comparing the rotation period to the convective overturning time deduced from stellar
evolution models \citep{Landin:2010fi}. Even in that case, there is some freedom in
choosing the convective over turning time, where one can consider either the base of the convective
envelope, the mid-depth value, or an averaged value. Another definition of the Rossby number can be
used, the so-called convective Rossby number as discussed in \citet{Glatzmaier:1982io}. This is
relevant in analyzing rotating convection simulations such as the ones presented in this study,
because it uses key nondimensional numbers such as the Rayleigh, Taylor and Prandtl numbers. It
evaluates the ratio between the driving buoyancy force and the rotation constraints, getting rid of
diffusion effects. A third definition that is not employed in this study, but that is found to be
relevant for the study of stellar dynamos, is the scale dependent Rossby number. It is evaluated by
comparing the kinetic energy contained in the first convective modes to the energy contained in all
the scales. Finally, there is the fluid Rossby number, which is used in most of this work. It
relates the turbulent vorticity to the planetary vorticity. It is an a postiori measurement of the
effect of the Coriolis force on the turbulence. We find it useful because it is quite
straightforward to compute, and it gives a robust assessment of the rotating dynamics. The
transition values given above and throughout the paper are based on the fluid Rossby
number. 

However, for the sake of completeness, we have also quoted in table \ref{tablenondimnb} the
convective and stellar Rossby numbers. What is clear from this table is that those numbers are
indeed different but their relative trends with respect to global stellar parameters follow a simple
quasi-linear relationship. Hence, what matters more than the magnitude of the Rossby number at the
point of the the transition is to distinguish the rotational regimes. For large Rossby numbers, we
expect to have anti-solar behavior, for intermediate values to be solar-like, and for small values
to have a cylindrical rotation profile that can be either monotonic with respect to latitude or
exhibit alternating zonal jets when the rotational constraint is very large.

Intrinsically, mixing length theory cannot predict differential rotation states. It can only
illustrate how the Rossby number changes with respect to global stellar parameters as demonstrated
in \S 2 and Figure \ref{om_rossby}. In contrast, the natural outcome of our 3-D numerical
simulations of rotating stellar convection is to provide, among other dynamical properties, the
state of rotation achieved in a model for a given set of stellar parameters. We can thus assess,
with a consistent definition of the Rossby number, what are the resulting states of differential
rotation for the simulations. In doing so, we can calibrate the nondimensional constant $c_1$ used
in plotting Figure \ref{om_rossby}.  Note that no attempts at fine tuning have been made to
precisely find the transitions, instead a systematic scan of the stellar parameters has been
performed so as to obtain the three differential rotation states identified.  With a multi-parameter
regression fit to our set of 15 numerical simulations, we obtain the following scaling relation
between the fluid Rossby, stellar rotation and mass:

\begin{equation}
    R_{of} = 0.89 \times \Omega_*^{-0.82 \pm 0.05} M_*^{1.53 \pm 0.22}.
\end{equation}

\begin{figure}[!htbp]
\begin{center}
\includegraphics[width=0.95\linewidth]{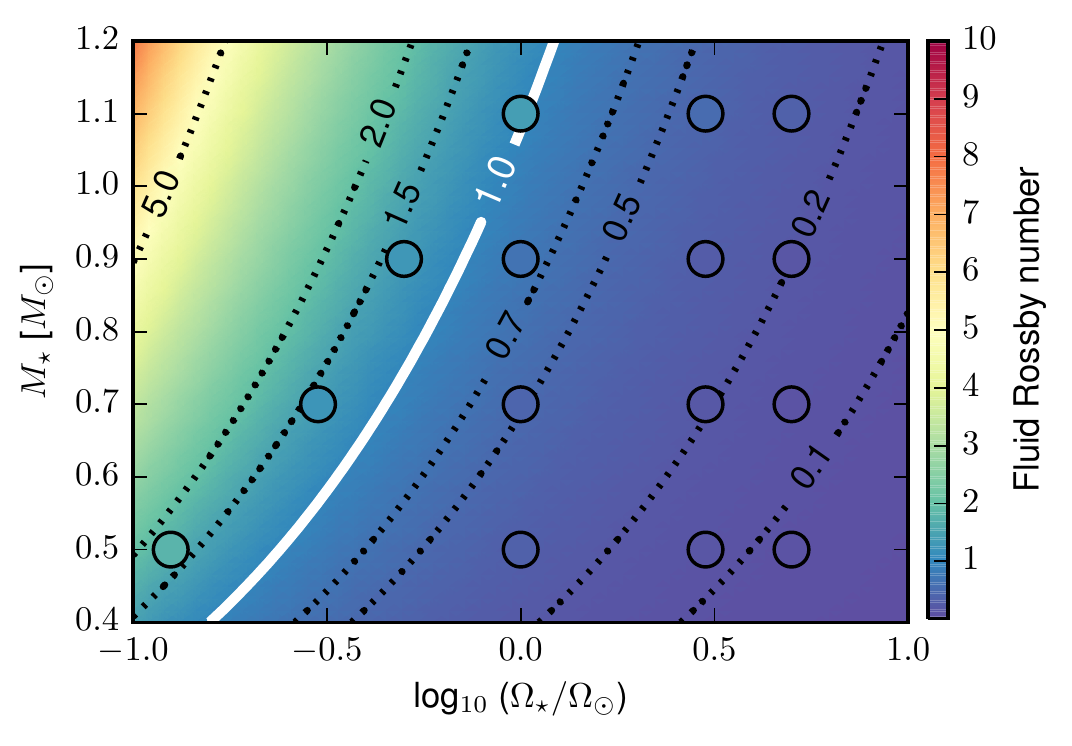}
\end{center}
\caption{Fluid Rossby number of the models.}
\label{Mass_Omega}
\end{figure}

We note that it is not far from the one we derived in \S2 from back of the envelope arguments using
MLT. The constant $c_1 \sim 0.89$ and the exponents are very close as well.  To further illustrate
this important result, we display this scaling relation based upon our set of 3-D simulations of
rotating stellar convection in Figure \ref{Mass_Omega} using color contours. What our study confirms
and Figure \ref{Mass_Omega} summarizes is that for a given observation of the stellar rotation rate
or $v \sin i$, the rotation state depends on the stellar mass of the solar-like star. In
  particular, for a given rotation rate, the more massive stars may be more likely to achieve an
  anti-solar-like rotational state. This trend could be observable and a systematic search for
  anti-solar stars should be undertaken, as it will greatly aid in the constraints upon our
  models. And indeed some attempts using Kepler data have already been started
  \citep{Reinhold:2015kx,Varela16}.

It is interesting to note that the transition of the state of differential rotation with Rossby number that we have indentified could have a direct correspondance 
with trends found in stellar X-ray luminosity studies \citep[see, e.g.][and references therein]{Pizzolato:2003ga,2011ApJ...743...48W}. 
Jupiter-like profile could correspond to the saturated X-ray regime, solar-like profiles to the linear regime and the anti-solar state to the well less defined regime found in these studies. 
Also this modification of the large differential rotation profile could lead to different dynamo regimes and magnetic field topology, impacting directly wind braking 
and stellar spin-down \citep{2014ApJ...789..101B,Vidotto:2016ej,2016ApJ...826L...2M}. We intend to verify these relations and changes of regime with dedicated dynamo simulations.

We also find that $\Delta \Omega$ varies significantly in amplitude with stellar mass, its amplitude being larger
for more luminous stars as observed by \citet{Barnes:2005eb}. It is also possible that the 
dependency (exponent) vs stellar mass becomes steeper for F-type stars than for G and K stars \citep[see, e.g.][]{Augustson:2012bn}.  
The trend with rotation rate is also to have a larger contrast for faster rotation rate, but the relative differential rotation $\Delta
\Omega/\Omega_*$ is found to decrease with $\Omega_*$.  The scaling relationships have a larger
dependency than is advocated by \citet{CollierCameron:2007ce}, but they are in close agreement with the
studies of \citep{Donahue:1996ch,Saar:2009tb}. Furthermore, we know that in MHD dynamo models the angular velocity
contrast has a weaker dependence on the rotation rate, as demonstrated in \citet{Varela16} which
utilizes MHD versions of the 3-D simulations discussed here. So, the scaling relationships given in
these paper with respect to the rotation rate should be considered as upper limits.

All our simulations include a stable radiative interior below the convective envelope. The seamless
nonlinear coupling of these regions greatly improves the realism of the bottom boundary condition of
the convective layer relative to an impenetrable one. The convective downdrafts can plumet through
the domain without hitting a solid wall, and so they can be buoyantly braked. A careful analysis of
the overshooting layer that results from the pummeling of the convective motions reveals interesting
trends.  We find that the amount of overshooting decreases with mass, when it is characterized by
latitudinally averaging the radial extent over which enthalpy flux is negative at the base of the
convective envelope. We further find that stars with prograde rotation have a smaller normalized
overshooting extent with increasing Rossby number, whereas it is the reverse for anti-solar-like
stars. The latitudinal variations are interesting too. We find that solar-like stars have prolate
overshooting layers and that in anti-solar-like stars they are oblate.

As for the overshooting layer, characterizing the shape and amplitude of stellar tachoclines at the
base of convective envelopes is of importance for stellar magnetism and chemical mixing. We find the
following trends in our study: the prolateness of the tachocline changes with rotation
rate. Anti-solar-like star have oblate tachoclines, thicker at the equator, whereas solar-like star
have prolate tachoclines. We also find that for faster rotation rates the overall shear is larger,
in agreement with $\Delta \Omega$ being larger in the convective envelope. However, our choices for
the radial profile of the thermal and viscous diffusivities influence the thickness and location of
the tachoclines. Indeed, since the pioneering work of \citet{Spiegel:1992tr}, it has been well known
that the viscous and radiative spreading of a tachocline in the radiative interior of a solar-like
star is at work, unless there is some yet-to-be-identified physical process that acts against
it. Several scenarios have been proposed: anisotropic turbulence, gravity waves, primordial or
cyclic dynamo-generated magnetic field. However, a specific numerical setup will likely be required
to be able to disentangle their various impacts, which is a task far beyond the scope of this
study. Nevertheless, we believe that the trends found for the shape of the tachoclines is robust as
all models have been built with the same physical ingredients and numerical accuracy.

Meridional circulation is found to change significantly with rotational influence. Its shape changes
from a monolithic unicellular poleward flow in the anti-solar cases, to multi-cellular flows, with
the number of cells increasing both in radius and latitude, for solar-like and Jupiter like states
of differential rotation. This can be understood by the ability or not of the meridional cells to
extend beyond the latitude corresponding to the tangent cylinder of each models. With stronger
rotational constraint, the meridional cells are more and more aligned with the rotation axis and are
confined to lower latitudes as discussed in \citet{Featherstone:2015bv}. Our study also confirms
previous findings that the meridional circulation weakens as the rotation rate is increased. This is
linked to the fact that more kinetic energy is being channeled to the longitudinal motions. However,
it could be the case that this decrease in the amplitude of the meridional circulation with respect
to the rotation rate is due to a lower level of supercriticality of the simulations, since it is
difficult to maintain it.  Indeed, we know that this large-scales flow is a direct response to any
net longitudinal torque applied to the envelope through the effect of gyroscopic pumping.  As a
consequence, if the amplitude of the Reynolds stresses weakens because of a less intense degree of
turbulence, then the meridional circulation will be weaker as well. Even taking into account the
decrease of amplitude of the Reynolds stresses, the global trend of weaker meridional circulation
for faster rotation rate is confirmed. As we clearly see in Table 4, the Reynolds number does not
decrease. On the contrary, it can increase.

Two cases exhibit active nests of convection (M05R3, M05R5), but
  they do not impact the conclusions derived in this study. Their
  co-existence with an underlying stably-stratified region in our
  models warrants further investigation regarding
  their formation, which we intend to explore in the near future with
  a specific parameter study.

To summarize, we have seen that many properties of stellar convection, with its associated mean flows
and transport mechanisms, are influenced by rotation and stellar mass. We have been able to find
useful trends and to anticipate the rotation regime of candidate stars thanks to MLT and 3-D
numerical simulations. This study was done with purely hydrodynamical models and lacks a
consideration for the nonlinear feed backs related to a dynamo generated magnetic field. Preliminary
studies by our group, and published in \citet{Varela16}, seem to confirm the main trends for the differential rotation states but with a
weaker sensitivity to global parameters. Work by
\citet{Kapyla:2014jr,Gastine:2014jr,Karak:2015dw,Guerrero:2016cz} also find some differences arising
from magnetic fields, but they do not change the global trends presented in this study. For instance we find that
the anti-solar cases presented in this hydrodynamical study remain anti-solar when magnetic field is taken into account \cite[see Figure 2 of][for a preliminary study of
our MHD antisolar cases]{Varela16}. This is likely due to the fact that their Rossby ($R_{os}$ or $R_{of}$) remain larger
than 1.0 even after the Lorentz force has influenced the convective flows.
We expect to publish a detailed analysis of the dynamo counterpart of the 15 hydrodynamical models discussed in this study in the near future.

Of course the turbulence degree used in the simulations discussed in this work are still limited 
by the current computer resources and one must be extremely careful in comparing directly numerical results 
with observations. Still we find these systematic parameters study useful to delineate the main trends and identify the key 
physical mechanisms and it is reconforting to see that the observational tendencies are recovered qualitatively. We are also convinced that anti-solar like 
differential rotation states are worth searching for observationaly by selecting stars with large fluid or stellar Rossby numbers.

\acknowledgements

A.S. Brun and A. Strugarek dedicate this paper to Professor Jean-Paul Zahn, whose continuous support
and advice over the years and great expertise in stellar fluid dynamics have been of invaluable help
in analyzing non linear 3-D simulations and in improving our understanding of 
stellar structure, evolution, and dynamics in general. We miss him dearly.  We also thank Paul Charbonneau, 
Steve Saar, Thomas Gastine, Michio Yamada, Shin-Ichi Takehiro and Rafael Garcia for
useful discussions.  We acknowledge funding by ERC STARS2 207430 grant, ANR Blanc Toupies SIMI5-6 020 01, INSU/PNST, CNES SolarOrbiter, 
PLATO and GOLF grants, FP7 SpaceInn 312844 grant, and NASA grants NNX11AJ36G, NNX13AG18G and NNX16AC92G.  
K.~C. Augustson is funded through the ERC SPIRE 647383 grant. A. Strugarek acknowledges support from 
the Canadian Institute of Theoretical Astrophysics (National Fellow), 
from CanadaÕs Natural Sciences and Engineering Research Council and from CNES postdoctoral fellowship.
Simulations have been performed on GENCI and PRACE
supercomputer infrastructures under grant 1623 and RA1964.  A.S. Brun wishes to thank the University
of Colorado and JILA as well as the University of Kyoto and RIMS for their hospitality. 

%For BibTex:

\bibliographystyle{aasjournal}
%\bibliography{gkstars}
%\bibliography{bibAS}

\begin{thebibliography}{}
\expandafter\ifx\csname natexlab\endcsname\relax\def\natexlab#1{#1}\fi

\bibitem[{Aigrain {et~al.}(2015)Aigrain, Llama, Ceillier, Chagas, Davenport,
  Garcia, Hay, Lanza, McQuillan, Mazeh, De~Medeiros, Nielsen, \&
  Reinhold}]{Aigrain:2015cc}
Aigrain, S., Llama, J., Ceillier, T., {et~al.} 2015, \mnras, 450, 3211

\bibitem[{Alvan {et~al.}(2014)Alvan, Brun, \& Mathis}]{Alvan:2014gx}
Alvan, L., Brun, A.~S., \& Mathis, S. 2014, \aap, 565, A42

\bibitem[{Augustson {et~al.}(2015)Augustson, Brun, Miesch, \&
  Toomre}]{Augustson:2015er}
Augustson, K., Brun, A.~S., Miesch, M., \& Toomre, J. 2015, \apj, 809, 149

\bibitem[{Augustson {et~al.}(2012)Augustson, Brown, Brun, Miesch, \&
  Toomre}]{Augustson:2012bn}
Augustson, K.~C., Brown, B.~P., Brun, A.~S., Miesch, M.~S., \& Toomre, J. 2012,
  \apj, 756, 169

\bibitem[{Augustson {et~al.}(2013)Augustson, Brun, \&
  Toomre}]{Augustson:2013jj}
Augustson, K.~C., Brun, A.~S., \& Toomre, J. 2013, \apj, 777, 153

\bibitem[{Augustson {et~al.}(2016)Augustson, Brun, \&
  Toomre}]{Augustson:2016wr}
---. 2016, to appear in ApJ, arXiv:1603.03659

\bibitem[{Aurnou {et~al.}(2007)Aurnou, Heimpel, \& Wicht}]{Aurnou:2007cp}
Aurnou, J., Heimpel, M., \& Wicht, J. 2007, Icarus, 190, 110

\bibitem[{Balbus {et~al.}(2009)Balbus, Bonart, Latter, \&
  Weiss}]{Balbus:2009id}
Balbus, S.~A., Bonart, J., Latter, H.~N., \& Weiss, N.~O. 2009, \mnras, 400,
  176

\bibitem[{Ballot {et~al.}(2007)Ballot, Brun, \&
  Turck-Chi{\`e}ze}]{Ballot:2007ea}
Ballot, J., Brun, A.~S., \& Turck-Chi{\`e}ze, S. 2007, \apj, 669, 1190

\bibitem[{{Balona} \& {Abedigamba}(2016)}]{2016MNRAS.461..497B}
{Balona}, L.~A., \& {Abedigamba}, O.~P. 2016, \mnras, 461, 497

\bibitem[{Barnes {et~al.}(2005)Barnes, Collier~Cameron, Donati, James, Marsden,
  \& Petit}]{Barnes:2005eb}
Barnes, J.~R., Collier~Cameron, A., Donati, J.-F., {et~al.} 2005, \mnras, 357,
  L1

\bibitem[{Barnes(2003)}]{Barnes:2003ga}
Barnes, S.~A. 2003, \apj, 586, 464

\bibitem[{{Barnes}(2010)}]{barnes2010}
{Barnes}, S.~A. 2010, \apj, 722, 222

\bibitem[{Bessolaz \& Brun(2011)}]{Bessolaz:2011ih}
Bessolaz, N., \& Brun, A.~S. 2011, \apj, 728, 115

\bibitem[{Bouvier(2013)}]{Bouvier:2013cz}
Bouvier, J. 2013, Role and Mechanisms of Angular Momentum Transport During the
  Formation and Early Evolution of Stars, 62, 143

\bibitem[{Brown {et~al.}(2008)Brown, Browning, Brun, Miesch, \&
  Toomre}]{Brown:2008ii}
Brown, B.~P., Browning, M.~K., Brun, A.~S., Miesch, M.~S., \& Toomre, J. 2008,
  \apjs, 689, 1354

\bibitem[{Brown {et~al.}(2010)Brown, Browning, Brun, Miesch, \&
  Toomre}]{Brown:2010cn}
---. 2010, \apj, 711, 424

\bibitem[{Brown {et~al.}(2011)Brown, Miesch, Browning, Brun, \&
  Toomre}]{Brown:2011fm}
Brown, B.~P., Miesch, M.~S., Browning, M.~K., Brun, A.~S., \& Toomre, J. 2011,
  \apj, 731, 69

\bibitem[{Brown {et~al.}(2012)Brown, Vasil, \& Zweibel}]{Brown:2012bd}
Brown, B.~P., Vasil, G.~M., \& Zweibel, E.~G. 2012, \apj, 756, 109

\bibitem[{Brown(2014)}]{Brown:2014el}
Brown, T.~M. 2014, \apj, 789, 101

\bibitem[{{Brown}(2014)}]{2014ApJ...789..101B}
{Brown}, T.~M. 2014, \apj, 789, 101

\bibitem[{Browning(2008)}]{Browning:2008dn}
Browning, M.~K. 2008, \apj, 676, 1262

\bibitem[{Browning {et~al.}(2004)Browning, Brun, \& Toomre}]{Browning:2004bx}
Browning, M.~K., Brun, A.~S., \& Toomre, J. 2004, \apj, 601, 512

\bibitem[{Browning {et~al.}(2006)Browning, Miesch, Brun, \&
  Toomre}]{Browning:2006ba}
Browning, M.~K., Miesch, M.~S., Brun, A.~S., \& Toomre, J. 2006, \apj, 648,
  L157

\bibitem[{Brummell {et~al.}(2002)Brummell, Clune, \& Toomre}]{Brummell:2002dj}
Brummell, N.~H., Clune, T.~L., \& Toomre, J. 2002, \apj, 570, 825

\bibitem[{Brun {et~al.}(2010)Brun, Antia, \& Chitre}]{Brun:2010ia}
Brun, A.~S., Antia, H.~M., \& Chitre, S.~M. 2010, \aap, 510, 33

\bibitem[{Brun {et~al.}(2002)Brun, Antia, Chitre, \& Zahn}]{Brun:2002cy}
Brun, A.~S., Antia, H.~M., Chitre, S.~M., \& Zahn, J.-P. 2002, \aap, 391, 725

\bibitem[{Brun {et~al.}(2015{\natexlab{a}})Brun, Browning, Dikpati, Hotta, \&
  Strugarek}]{Brun:2015kca}
Brun, A.~S., Browning, M.~K., Dikpati, M., Hotta, H., \& Strugarek, A.
  2015{\natexlab{a}}, \ssr, 196, 101

\bibitem[{Brun {et~al.}(2015{\natexlab{b}})Brun, Garcia, Houdek, Nandy, \&
  Pinsonneault}]{Brun:2015co}
Brun, A.~S., Garcia, R.~A., Houdek, G., Nandy, D., \& Pinsonneault, M.
  2015{\natexlab{b}}, \ssr, 196, 303

\bibitem[{Brun {et~al.}(2004)Brun, Miesch, \& Toomre}]{Brun:2004ji}
Brun, A.~S., Miesch, M.~S., \& Toomre, J. 2004, \apj, 614, 1073

\bibitem[{Brun {et~al.}(2011)Brun, Miesch, \& Toomre}]{Brun:2011bl}
---. 2011, \apj, 742, 79

\bibitem[{Brun \& Palacios(2009)}]{2009ApJ...702.1078B}
Brun, A.~S., \& Palacios, A. 2009, \apj, 702, 1078

\bibitem[{Brun \& Rempel(2009)}]{Brun:2009by}
Brun, A.~S., \& Rempel, M. 2009, \ssr, 144, 151

\bibitem[{Brun \& Toomre(2002)}]{Brun:2002gi}
Brun, A.~S., \& Toomre, J. 2002, \apj, 570, 865

\bibitem[{{Cattaneo} {et~al.}(1991){Cattaneo}, {Brummell}, {Toomre},
  {Malagoli}, \& {Hurlburt}}]{1991ApJ...370..282C}
{Cattaneo}, F., {Brummell}, N.~H., {Toomre}, J., {Malagoli}, A., \& {Hurlburt},
  N.~E. 1991, \apj, 370, 282

\bibitem[{{Chan} {et~al.}(2011){Chan}, {Cai}, \& {Singh}}]{2011IAUS..271..317C}
{Chan}, K.~L., {Cai}, T., \& {Singh}, H.~P. 2011, in IAU Symposium, Vol. 271,
  Astrophysical Dynamics: From Stars to Galaxies, ed. N.~H. {Brummell}, A.~S.
  {Brun}, M.~S. {Miesch}, \& Y.~{Ponty}, 317--325

\bibitem[{Chandrasekhar(1961)}]{Chandrasekhar:1961gu}
Chandrasekhar, S. 1961, \apj, 134, 662

\bibitem[{Charbonneau(2005)}]{Charbonneau:2005il}
Charbonneau, P. 2005, \lrsp, 2, doi:10.12942/lrsp-2005-2

\bibitem[{Christensen \& Aubert(2006)}]{Christensen:2006jo}
Christensen, U.~R., \& Aubert, J. 2006, Geophysical Journal International, 166,
  97

\bibitem[{Clune {et~al.}(1999)Clune, Elliott, Miesch, Toomre, \&
  Glatzmaier}]{Clune:1999vd}
Clune, T.~C., Elliott, J.~R., Miesch, M.~S., Toomre, J., \& Glatzmaier, G.~A.
  1999, Parallel Computing, 25, 361

\bibitem[{Collier~Cameron(2007)}]{CollierCameron:2007ce}
Collier~Cameron, A. 2007, Astro. Nach., 328, 1030

\bibitem[{{Dikpati}(2011)}]{2011ApJ...733...90D}
{Dikpati}, M. 2011, \apj, 733, 90

\bibitem[{do~Nascimento {et~al.}(2014)do~Nascimento, Garcia, Mathur, Anthony,
  Barnes, Meibom, da~Costa, Castro, Salabert, \&
  Ceillier}]{doNascimento:2014is}
do~Nascimento, J. D.~J., Garcia, R.~A., Mathur, S., {et~al.} 2014, \apjl, 790,
  L23

\bibitem[{Dobler {et~al.}(2006)Dobler, Stix, \& Brandenburg}]{Dobler:2006ha}
Dobler, W., Stix, M., \& Brandenburg, A. 2006, \apj, 638, 336

\bibitem[{Donahue {et~al.}(1996)Donahue, Saar, \& Baliunas}]{Donahue:1996ch}
Donahue, R.~A., Saar, S.~H., \& Baliunas, S.~L. 1996, Astrophysical Journal
  v.466, 466, 384

\bibitem[{Durney(1989)}]{Durney:1989kq}
Durney, B.~R. 1989, \apj, 338, 509

\bibitem[{Durney(1999)}]{Durney:1999du}
---. 1999, \apj, 511, 945

\bibitem[{Ekstr{\"o}m {et~al.}(2012)Ekstr{\"o}m, Georgy, Eggenberger, Meynet,
  Mowlavi, Wyttenbach, Granada, Decressin, Hirschi, Frischknecht, Charbonnel,
  \& Maeder}]{Ekstrom:2012ke}
Ekstr{\"o}m, S., Georgy, C., Eggenberger, P., {et~al.} 2012, \aap, 537, A146

\bibitem[{Elliott {et~al.}(2000)Elliott, Miesch, \& Toomre}]{Elliott:2000kp}
Elliott, J.~R., Miesch, M.~S., \& Toomre, J. 2000, \apj, 533, 546

\bibitem[{Fan \& Fang(2014)}]{Fan:2014ct}
Fan, Y., \& Fang, F. 2014, \apj, 789, 35

\bibitem[{Featherstone {et~al.}(2009)Featherstone, Browning, Brun, \&
  Toomre}]{Featherstone:2009ft}
Featherstone, N.~A., Browning, M.~K., Brun, A.~S., \& Toomre, J. 2009, \apj,
  705, 1000

\bibitem[{Featherstone \& Miesch(2015)}]{Featherstone:2015bv}
Featherstone, N.~A., \& Miesch, M.~S. 2015, \apj, 804, 67

\bibitem[{Folsom {et~al.}(2016)Folsom, Petit, Bouvier, L{\`e}bre, Amard,
  Palacios, Morin, Donati, Jeffers, Marsden, \& Vidotto}]{Folsom:2016dl}
Folsom, C.~P., Petit, P., Bouvier, J., {et~al.} 2016, \mnras, 457, 580

\bibitem[{Garaud \& Bodenheimer(2010)}]{Garaud:2010kd}
Garaud, P., \& Bodenheimer, P. 2010, \apj, 719, 313

\bibitem[{Garcia {et~al.}(2014)Garcia, Ceillier, Salabert, Mathur, van Saders,
  Pinsonneault, Ballot, Beck, Bloemen, Campante, Davies, do~Nascimento, Mathis,
  Metcalfe, Nielsen, Su{\'a}rez, Chaplin, Jim{\'e}nez, \&
  Karoff}]{Garcia:2014ds}
Garcia, R.~A., Ceillier, T., Salabert, D., {et~al.} 2014, \aap, 572, A34

\bibitem[{Gastine {et~al.}(2014{\natexlab{a}})Gastine, Heimpel, \&
  Wicht}]{Gastine:2014da}
Gastine, T., Heimpel, M., \& Wicht, J. 2014{\natexlab{a}}, Physics of the Earth
  and Planetary Interiors, 232, 36

\bibitem[{Gastine {et~al.}(2013)Gastine, Wicht, \& Aurnou}]{Gastine:2013fg}
Gastine, T., Wicht, J., \& Aurnou, J.~M. 2013, Icarus, 225, 156

\bibitem[{Gastine {et~al.}(2014{\natexlab{b}})Gastine, Yadav, Morin, Reiners,
  \& Wicht}]{Gastine:2014jr}
Gastine, T., Yadav, R.~K., Morin, J., Reiners, A., \& Wicht, J.
  2014{\natexlab{b}}, \mnras, 438, L76

\bibitem[{Ghizaru {et~al.}(2010)Ghizaru, Charbonneau, \&
  Smolarkiewicz}]{Ghizaru:2010im}
Ghizaru, M., Charbonneau, P., \& Smolarkiewicz, P.~K. 2010, \apjl, 715, L133

\bibitem[{Gilman(1983)}]{Gilman:1983dx}
Gilman, P.~A. 1983, \apjs, 53, 243

\bibitem[{Gilman \& Glatzmaier(1981)}]{Gilman:1981eh}
Gilman, P.~A., \& Glatzmaier, G.~A. 1981, \apjs, 45, 335

\bibitem[{Gilman \& Miller(1981)}]{Gilman:1981cf}
Gilman, P.~A., \& Miller, J. 1981, \apjs, 46, 211

\bibitem[{Glatzmaier(1984)}]{Glatzmaier:1984jh}
Glatzmaier, G.~A. 1984, J. Comp. Phys., 55, 461

\bibitem[{Glatzmaier \& Gilman(1982)}]{Glatzmaier:1982io}
Glatzmaier, G.~A., \& Gilman, P.~A. 1982, \apj, 256, 316

\bibitem[{Guerrero {et~al.}(2016)Guerrero, Smolarkiewicz, de~Gouveia Dal~Pino,
  Kosovichev, \& Mansour}]{Guerrero:2016cz}
Guerrero, G., Smolarkiewicz, P.~K., de~Gouveia Dal~Pino, E.~M., Kosovichev,
  A.~G., \& Mansour, N.~N. 2016, \apj, 819, 104

\bibitem[{Guerrero {et~al.}(2013)Guerrero, Smolarkiewicz, Kosovichev, \&
  Mansour}]{Guerrero:2013hb}
Guerrero, G., Smolarkiewicz, P.~K., Kosovichev, A.~G., \& Mansour, N.~N. 2013,
  \apj, 779, 176

\bibitem[{Haber {et~al.}(2002)Haber, Hindman, Toomre, Bogart, Larsen, \&
  Hill}]{Haber:2002ib}
Haber, D.~A., Hindman, B.~W., Toomre, J., {et~al.} 2002, \apj, 570, 855

\bibitem[{Jones {et~al.}(2011)Jones, Boronski, Brun, Glatzmaier, Gastine,
  Miesch, \& Wicht}]{Jones:2011in}
Jones, C.~A., Boronski, P., Brun, A.~S., {et~al.} 2011, Icarus, 216, 120

\bibitem[{Jones {et~al.}(2009)Jones, Kuzanyan, \& Mitchell}]{Jones:2009ko}
Jones, C.~A., Kuzanyan, K.~M., \& Mitchell, R.~H. 2009, \jfm, 634, 291

\bibitem[{Jouve {et~al.}(2010)Jouve, Brown, \& Brun}]{Jouve:2010jl}
Jouve, L., Brown, B.~P., \& Brun, A.~S. 2010, \aap, 509, 32

\bibitem[{K{\"a}pyl{\"a} {et~al.}(2014)K{\"a}pyl{\"a}, K{\"a}pyl{\"a}, \&
  Brandenburg}]{Kapyla:2014jr}
K{\"a}pyl{\"a}, P.~J., K{\"a}pyl{\"a}, M.~J., \& Brandenburg, A. 2014, \aap,
  570, A43

\bibitem[{K{\"a}pyl{\"a} {et~al.}(2012)K{\"a}pyl{\"a}, Mantere, \&
  Brandenburg}]{Kapyla:2012dg}
K{\"a}pyl{\"a}, P.~J., Mantere, M.~J., \& Brandenburg, A. 2012, \apj, 755, L22

\bibitem[{K{\"a}pyl{\"a} {et~al.}(2013)K{\"a}pyl{\"a}, Mantere, Cole, Warnecke,
  \& Brandenburg}]{Kapyla:2013gr}
K{\"a}pyl{\"a}, P.~J., Mantere, M.~J., Cole, E., Warnecke, J., \& Brandenburg,
  A. 2013, \apj, 778, 41

\bibitem[{K{\"a}pyl{\"a} {et~al.}(2011)K{\"a}pyl{\"a}, Mantere, Guerrero,
  Brandenburg, \& Chatterjee}]{Kapyla:2011kr}
K{\"a}pyl{\"a}, P.~J., Mantere, M.~J., Guerrero, G., Brandenburg, A., \&
  Chatterjee, P. 2011, \aap, 531, A162

\bibitem[{Karak {et~al.}(2015)Karak, K{\"a}pyl{\"a}, K{\"a}pyl{\"a},
  Brandenburg, Olspert, \& Pelt}]{Karak:2015dw}
Karak, B.~B., K{\"a}pyl{\"a}, P.~J., K{\"a}pyl{\"a}, M.~J., {et~al.} 2015,
  \aap, 576, A26

\bibitem[{Kawaler(1988)}]{Kawaler:1988fi}
Kawaler, S.~D. 1988, \apj, 333, 236

\bibitem[{Kippenhahn \& Weigert(1994)}]{Kippenhahn:1994tm}
Kippenhahn, R., \& Weigert, A. 1994, {Stellar structure and evolution},
  springer verlag edn. (Springer Verlag)

\bibitem[{{K{\"u}ker} \& {R{\"u}diger}(2007)}]{2007AN....328.1050K}
{K{\"u}ker}, M., \& {R{\"u}diger}, G. 2007, Astronomische Nachrichten, 328,
  1050

\bibitem[{{K{\"u}ker} {et~al.}(2011){K{\"u}ker}, {R{\"u}diger}, \&
  {Kitchatinov}}]{2011A&A...530A..48K}
{K{\"u}ker}, M., {R{\"u}diger}, G., \& {Kitchatinov}, L.~L. 2011, \aap, 530,
  A48

\bibitem[{Landin {et~al.}(2010)Landin, Mendes, \& {Vaz, L. P.
  R.}}]{Landin:2010fi}
Landin, N.~R., Mendes, L. T.~S., \& {Vaz, L. P. R.} 2010, \aap, 510, A46

\bibitem[{Lawson {et~al.}(2015)Lawson, Strugarek, \&
  Charbonneau}]{Lawson:2015fq}
Lawson, N., Strugarek, A., \& Charbonneau, P. 2015, \apj, 813, 95

\bibitem[{{Maeder} \& {Meynet}(1989)}]{1989A&A...210..155M}
{Maeder}, A., \& {Meynet}, G. 1989, \aap, 210, 155

\bibitem[{Masada {et~al.}(2013)Masada, Yamada, \& Kageyama}]{Masada:2013fc}
Masada, Y., Yamada, K., \& Kageyama, A. 2013, \apj, 778, 11

\bibitem[{Matt {et~al.}(2015)Matt, Brun, Baraffe, Bouvier, \&
  Chabrier}]{Matt:2015cb}
Matt, S.~P., Brun, A.~S., Baraffe, I., Bouvier, J., \& Chabrier, G. 2015,
  \apjl, 799, L23

\bibitem[{Matt {et~al.}(2011)Matt, Do~Cao, Brown, \& Brun}]{Matt:2011jl}
Matt, S.~P., Do~Cao, O., Brown, B.~P., \& Brun, A.~S. 2011, Astro. Nach., 332,
  897

\bibitem[{Matt {et~al.}(2012)Matt, MacGregor, Pinsonneault, \&
  Greene}]{Matt:2012ib}
Matt, S.~P., MacGregor, K.~B., Pinsonneault, M.~H., \& Greene, T.~P. 2012,
  \apjl, 754, L26

\bibitem[{McIntyre(2007)}]{McIntyre:2007ww}
McIntyre, M.~E. 2007, The Solar Tachocline, 183

\bibitem[{Meibom {et~al.}(2015)Meibom, Barnes, Platais, Gilliland, Latham, \&
  Mathieu}]{Meibom:2015if}
Meibom, S., Barnes, S.~A., Platais, I., {et~al.} 2015, \nat, 517, 589

\bibitem[{{Messina} \& {Guinan}(2003)}]{2003A&A...409.1017M}
{Messina}, S., \& {Guinan}, E.~F. 2003, \aap, 409, 1017

\bibitem[{{Metcalfe} {et~al.}(2016){Metcalfe}, {Egeland}, \& {van
  Saders}}]{2016ApJ...826L...2M}
{Metcalfe}, T.~S., {Egeland}, R., \& {van Saders}, J. 2016, \apjl, 826, L2

\bibitem[{{Meynet} {et~al.}(1993){Meynet}, {Mermilliod}, \&
  {Maeder}}]{1993A&AS...98..477M}
{Meynet}, G., {Mermilliod}, J.-C., \& {Maeder}, A. 1993, \aaps, 98, 477

\bibitem[{Miesch(2005)}]{Miesch:2005wz}
Miesch, M.~S. 2005, \lrsp, 2, 1

\bibitem[{Miesch {et~al.}(2006)Miesch, Brun, \& Toomre}]{Miesch:2006iz}
Miesch, M.~S., Brun, A.~S., \& Toomre, J. 2006, \apj, 641, 618

\bibitem[{Miesch {et~al.}(2000)Miesch, Elliott, Toomre, Clune, Glatzmaier, \&
  Gilman}]{Miesch:2000gs}
Miesch, M.~S., Elliott, J.~R., Toomre, J., {et~al.} 2000, \apj, 532, 593

\bibitem[{Miesch \& Hindman(2011)}]{Miesch:2011cg}
Miesch, M.~S., \& Hindman, B.~W. 2011, \apj, 743, 79

\bibitem[{Mitra-Kraev \& Thompson(2007)}]{MitraKraev:2007ej}
Mitra-Kraev, U., \& Thompson, M.~J. 2007, Astro. Nach., 328, 1009

\bibitem[{Moffatt(1978)}]{Moffatt:1978tc}
Moffatt, H.~K. 1978, {Magnetic field generation in electrically conducting
  fluids} (Cambridge)

\bibitem[{Morel(1997)}]{Morel:1997gy}
Morel, P. 1997, A{\&}A Supp. Series, 124, 597

\bibitem[{Nelson {et~al.}(2013)Nelson, Brown, Brun, Miesch, \&
  Toomre}]{Nelson:2013fa}
Nelson, N.~J., Brown, B.~P., Brun, A.~S., Miesch, M.~S., \& Toomre, J. 2013,
  \apj, 762, 73

\bibitem[{Noyes {et~al.}(1984)Noyes, Weiss, \& Vaughan}]{Noyes:1984bp}
Noyes, R.~W., Weiss, N.~O., \& Vaughan, A.~H. 1984, \apj, 287, 769

\bibitem[{Parker(1955{\natexlab{a}})}]{Parker:1955km}
Parker, E.~N. 1955{\natexlab{a}}, \apj, 122, 293

\bibitem[{Parker(1955{\natexlab{b}})}]{Parker:1955gc}
---. 1955{\natexlab{b}}, \apj, 121, 491

\bibitem[{Parker(1958)}]{Parker:1958dn}
---. 1958, \apj, 128, 664

\bibitem[{Pedlosky(1987)}]{Pedlosky:1987vt}
Pedlosky, J. 1987, {Geophysical fluid dynamics} (Springer)

\bibitem[{Pizzolato {et~al.}(2003)Pizzolato, Maggio, Micela, Sciortino, \&
  Ventura}]{Pizzolato:2003ga}
Pizzolato, N., Maggio, A., Micela, G., Sciortino, S., \& Ventura, P. 2003,
  \aap, 397, 147

\bibitem[{Racine {et~al.}(2011)Racine, Charbonneau, Ghizaru, Bouchat, \&
  Smolarkiewicz}]{Racine:2011gh}
Racine, {\'E}., Charbonneau, P., Ghizaru, M., Bouchat, A., \& Smolarkiewicz,
  P.~K. 2011, \apj, 735, 46

\bibitem[{{Rauer} {et~al.}(2014){Rauer}, {Catala}, {Aerts}, {Appourchaux},
  {Benz}, {Brandeker}, {Christensen-Dalsgaard}, {Deleuil}, {Gizon}, {Goupil},
  {G{\"u}del}, {Janot-Pacheco}, {Mas-Hesse}, {Pagano}, {Piotto}, {Pollacco},
  {Santos}, {Smith}, {Su{\'a}rez}, {Szab{\'o}}, {Udry}, {Adibekyan}, {Alibert},
  {Almenara}, {Amaro-Seoane}, {Eiff}, {Asplund}, {Antonello}, {Barnes},
  {Baudin}, {Belkacem}, {Bergemann}, {Bihain}, {Birch}, {Bonfils}, {Boisse},
  {Bonomo}, {Borsa}, {Brand{\~a}o}, {Brocato}, {Brun}, {Burleigh}, {Burston},
  {Cabrera}, {Cassisi}, {Chaplin}, {Charpinet}, {Chiappini}, {Church},
  {Csizmadia}, {Cunha}, {Damasso}, {Davies}, {Deeg}, {D{\'{\i}}az}, {Dreizler},
  {Dreyer}, {Eggenberger}, {Ehrenreich}, {Eigm{\"u}ller}, {Erikson}, {Farmer},
  {Feltzing}, {de Oliveira Fialho}, {Figueira}, {Forveille}, {Fridlund},
  {Garc{\'{\i}}a}, {Giommi}, {Giuffrida}, {Godolt}, {Gomes da Silva},
  {Granzer}, {Grenfell}, {Grotsch-Noels}, {G{\"u}nther}, {Haswell}, {Hatzes},
  {H{\'e}brard}, {Hekker}, {Helled}, {Heng}, {Jenkins}, {Johansen},
  {Khodachenko}, {Kislyakova}, {Kley}, {Kolb}, {Krivova}, {Kupka}, {Lammer},
  {Lanza}, {Lebreton}, {Magrin}, {Marcos-Arenal}, {Marrese}, {Marques},
  {Martins}, {Mathis}, {Mathur}, {Messina}, {Miglio}, {Montalban}, {Montalto},
  {Monteiro}, {Moradi}, {Moravveji}, {Mordasini}, {Morel}, {Mortier},
  {Nascimbeni}, {Nelson}, {Nielsen}, {Noack}, {Norton}, {Ofir}, {Oshagh},
  {Ouazzani}, {P{\'a}pics}, {Parro}, {Petit}, {Plez}, {Poretti}, {Quirrenbach},
  {Ragazzoni}, {Raimondo}, {Rainer}, {Reese}, {Redmer}, {Reffert},
  {Rojas-Ayala}, {Roxburgh}, {Salmon}, {Santerne}, {Schneider}, {Schou},
  {Schuh}, {Schunker}, {Silva-Valio}, {Silvotti}, {Skillen}, {Snellen}, {Sohl},
  {Sousa}, {Sozzetti}, {Stello}, {Strassmeier}, {{\v S}vanda}, {Szab{\'o}},
  {Tkachenko}, {Valencia}, {Van Grootel}, {Vauclair}, {Ventura}, {Wagner},
  {Walton}, {Weingrill}, {Werner}, {Wheatley}, \&
  {Zwintz}}]{2014ExA....38..249R}
{Rauer}, H., {Catala}, C., {Aerts}, C., {et~al.} 2014, Experimental Astronomy,
  38, 249

\bibitem[{Reiners(2012)}]{Reiners:2012ug}
Reiners, A. 2012, \lrsp, 9, 1

\bibitem[{Reinhold \& Arlt(2015)}]{Reinhold:2015kx}
Reinhold, T., \& Arlt, R. 2015, \aap, 576, A15

\bibitem[{{Reinhold} \& {Gizon}(2015)}]{2015A&A...583A..65R}
{Reinhold}, T., \& {Gizon}, L. 2015, \aap, 583, A65

\bibitem[{Reinhold \& Reiners(2013)}]{Reinhold:2013eo}
Reinhold, T., \& Reiners, A. 2013, \aap, 557, A11

\bibitem[{Reinhold {et~al.}(2013)Reinhold, Reiners, \& Basri}]{Reinhold:2013iz}
Reinhold, T., Reiners, A., \& Basri, G. 2013, \aap, 560, A4

\bibitem[{Rempel(2004)}]{Rempel:2004ek}
Rempel, M. 2004, \apj, 607, 1046

\bibitem[{Rempel(2005)}]{Rempel:2005fk}
---. 2005, \apj, 631, 1286

\bibitem[{Rempel(2006)}]{Rempel:2006eh}
---. 2006, \apj, 647, 662

\bibitem[{R{\'e}ville {et~al.}(2015)R{\'e}ville, Brun, Matt, Strugarek, \&
  Pinto}]{Reville:2015bb}
R{\'e}ville, V., Brun, A.~S., Matt, S.~P., Strugarek, A., \& Pinto, R.~F. 2015,
  \apj, 798, 116

\bibitem[{Rogers {et~al.}(2006)Rogers, Glatzmaier, \& Jones}]{Rogers:2006ks}
Rogers, T.~M., Glatzmaier, G.~A., \& Jones, C.~A. 2006, \apjs, 653, 765

\bibitem[{{Roxburgh}(1978)}]{1978A&A....65..281R}
{Roxburgh}, I.~W. 1978, \aap, 65, 281

\bibitem[{Saar(2009)}]{Saar:2009tb}
Saar, S.~H. 2009, Solar-Stellar Dynamos as Revealed by Helio- and
  Asteroseismology: GONG 2008/SOHO 21 ASP Conference Series, 416, 375

\bibitem[{Saar \& Brandenburg(1999)}]{Saar:1999dg}
Saar, S.~H., \& Brandenburg, A. 1999, \apj, 524, 295

\bibitem[{Schatzman(1962)}]{Schatzman:1962vc}
Schatzman, E. 1962, Annales d'Astrophysique, 25, 18

\bibitem[{{Schrijver}(2001)}]{2001ApJ...547..475S}
{Schrijver}, C.~J. 2001, \apj, 547, 475

\bibitem[{Simitev {et~al.}(2015)Simitev, Kosovichev, \& Busse}]{Simitev:2015fc}
Simitev, R.~D., Kosovichev, A.~G., \& Busse, F.~H. 2015, \apj, 810, 80

\bibitem[{Skumanich(1972)}]{Skumanich:1972fq}
Skumanich, A. 1972, \apj, 171, 565

\bibitem[{Spiegel \& Zahn(1992)}]{Spiegel:1992tr}
Spiegel, E.~A., \& Zahn, J.-P. 1992, \aap, 265, 106

\bibitem[{Takehiro {et~al.}(2013)Takehiro, Sasaki, Hayashi, \&
  Yamada}]{2013ASPC..479..285T}
Takehiro, S., Sasaki, Y., Hayashi, Y.~Y., \& Yamada, M. 2013, in Progress in
  Physics of the Sun and Stars: A New Era in Helio- and Asteroseismology.
  Proceedings of a Fujihara Seminar held 25-29 November, 285--

\bibitem[{van Saders {et~al.}(2016)van Saders, Ceillier, Metcalfe, Aguirre,
  Pinsonneault, Garcia, Mathur, \& Davies}]{vanSaders:2016cr}
van Saders, J.~L., Ceillier, T., Metcalfe, T.~S., {et~al.} 2016, \nat, 529, 181

\bibitem[{{Varela} {et~al.}(2016){Varela}, {Strugarek}, \& {Brun}}]{Varela16}
{Varela}, J., {Strugarek}, A., \& {Brun}, A.~S. 2016, Advances in Space
  Research, 58, 1507

\bibitem[{Vasil {et~al.}(2013)Vasil, Lecoanet, Brown, Wood, \&
  Zweibel}]{Vasil:2013ij}
Vasil, G.~M., Lecoanet, D., Brown, B.~P., Wood, T.~S., \& Zweibel, E.~G. 2013,
  \apj, 773, 169

\bibitem[{Vidotto {et~al.}(2014)Vidotto, Gregory, Jardine, Donati, Petit,
  Morin, Folsom, Bouvier, Cameron, Hussain, Marsden, Waite, Fares, Jeffers, \&
  do~Nascimento}]{Vidotto:2014ba}
Vidotto, A.~A., Gregory, S.~G., Jardine, M., {et~al.} 2014, \mnras, 441, 2361

\bibitem[{Vidotto {et~al.}(2016)Vidotto, Donati, Jardine, See, Petit, Boisse,
  Boro~Saikia, H{\'e}brard, Jeffers, Marsden, \& Morin}]{Vidotto:2016ej}
Vidotto, A.~A., Donati, J.-F., Jardine, M., {et~al.} 2016, \mnras, 455, L52

\bibitem[{{Wang} {et~al.}(1989){Wang}, {Nash}, \&
  {Sheeley}}]{1989Sci...245..712W}
{Wang}, Y.-M., {Nash}, A.~G., \& {Sheeley}, Jr., N.~R. 1989, Science, 245, 712

\bibitem[{Weber \& Davis(1967)}]{Weber:1967kx}
Weber, E.~J., \& Davis, L.~J. 1967, \apjs, 148, 217

\bibitem[{Weiss(1994)}]{Weiss:1994tz}
Weiss, N.~O. 1994, Lectures on Solar and Planetary Dynamos. Edited by M. R. E.
  Proctor and A. D. Gilbert. ISBN 0 521 46142 1 and ISBN 0 521 46704 7.
  Published by Cambridge University Press, 59

\bibitem[{Wilson(1978)}]{Wilson:1978is}
Wilson, O.~C. 1978, \apj, 226, 379

\bibitem[{{Wright} {et~al.}(2011){Wright}, {Drake}, {Mamajek}, \&
  {Henry}}]{2011ApJ...743...48W}
{Wright}, N.~J., {Drake}, J.~J., {Mamajek}, E.~E., \& {Henry}, G.~W. 2011,
  \apj, 743, 48

\bibitem[{Zahn(1991)}]{Zahn:1991uz}
Zahn, J.-P. 1991, \aap, 252, 179

\bibitem[{Zahn(1992)}]{Zahn:1992vi}
---. 1992, \aap, 265, 115

\bibitem[{Zahn {et~al.}(1997)Zahn, Talon, \& Matias}]{Zahn:1997ui}
Zahn, J.-P., Talon, S., \& Matias, J. 1997, \aap, 322, 320

\bibitem[{Zhao {et~al.}(2013)Zhao, Bogart, Kosovichev, Duvall, \&
  Hartlep}]{Zhao:2013kz}
Zhao, J., Bogart, R.~S., Kosovichev, A.~G., Duvall, T. L.~J., \& Hartlep, T.
  2013, \apjl, 774, L29

\end{thebibliography}
%\bibliographystyle{/Users/smatt/latex/bibstyles/apj}

\appendix

\section{Model ingredient parameters}

In Table \ref{tablediff} we list some of the parameters used in the simulations.
\begin{table*}[!ht]
\begin{center}
\caption{Diffusivity profile and nuclear heating source parameters}\label{tablediff}
\vspace{0.2cm}
%\begin{tabular}{||p{1.8cm}*{1}{||c} cccc ||}
\begin{tabular}{ccccccccc}
\tableline
\tableline
\\ [-1.5ex]
 Mass & Name & $\nu_{top}$ & $r_t$ & $\sigma_t$ & $\epsilon_0$ & $n_c$ & a & b\\ [0.5ex]
 $(M_{\odot})$ &  & $(cm^2 s^{-1})$ & $(cm)$ & $(cm)$ & & & &\\   [0.8ex]
\tableline
\tableline
\\ [-1.5ex]
  0.5 & M05 S & $10.5\times10^{11}$ & $1.65\times10^{10}$ & $4.0\times10^{8}$ & $2.20\times10^{-7}$ & 7.3 & 9.78e-04 & -3.57e-09 \\
  & M05 R1 & $4.47\times10^{11}$ & & & & & &\\
  & M05 R3 & $2.58\times10^{11}$ & & & & & &\\
  & M05 R5 & $2.00\times10^{11}$ & & & & & &\\ [0.5ex]
\hline
\\ [-2ex]
  0.7 & M07 S & $5.31\times10^{12}$ & $2.67\times10^{10}$ & $1.0\times10^{9}$ & $5.56\times10^{-9}$ & 8.8 & 9.59e-03 & -9.56e-09 \\
  & M07 R1 & $2.91\times10^{12}$ & & & & & &\\
  & M07 R3 & $1.68\times10^{12}$ & & & & & &\\
  & M07 R5 & $1.30\times10^{12}$ & & & & & &\\ [0.5ex]
\hline
\\ [-2ex]
  0.9 & M09 S & $1.43\times10^{13}$ & $3.94\times10^{10}$ & $7.0\times10^{8}$ & & & 1.37e-02 & -3.78e-08 \\
  & M09 R1 & $1.01\times10^{13}$ & & & & & &\\
  & M09 R3 & $5.81\times10^{12}$ & & & & & &\\
  & M09 R5 & $4.50\times10^{12}$ & & & & & &\\ [0.5ex]
 \hline
 \\ [-2ex]
  1.1 & M11 R1 & $3.80\times10^{13}$ & $6.20\times10^{10}$ & $8.0\times10^{8}$ & & & 1.09e-02 & -3.01e-07 \\ 
  & M11 R3 & $2.20\times10^{13}$ & & & & & &\\
  & M11 R5 & $1.70\times10^{13}$ & & & & & &\\ [0.5ex]
\hline
 \tableline
 \tableline
\end{tabular}
\end{center}
\end{table*}

\section{Rossby numbers}
\label{sec:rossby-numbers}

There are multiple definitions of the Rossby number in the literature that quantify the influence of
rotation on the dynamics of the system. We chose to use the \textit{fluid} Rossby number $R_{of}$,
which is a direct comparison of the advection term and the Coriolis force in the Navier-Stokes
equation. It is defined by

\begin{equation}
  \label{eq:rossby_fluid_def}
  R_{\rm of} = \frac{\tilde{\omega}}{2\Omega_\star} \sim \frac{\tilde{v}}{2\Omega_\star R_\star}  \, ,
\end{equation}
where $\tilde{\omega}$ is the rms vorticity at mid-depth in the
convection zone. We choose to evaluate the Rossby number at
  mid-depth of the convection zone as it is close to the location of
  the maximum angular momentum transfers shown in Figure
  \ref{Amombal}. We also tested a different definition of the Rossby
  number for which $\tilde{v}$ is averaged over the all convection
  zone rather evaluated at mid-depth and did not find any significant
  difference in the analysis reported in this work.

The \textit{stellar} Rossby number, which is often used in the literature, corresponds to the ratio
of the rotation period of the star ($P_{\rm rot}=2\pi/\Omega_*$), to the convective over-turning
time of convection ($\tau_{\rm conv} = d_{\rm CZ} /\tilde{v}_r$, with $d_{\rm CZ}$ the thickness of
the convective envelope). With the definition of the convective over-turning time being different
for various authors, it is generally deduced from stellar structure models using mixing-length at
various locations in the convective envelope, leading to some confusion in its exact definition
\citep{Landin:2010fi}. In our simulations, we directly use the mid-depth value of the radial
velocity. It is defined by

\begin{equation}
  \label{eq:rossby_stellar_def}
  R_{\rm os} = \frac{P_{\rm rot}}{\tau_{\rm conv}}\, .
\end{equation}

The \textit{convective} Rossby number $R_{\rm oc}$, first introduced by \citet{Gilman:1981eh}, is a combination of the
Taylor, Rayleigh and Prandtl numbers and is defined by

\begin{equation}
  \label{eq:rossby_convective_def}
  R_{\rm oc} = \sqrt{\frac{R_a}{T_aP_r}}\, .
\end{equation}

Finally, a \textit{modified} Rossby number $R_{\rm ol}$ was introduced by \citet{Christensen:2006jo} to take into
account the characteristic length scale of the flow rather than the shell thickness $d_{\rm CZ}$. It is defined by

\begin{equation}
  \label{eq:rossby_modified_def}
  R_{\rm ol} = \frac{U}{2\Omega_\star L} \frac{\bar{l}_u}{\pi}\, ,
\end{equation}
where $U$ is the rms velocity at mid-depth, $L$ the size of the
convective enveloppe, and
the characteristic length scale $\bar{l}_u$ is defined as
\begin{equation}
  \label{eq:lu_def}
  \bar{l}_u = \frac{\sum_l l \left\langle {\bf v}_l \cdot {\bf
        v}_l\right\rangle}{\sum_l  \left\langle {\bf v}_l \cdot {\bf
        v}_l\right\rangle}\, ,
\end{equation}
where ${\bf v}_l$ is the velocity field at scale $l$ in the spherical
harmonics spectral space.

\section{Thermal wind balance}
\label{sec:thermal-wind-balance}

The complete thermal wind balance equation can be derived from the vorticity equation
\begin{eqnarray}\label{eq:vort}
\frac{\p \vort}{\p t}&= &(\vort_a\cdot\nab){\bf v} - ({\bf v}\cdot\nab)\vort_a - \vort_a(\nab\cdot{\bf v})\\\nonumber
 & + &\frac{1}{\rb^2}\nab\rb\times\nab P - \curl\left(\frac{\rho g}{\rb}\uvr\right) - \curl(\rbi\nab\cdot\mbox{\boldmath $\cal D$}), 
\end{eqnarray}
with $\vort_a=\curl {\bf v} + 2\Om$ the absolute vorticity and $\vort=\curl {\bf v}$ the vorticity in the rotating
frame.  Averaging the zonal component of this vorticity equation over longitude and time and assuming a statistically
stationary state yields the general equation for force balance in the meridional plane:
\begin{eqnarray}\label{eq:TWfull}
2\Omega_*\frac{\p \langle v_{\phi}\rangle}{\p
  z}&=&\underbrace{-\langle (\vort\cdot\nab)v_{\phi} -
        \frac{\omega_{\phi}v_r}{r} -
        \frac{\omega_{\phi}v_{\theta}\cot\theta}{r}\rangle}_{{\rm
        Stretching}\; \mathcal{S}}\nonumber \\
& +& \underbrace{\langle({\bf v}\cdot\nab)\omega_{\phi}
     +\frac{v_{\phi}\omega_r}{r} +
     \frac{v_{\phi}\omega_{\theta}\cot\theta}{r}\rangle}_{{\rm
     Advection}\; \mathcal{A}} \nonumber\\ 
&-& \underbrace{\langle \omega_{\phi}v_r \rangle\frac{d\ln \rb}{d
    r}}_{{\rm Compressibility}\; \mathcal{C}}
    +\underbrace{\frac{1}{r}\left[\frac{\p}{\p r}(r \langle {\cal
    A}_{\theta}\rangle) - \frac{\p}{\p \theta}\langle {\cal A}_r
    \rangle \right]}_{{\rm Viscous\; stresses}\; \mathcal{V}} \\
&+& \underbrace{ \frac{g}{r c_p}\frac{\p \langle S\rangle}{\p
    \theta}+\frac{1}{r \rb c_p}\frac{d\bar{S}}{dr}\frac{\p \langle
    P\rangle}{\p\theta}}_{{\rm Baroclinicity}\; \mathcal{B}} \nonumber
\end{eqnarray}
where $\displaystyle \frac{\p}{\p z}=\cos\theta\frac{\p}{\p r}-\frac{\sin\theta}{r}\frac{\p}{\p \theta}$ and 
\begin{eqnarray}
%\begin{multline}
\langle{\cal A}_r\rangle&=& \rbi\langle\left[\frac{1}{r^2}\frac{\p(r^2{\cal D}_{rr})}{\p r}+\frac{1}{r\sin\theta}\frac{\p(\sin\theta{\cal D}_{\theta r})}{\p\theta} - \frac{ {\cal D}_{\theta \theta} + {\cal D}_{\phi \phi}}{r} \right]\rangle , \nonumber \\
\langle{\cal A}_{\theta}\rangle&=& \rbi\langle\left[\frac{1}{r^2}\frac{\p(r^2{\cal D}_{r\theta})}{\p r}+\frac{1}{r\sin\theta}\frac{\p(\sin\theta{\cal D}_{\theta \theta})}{\p\theta} \right] \\
&+& \rbi\left[ \frac{ {\cal D}_{\theta r} - cot \theta {\cal D}_{\phi \phi}}{r} \right]\rangle . \nonumber
%\end{multline}
\end{eqnarray}

\section{Kinetic energy balance for the differential rotation}
\label{sec:kinet-energy-balance}

We recall the evolution equation of the kinetic energy associated with the differential rotation:
\begin{equation}
  \label{eq:ke_balance_appendix}
  \partial_t\left({\rm DRKE}\right)= Q_C^r + Q_C^l + Q_{R}^r + Q_{R}^l + Q_A
  + Q_V + Q_{\rm curv}\, .
\end{equation}

The various terms of Equation \ref{eq:ke_balance_appendix} are defined by

\begin{eqnarray}
  \label{eq:ke_bal_qcr}
  Q_C^r &=& -2 \left\langle \Omega \overline{\rho}\, 
  \overline{v_r}\overline{v_\varphi}\sin\theta \right
            \rangle_{r,\theta}\, , \\ 
  \label{eq:ke_bal_qcl}
  Q_C^l &=& -2 \left\langle \Omega \overline{\rho}\,
  \overline{v_\theta}\overline{v_\varphi}\cos\theta \right
            \rangle_{r,\theta} \, , \\
  \label{eq:ke_bal_qrr}
  Q_R^r &=& \left\langle\overline{\rho}\, \overline{v_r
            v_\varphi}\partial_r\overline{v_\varphi}\right\rangle_{r,\theta}
            \, , \\ 
  \label{eq:ke_bal_qrl}
  Q_R^l &=& \left\langle\overline{\rho}\, \overline{v_\theta
            v_\varphi}\partial_\theta\overline{v_\varphi}\right\rangle_{r,\theta}
            \, , \\ 
  \label{eq:ke_bal_qadv}
  Q_A &=& -\left\langle \frac{1}{r^2}\partial_r \left[
          r^2\overline{\rho}\,\overline{v_\varphi}\,\overline{v_r
          v_\varphi} \right] \right\rangle_{r,\theta} \nonumber \\ && -
         \left\langle \frac{1}{r\sin\theta}\partial_\theta\left[
          \sin\theta\overline{\rho}\,\overline{v_\varphi}\,\overline{v_\theta
          v_\varphi} \right] \right\rangle_{r,\theta} \, , \\ 
  Q_V &=& \left\langle \frac{1}{r^2}\partial_r\left(r^3\nu\bar{\rho}\,\overline{v_\varphi}\partial_r\frac{\overline{v_\varphi}}{r}\right) 
\right. \nonumber \\ && \left. + \frac{1}{r^2\sin\theta}\partial_\theta\left( \sin^2\theta \nu \bar{\rho}\,\overline{v_\varphi} \partial_\theta\frac{\overline{v_\varphi}}{\sin\theta} \right)
 \right. \nonumber \\ && \left. - \nu\bar{\rho}\left[ \left(
                         r\partial_r\frac{\overline{v_\varphi}}{r}
                         \right)^2 + \left(
                         \frac{\sin\theta}{r}\partial_\theta\frac{\overline{v_\varphi}}{\sin\theta}
                         \right)^2  \right] \right\rangle_{r,\theta}\,
                         , 
  \label{eq:ke_bal_qv} \\
Q_{\rm curv} &=& -\left\langle
                 \frac{\overline{v_\varphi}}{r\sin\theta} \left(
                 \bar{\rho}\, \overline{v_rv_\varphi}\sin\theta +
                 \bar{\rho}\,\overline{v_\theta v_\varphi}\cos\theta \right) \right\rangle_{r,\theta} \, , \label{eq:ke_bal_curv}
\end{eqnarray}
where $\left\langle . \right\rangle_{r,\theta}$ stands for the average over the meridional plane, and the overbar stands for the azimutal average.

\end{document}